\documentclass[12pt,a4paper]{article}

\usepackage{afterpage}
\usepackage{pdflscape}

\usepackage[bottom]{footmisc}
\usepackage{caption}
\usepackage[text={450pt,650pt},headheight={14.5pt},centering]{geometry}

\usepackage{lmodern}
\usepackage{graphicx}
\usepackage{subfig}
\usepackage{booktabs}
\usepackage{multirow}
\usepackage{dsfont}
\usepackage[numbers,merge]{natbib}
\usepackage{floatpag}
\usepackage{diagbox}

\usepackage[svgnames]{xcolor}
\usepackage{tikz}
\usetikzlibrary{plotmarks,calc,decorations.markings,decorations.pathmorphing,decorations.pathreplacing,patterns,matrix}

\newcommand{\kbar}{\mathchar'26\mkern-9mu k}

\usepackage[intlimits]{amsmath}
\usepackage{amssymb}
\usepackage{mathtools}
\usepackage{slashed}
\usepackage{braket}

\usepackage{hyperref}

\usepackage{xspace}

\usepackage{accents}

\numberwithin{equation}{section}

\makeatletter
 \let\old@startsection=\@startsection
 \let\oldl@section=\l@section
 \renewcommand{\@startsection}[6]{\old@startsection{#1}{#2}{#3}{#4}{#5}{#6\mathversion{bold}}}
 \renewcommand{\l@section}[2]{\oldl@section{\mathversion{bold}#1}{#2}}
\makeatother

\DeclareMathOperator{\Sym}{Sym}

\def\XXint#1#2#3{{\setbox0=\hbox{$#1{#2#3}{\int}$}
    \vcenter{\hbox{$#2#3$}}\kern-.5\wd0}}

\newcommand{\AdS}{\text{AdS}}
\newcommand{\CFT}{\text{CFT}}
\newcommand{\Sphere}{\mathrm{S}}
\newcommand{\Torus}{\mathrm{T}}

\newcommand{\Smat}{\mathcal{S}}

\newcommand{\alg}[1]{\mathrm{#1}}
\newcommand{\grp}[1]{\mathrm{#1}}

\newcommand{\grpSL}{\grp{SL}}

\newcommand{\algSU}{\alg{su}}
\newcommand{\grpSU}{\grp{SU}}

\newcommand{\grpU}{\grp{U}}

\newcommand{\algPSU}{\alg{psu}}

\newcommand{\Integers}{\mathbb{Z}}
\newcommand{\Naturals}{\mathbb{N}}
\newcommand{\Reals}{\mathbb{R}}

\newcommand{\superN}{\mathcal{N}}

\newcommand{\ie}{\textit{i.e.}\xspace}
\newcommand{\eg}{\textit{e.g.}\xspace}
\newcommand{\cf}{\textit{cf.}\xspace}
\newcommand{\ce}{\text{ce}}

\renewcommand{\Re}{\mathop{\mathrm{Re}}}
\renewcommand{\Im}{\mathop{\mathrm{Im}}}
\DeclareMathOperator{\sign}{sign}

\newcommand{\sL}{\mbox{\tiny L}}
\newcommand{\sR}{\mbox{\tiny R}}
\newcommand{\sI}{\mbox{\tiny I}}
\newcommand{\sJ}{\mbox{\tiny J}}
\newcommand{\sLL}{\mathrlap{\sL\sL}\protect\phantom{\sR\sR}}
\newcommand{\sLR}{\mathrlap{\sL\sR}\protect\phantom{\sR\sR}}
\newcommand{\sRL}{\mathrlap{\sR\sL}\protect\phantom{\sR\sR}}
\newcommand{\sRR}{\sR\sR}
\newcommand{\sIJ}{\sI\sJ}

\newcommand{\BES}{\text{BES}}
\newcommand{\HL}{\text{HL}}

\newcommand{\curvearrowurl}{\curvearrowleft}

\newcommand{\curvearrowdrl}{\reflectbox{\rotatebox[origin=c]{180}{$\curvearrowleft$}}}
\newcommand{\curvearrowudrl}{\mathrlap{\raisebox{-3pt}{\curvearrowdrl}}\raisebox{3pt}{$\curvearrowurl$}}

\newcommand{\intudrl}{\;{\int \negthickspace \negthickspace \negthickspace \negthinspace \negthinspace \curvearrowudrl}\,}

\usepackage{floatpag}
\floatpagestyle{plain}

\begin{document}

\thispagestyle{empty}

% \begin{flushright}\footnotesize\ttfamily
% \end{flushright}
\vspace*{2em}

\begin{center}
  \textbf{\Large\mathversion{bold} 
  Worldsheet kinematics, dressing factors and odd crossing in mixed-flux $\AdS_3$ backgrounds}

  \vspace{2em}

  \textrm{\large 
    Olof Ohlsson Sax${}^1$, Dmitrii Riabchenko${}^{2}$
    and Bogdan Stefa\'nski, jr.${}^{2}$ 
  } 

  \vspace{4em}

  \begingroup\itshape
  ${}^1$Nordita, Stockholm University and KTH Royal Institute of Technology,\\
  Hannes Alfv\'ens v\"ag 12, 114 19 Stockholm, Sweden\\[0.2cm]
  ${}^2$Centre for Mathematical Science, City, University of London,\\ Northampton Square, EC1V 0HB, London, UK\\[0.2cm]
  \par\endgroup

  \vspace{1em}

  \texttt{olof.ohlsson.sax@gmail.com, \\ dmitrii.riabchenko@city.ac.uk, \\ bogdan.stefanski.1@city.ac.uk}

  %%%%%%%% 

\end{center}

\vspace{3em}

\begin{abstract}\noindent 
String theory on $\AdS_3\times \Sphere^3\times \Torus^4$ geometries supported by a combination of NS-NS and R-R charges is believed to be integrable. We elucidate the kinematics and analytic structure of worldsheet excitations in mixed charge and pure NS-NS backgrounds, when expressed in momentum, Zhukovsky variables and the rapidity $u$ which appears in the quantum spectral curve. We discuss the relations between fundamental and bound state excitations and the role of fusion in constraining and determining the S matrices of these theories. We propose a scalar dressing factor consistent with a novel $u$-plane periodicity and comment on its close relation to the XXZ model at roots of unity. We solve the odd part of crossing and show that our solution is consistent with fusion and reduces in the relativistic limit to dressing phases previously found in the literature.
\end{abstract}

%%%%%%%%%%%%%%%%%%%%%%%%%%%%%%%%%%%%%%%%%%%%%%%%%%%%%%%%%%%%%%%%%%%%%%%%%%% 
\newpage

\tableofcontents

\newpage

\section{Introduction}
\label{sec:introduction}

The spectral problem of string theory on $\AdS_3\times \Sphere^3\times \Torus^4$ backgrounds is believed to  be integrable~\cite{Babichenko:2009dk,OhlssonSax:2011ms,Cagnazzo:2012se}.\footnote{For earlier work in this direction see~\cite{David:2008yk,David:2010yg}.} Given any set of NS-NS and R-R charges and any value of moduli~\cite{OhlssonSax:2018hgc}, an exact worldsheet S matrix satisfying the Yang-Baxter equation was constructed~\cite{Lloyd:2014bsa}, by using the centrally-extended symmetries of the magnons in this theory. This generalises the  findings for backgrounds with only R-R charge~\cite{Borsato:2013qpa,Borsato:2014exa,Borsato:2014hja}. In all  backgrounds supported by a mixture of NS-NS and R-R charge, the elementary worldsheet magnons have mass $m=1$ or $m=0$ and come in two types, denoted by L and R, in reference to their chirality in the dual $\CFT_2$. Their exact dispersion relations~\cite{Hoare:2013lja,Lloyd:2014bsa} are\footnote{Massless L and R representations are physically equivalent.}
\begin{equation}\label{eq:disp-rels-mf}
  E_{\sL}(p) = \sqrt{(m + \kbar p)^2 + 4h^2\sin^2\frac{p}{2}} , \qquad
  E_{\sR}(p) = \sqrt{(m - \kbar p)^2 + 4h^2\sin^2\frac{p}{2}} ,
\end{equation}
where 
\begin{equation} 
\kbar=\frac{k}{2\pi},
\end{equation}
with $k\in\mathbb{Z}$ the level of the  WZW term in the action. Above, $h$ is a function of the moduli of the mixed NS-NS and R-R charges, whose exact form remains to be determined, but whose leading order expressions are known~\cite{OhlssonSax:2018hgc}.\footnote{For each (supersymmetric) configuration of NS-NS and R-R charges, there is an exact string theory background. The moduli of each background are different combinations of supergravity fields determined by the attractor mechanism~\cite{Larsen:1999uk}. Of these moduli, only 4 have an effect on the closed string spectrum with zero winding and momentum along $\Torus^4$ that is under consideration here. The dependence of the spectrum on thrse moduli is entirely contained in the function $h$~\cite{OhlssonSax:2018hgc}.} When $\kbar$ is non-zero, the backgrounds have been referred to in the literature as \textit{mixed-flux}, because at a generic point in moduli space both R-R and NS-NS fluxes are turned on in the supergravity solutions. As was clarified in~\cite{OhlssonSax:2018hgc}, mixed-flux backgrounds include the theory supported by only NS-NS \textit{charge}, for which at a generic point in moduli space the R-R three form \textit{flux} is also non-zero due to the coupling to R-R axions. Turning off such axion moduli corresponds to setting $h=0$ and corresponds to the WZW point~\cite{Maldacena:2000hw}. With this clarification implict, we will continue to use the term mixed-flux to refer to all backgrounds with  $\kbar\neq 0$.

Unlike in the R-R theory~\cite{Borsato:2014exa}, the mixed-flux dispersion relations~\eqref{eq:disp-rels-mf} are no longer periodic in $p$ and physical momenta can take any real value. As a result, the kinematics of the excitations in mixed-flux theories is very different from the R-R cases. While momentum periodicity is broken, it is nevertheless useful to think of distinct momentum intervals, each with their own Zhukovsky sheets that can be reached through analytic continuation. While most of the Zhukovsky sheets share some similarities with the familiar R-R ones, they have important differences including a more complicated set of physical regions. In addition, there is one set of sheets which exhibits a novel type of periodicity. We explore these new features, in part using a visualisation programme~\cite{pxu-gui} that is released together with this paper. For example, we find that the bound-state structure is novel compared to R-R theories, with bound states whose mass is larger than $k$ being identified with fundamental excitations with momentum on a different interval or equivalently on a different set of Zhukovsky sheets. This novel behaviour places important restrictions on the S matrices through fusion, relating for example the RL S matrix to the LL one.

This paper is organised as follows. In section~\ref{sec:mixed-flux-kinematics}, we disccuss in detail the mixed-flux Zhukovsky variables in terms of which the S matrix is most easily expressed, as well as the mixed-flux generalisation of the rapidity $u$, which has featured prominently in the quantum spectral curve (QSC) of R-R backgrounds~\cite{Gromov:2014caa,Bombardelli:2017vhk,Cavaglia:2021eqr,Ekhammar:2021pys}. In section~\ref{sec:cross-transformation} we discuss the crossing transformations that place restricitions on the scalar factors of the S matrix and summarise our normalization conventions in section~\ref{sec:two-particle-s-matrix}. In section~\ref{sec:r-r-scalar-factors}, we review the R-R dressing factors proposed as solutions to the massive crossing equations in~\cite{Borsato:2014hja}. It was pointed out in~\cite{Frolov:2021fmj} that these factors have unphysical logarithmic cuts in the Zhukovsky plane. Cavagli\`a and Ekhammar showed that these unwanted cuts can be removed in a straightforward way without spoiling the crossing or unitarity properites. This simple correction to~\cite{Borsato:2014hja} matches the dressing factors found in~\cite{Frolov:2021fmj}. We review these findings in section~\ref{sec:r-r-scalar-factors}, before 
solving the odd part~\cite{Beisert:2006ib} of the mixed-flux crossing equations in section~\ref{sec:mixed-flux-scalar-factors}. There, we also make proposals for the even \textit{scalar} factors in the theory, which give rise to the expected bound state poles, but are trivial under crossing. Exploiting the periodicity on one of the sheets, we propose a scalar factor closely related to that of the XXZ model at a root of unity, \ie, with anisotropy of the form $\Delta = \cos\frac{\pi}{k}$.  We show how in a relativistic limit, our results reduce to those found in~\cite{Frolov:2023lwd}. Finally, in section~\ref{sec:bound-state-s-mat-and-phases} we discuss the bound-state analytic structure and fusion and show their compatibility with the odd dressing factors and the scalar factors we found. Finding the full even \textit{dressing} factor which generalises the even part of the Beisert-Eden-Staudacher (BES) phase~\cite{Beisert:2006ib,Beisert:2006ez} is beyond the scope of this paper and we hope to return to it in the near future.

\section{Mixed-flux kinematics}
\label{sec:mixed-flux-kinematics}

The global symmetry of string theory on $\AdS_3\times \Sphere^3\times \Torus^4$ backgrounds is $\algPSU(1,1|2)^2$. When quantising the worldsheet theory in uniform light-cone gauge around the BMN vacuum, this  is broken to $\algPSU(1|1)^4$. Each fundamental world-sheet excitation transforms in a short representation of a particular central extension of this algebra, conventionally denoted 	$\algPSU(1|1)^4_{\ce}$.\footnote{Upon imposing the level-matching condition on physical states the central extensions trivialise.} The dispersion relations~\eqref{eq:disp-rels-mf} follow from the shoretening conditions, just as they do in integrable R-R backgrounds, see~\cite{Beisert:2010jr,Arutyunov:2009ga} and references therein. In the pure R-R charge world-sheet theory ($\kbar=0$) the dispersion relation is periodic
\begin{equation}\label{eq:R-R-disp-rel}
  E_{\text{R-R}}(p) = \sqrt{m^2 + 4h^2\sin^2\frac{p}{2}} ,
\end{equation}
while in the pure NS-NS charge theory at the WZW point $(h=0)$, the dispersion relation reduces to a linear relativistic one
\begin{equation}
  E_{\text{NS-NS}}(p) = | m + \kbar p | .
\end{equation}
In contrast to the case of R-R backgrounds (for which $\kbar=0$), the mixed-flux dispersion relations are not periodic. There is instead a shift symmetry 
\begin{equation}\label{eq:mom-m-shift-sym}
  p \to p \pm 2\pi,
  \qquad
  m \to m + k,
\end{equation}
with $\pm = -$ for L representations and $\pm=+$ for R representations, discussed below. In this paper we are mainly concerned with fundamental excitations with mass $m=1$. Since the dispersion relation is no longer periodic, we allow for momenta of physical excitations to take any real value, rather than being restricted to the $\left[0,2\pi\right]$ interval familiar from integrable string theories with R-R flux. Analytic continuation in $p$ will then lead to a complex $p$ plane with cuts. 

\begin{figure}
  \centering
  \includegraphics{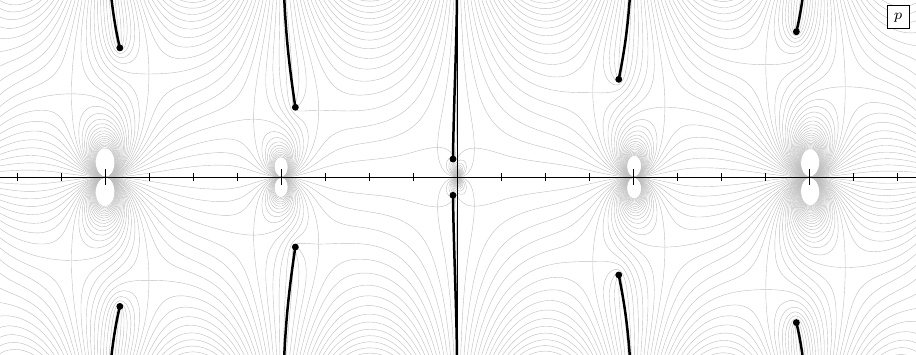}
  \caption{\label{fig:Ep-branch-cuts}Location of branch points and branch cuts of $E_L(p)$. They come in complex conjugate pairs and we label each pair by an integer $n$, with the $n=0$ pair closest to the imaginary $p$-axis.}
  \end{figure}

The L and R dispersion relations have branch points at complex values of momenta for which $E_{\sL}(p)=0$, respectively $E_{\sR}(p)=0$. The solutions of these equations, labelled $p^{(n)}_{b.p.}$, with $n\in\Integers$ can be found numerically and come in complex conjugate pairs, with 
\begin{equation}
\mbox{Re}\left( p_{b.p.}^{(n)}\right)\in \bigl(2\pi (n-\tfrac{1}{2}),2\pi (n+\tfrac{1}{2})\bigr).
\end{equation}
We choose the cuts to lie along the contours where $E_{\sL}(p)$ is imaginary, so the cuts do not cross the real $p$ axis as shown in figure~\ref{fig:Ep-branch-cuts}.\footnote{Unless otherwise indicated the figures are drawn with the coupling constants set to $h=2$ and $k=5$.}  It is natural to think about these two sheets of the $p$ plane as a \emph{physical} sheet, which contains the line where $p \in \Reals$ and $\Re E > 0$, and a \emph{crossed} sheet, containing the line $p \in \Reals$ and $\Re E < 0$. However, as we will see later, once we consider bound states, not all physical (crossed) states fully sit on the physical (crossed) sheet of the $m=1$ $p$ plane.

\subsection{Zhukovsky variables}
\label{sec:zhukovsky-variables}

The mixed-flux dispersion relation and world-sheet S matrix take a simple form when expressed in terms of Zhukovsky variables. Since the theory is not parity invariant, these come in two types: $x_{\sL}^{\pm}$ and $x_{\sR}^{\pm}$, corresponding to two types of short representations L and R. Worldsheet energy and momentum can be expressed in terms of Zhukovsky variables in the same way as in the R-R case
\begin{equation}\label{eq:energy-momentum-zhukovksy}
  e^{ip} = \frac{x_{\sI}^+}{x_{\sI}^-} , \qquad
  E_{\sI}(p) = -\frac{ih}{2} \Bigl( x_{\sI}^+ - \frac{1}{x_{\sI}^+} - x_{\sI}^- + \frac{1}{x_{\sI}^-} \Bigr), \qquad
  \mbox{I} = \mbox{L}, \mbox{R} .
\end{equation}
However, the L and R variables satisfy  different shortening conditions
\begin{equation}\label{eq:mixed-shortening-cond}
  \begin{aligned}
    x_{\sL}^+ + \frac{1}{x_{\sL}^+} - x_{\sL}^- - \frac{1}{x_{\sL}^-} &= \frac{2i (m + \kbar p)}{h} ,
    \\
    x_{\sR}^+ + \frac{1}{x_{\sR}^+} - x_{\sR}^- - \frac{1}{x_{\sR}^-} &= \frac{2i (m - \kbar p)}{h} .
  \end{aligned}
\end{equation}
Focussing on the L representations, from~\eqref{eq:energy-momentum-zhukovksy} we find that an excitation with real momentum and positive energy has $x_{\sL}^+$ in the upper half-plane and $x_{\sL}^- = ( x_{\sL}^+ )^*$.
The shortening condition is then solved by
\begin{equation}\label{eq:xpm-of-p-L-outer}
  x_{\sL}^{\pm}(p,m) = \Xi_{\sL}^{\pm}(p,m) \equiv \frac{m + \kbar p + \sqrt{(m + \kbar p)^2 + 4h^2\sin^2\frac{p}{2}}}{2h\sin\frac{p}{2}} e^{\pm \frac{ip}{2}} .
\end{equation}
We often drop the arguments of $x_{\sL}^{\pm}$ and, unless otherwise specified, we then consider the case of a fundamental excitation with $m=1$.

\begin{figure}
  \centering
  \subfloat[\label{fig:xpL-cover}$\Xi_{\sL}^+(p,m)$]{\includegraphics{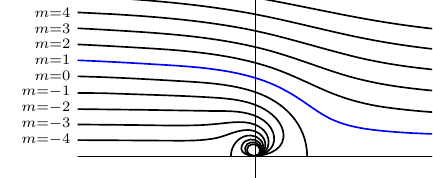}}
  \hspace{0.25cm}
  \subfloat[\label{fig:xmL-cover}$\Xi_{\sL}^-(p,m)$]{\includegraphics{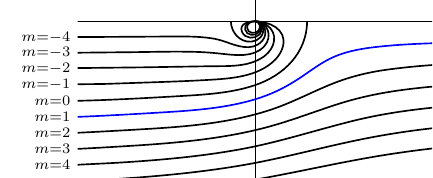}}
  
  \caption{\label{fig:x-cover} Plots of the functions $\Xi_{\sL}^{\pm}(p,m)$ with $m=-8,\dots,6$ and $p\in[0,2\pi]$. In the $\Xi_{\sL}^+(p,m)$ ($\Xi_{\sL}^-(p,m)$) plot the top (bottom) curve has $m=6$ and $m$ decreases as one moves to lower (higher) curves, with the smallest oval curve shown having $m=-8$. Non-integer values of $m$ will fill the remaining parts of each half-plane. The blue curves have $m=1$ and shows the location of fundamental excitations with $p\in[0,2\pi]$.}	 
\end{figure}
\begin{figure}
  \centering

  \subfloat[$h=7$, $k=3$]{\includegraphics{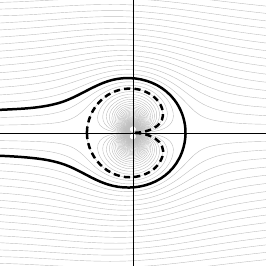}}
  \hspace{0.5cm}
  \subfloat[$h=2$, $k=5$]{\includegraphics{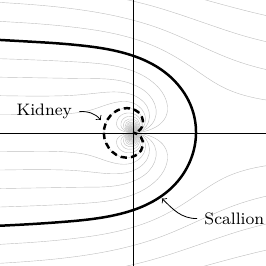}}
  \hspace{0.5cm}
  \subfloat[$h=1$, $k=7$]{\includegraphics{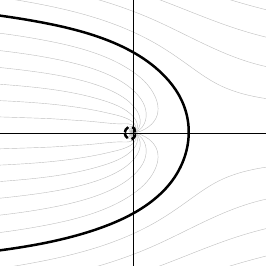}}
  
  \caption{\label{fig:scallion-and-kidney}The scallion and kidney contours for various values of the couplings.}
\end{figure}
The functions $\Xi_{\sL}^{\pm}(p,m)$ appearing on the right hand side of~\eqref{eq:xpm-of-p-L-outer} can also be used to cover the Zhukovsky planes. To do this we consider $m\in\mathbb{R}$ and because of the symmetry~\eqref{eq:mom-m-shift-sym}, we can restrict to $p\in[0,2\pi]$. In figure~\ref{fig:x-cover} we plot $\Xi_{\sL}^{\pm}(p,m)$ for $m=-8,\dots,6$ and $p\in[0,2\pi]$. From the figure it should be clear that with $m\in\mathbb{R}$, $\Xi_{\sL}^{+}(p,m)$ covers the UHP while $\Xi_{\sL}^{-}(p,m)$ covers the LHP. We may view the parameter $m$ as a type of radial ordering of the Zhukovsky planes.

For later convenience we also introduce the related function
\begin{equation}\label{eq:def-xi-tilde}
  \tilde{\Xi}_{\sL}^{\pm}(p,m) = \frac{m + \kbar p - \sqrt{(m + \kbar p)^2 + 4h^2\sin^2\frac{p}{2}}}{2h\sin\frac{p}{2}} e^{\pm \frac{ip}{2}} ,
\end{equation}
which solves the L shortening condition when $\Re E_{\sL}(p) < 0$, as well as the functions
\begin{equation}\label{eq:def-xi-R-xi-R-tilde}
  \begin{aligned}
    \Xi_{\sR}^{\pm}(p,m) &= \frac{m - \kbar p + \sqrt{(m - \kbar p)^2 + 4h^2\sin^2\frac{p}{2}}}{2h\sin\frac{p}{2}} e^{\pm \frac{ip}{2}} , \\
    \tilde{\Xi}_{\sR}^{\pm}(p,m) &= \frac{m - \kbar p - \sqrt{(m - \kbar p)^2 + 4h^2\sin^2\frac{p}{2}}}{2h\sin\frac{p}{2}} e^{\pm \frac{ip}{2}} ,
  \end{aligned}
\end{equation}
which solve the R shortening conditions when $\Im E_{\sR}(p) > 0$ and $\Im E_{\sR}(p) < 0$, respectively.

The contours $\Xi_{\sL}^{\pm}(p,m=0)$ and $\Xi_{\sL}^{\pm}(p,m=-k)$, which are shown in figure~\ref{fig:scallion-and-kidney}, play a distinguished role in the Zhukovsky planes: as we will see shortly they correspond to the branch cuts in the $u$ rapidity plane. We find their form reminiscent of \emph{scallion} and a \emph{kidney} and we will often refer to these curves as such.

\begin{figure}[!t]
    \centering
    \subfloat[\label{fig:p-xpL-preimage}$x_{\sL}^+(p,m=1) = \Xi^{\pm}(p',m')$]{\includegraphics{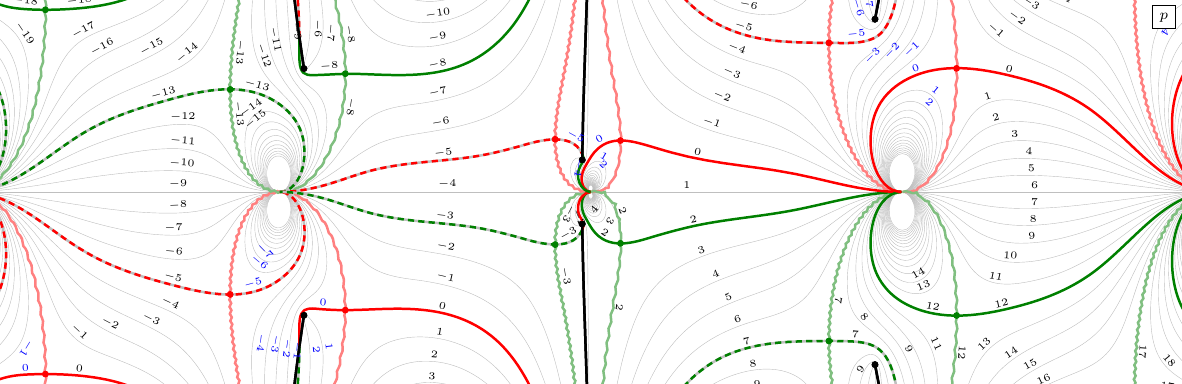}}
    
    \subfloat[\label{fig:p-xmL-preimage}$x_{\sL}^+(p,m=1) = \Xi^{\pm}(p',m')$]{\includegraphics{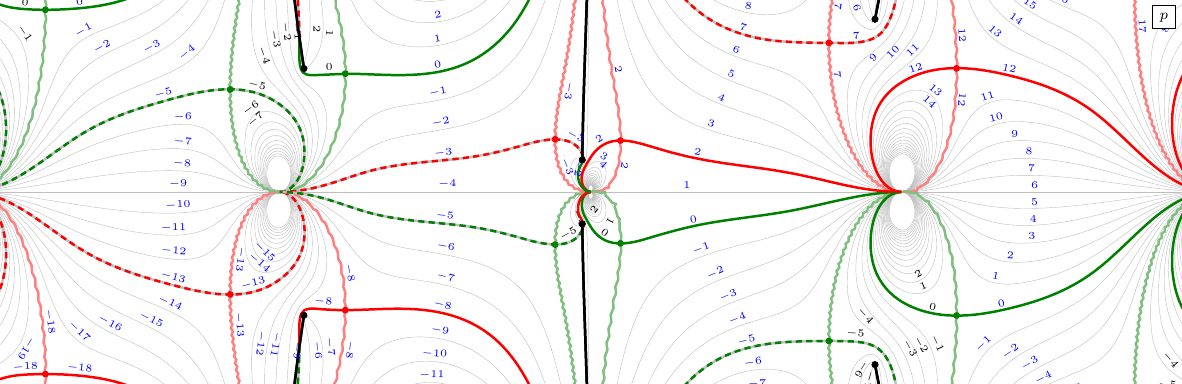}}
      
    \caption{\label{fig:xpm-preimages} Plots of the complex $p$ plane with contours labelled by $m' \in \Integers$ such that $x_{\sL}^{\pm}(p,m=1) = \Xi_{\sL}^+(p',m')$ (in black) or $x_{\sL}^{\pm}(p,m=1) = \Xi_{\sL}^-(p',m')$ (in blue). The wavy red and green lines show the pre-image of the real line in the $x_{\sL}^+$ and $x_{\sL}^-$ lines. We also indicate the location of the $x_{\sL}^+$/$x_{\sL}^-$ scallion with solid red/green curves, and the kidneys with corresponding dashed curves.}
  \end{figure}

For a given real momentum $p' \in [0,2\pi]$ and real mass $m'$ there is an infinite number of complex momenta $p$ such that
$x_{\sL}^+(p,m=1) = \Xi_{\sL}^+(p',m')$ and also an infinite number of (separate) complex momenta $p$ such that $x_{\sL}^+(p,m=1) = \Xi_{\sL}^-(p',m')$. The same holds for the equations $x_{\sL}^-(p,m=1) = \Xi_{\sL}^{\pm}(p',m')$. To illustrate this, figure~\ref{fig:p-xpL-preimage} on page~\pageref{fig:p-xpL-preimage} shows the $p$ plane with contours labelled by $m' \in \Integers$ such that $x_{\sL}^+(p,m=1) = \Xi_{\sL}^+(p',m')$ or $x_{\sL}^+(p,m=1) = \Xi_{\sL}^-(p',m')$ for some real $p' \in [0,2\pi]$. Figure~\ref{fig:p-xmL-preimage} shows the corresponding plot for $x_{\sL}^-(p,m=1)$. In the figures we have also indicated the curves mapping to the real line of the $x_{\sL}^{\pm}$ planes, as well as the pre-images of the scallion and kidney contours. The distinguished role of these contours can be seen in the figures: in the $x_{\sL}^+$ plot in figure~\ref{fig:p-xpL-preimage} there is a branch point where the $x_{\sL}^-$ scallion or kidney intersects with the $x_{\sL}^-$ real line. The nature of these branch points is best understood through the $u$ plane which we will introduce next.
\pagebreak
%\begin{landscape}
%\end{landscape}

\subsection{The \texorpdfstring{$u$-plane}{u-plane}}
\label{sec:the-u-plane}

Another variable that is often useful in integrable holographic models is the rapidity $u$. In the mixed-flux setting this is defined as~\cite{Hoare:2013lja}
\begin{equation}\label{eq:mixed-u-of-x}
  u_{\sL}(x) = x + \frac{1}{x} - \frac{2\kbar}{h} \log x ,
  \qquad
  u_{\sR}(x) = x + \frac{1}{x} + \frac{2\kbar}{h} \log x .
\end{equation}
The shortening conditions~\eqref{eq:mixed-shortening-cond} can then be written as\footnote{This is clearly true for $p \in [0,2\pi]$ with the principal branch of the log. We will discuss the role of the log cut further below.}
\begin{equation}\label{eq:mixed-shortening-cond-u}
  u_{\sL}(x_{\sL}^+) - u_{\sL}(x_{\sL}^-) = \frac{2im}{h} , \qquad
  u_{\sR}(x_{\sR}^+) - u_{\sR}(x_{\sR}^-) = \frac{2im}{h} .
\end{equation}
It is then natural the introduce the rapidities $u_{\sL}$ and $u_{\sR}$ such that
\begin{equation}
  u_{\sL}(x_{\sL}^{\pm}) = u_{\sL} \pm \frac{im}{h} , \qquad
  u_{\sR}(x_{\sR}^{\pm}) = u_{\sR} \pm \frac{im}{h} .
\end{equation}
By solving $\frac{du}{dx}=0$ we find that the branch points of $x_{\sL}(u_{\sL})$ are at 
\begin{equation}\label{eq:xl-bps}
x^{\mbox{\scriptsize b.p.}}_L=s,\,-s^{-1},
\end{equation}
while those of $x_{\sR}(u_{\sR})$ are at 
\begin{equation}\label{eq:xr-bps}
x^{\mbox{\scriptsize b.p.}}_R=-s,\,s^{-1},
\end{equation}
where, following~\cite{Babichenko:2014yaa}, we have introduced the parameter $s$  
\begin{equation}
  \frac{\kbar}{h} = \frac{s - s^{-1}}{2} ,
\end{equation}
with the convention that $s \ge 1$, in which case
\begin{equation}\label{eq:explicit-s-and-sinv}
  s=\frac{\kbar+\sqrt{h^2+\kbar^2}}{h},\qquad\qquad -s^{-1}=\frac{\kbar-\sqrt{h^2+\kbar^2}}{h}.
\end{equation}

\subsubsection{The analytical structure of \texorpdfstring{$u_{\sL}(x_{\sL})$}{u(x)}}

The function $u_{\sL}(x_{\sL})$ has a fairly complicated analytical structure. The inverse function $x_{\sL}(u_{\sL})$ has branch points at $x_{\sL} = s$ and $x_{\sL} = -1/s$. Additionally $u_{\sL}(x_{\sL})$ itself has a log cut which sits on the negative real axis of the $x_{\sL}$ plane. Together this means that we will need an infinite number of sheets in both the $x_{\sL}$ plane and the $u_{\sL}$ plane to describe the full function.

\begin{figure}
  \centering
  \subfloat[\label{fig:u-regions-long-upper}Sheet covering the upper half of the $x_{\sL}$ plane.]{\includegraphics{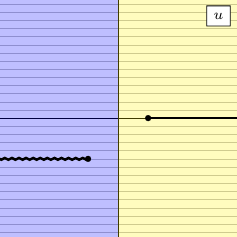}}
  \hspace{0.5cm}
  \subfloat[\label{fig:u-regions-long-lower}Sheet covering the lower half of the $x_{\sL}$ plane.]{\includegraphics{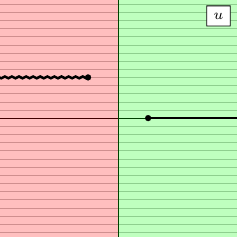}}
    \hspace{0.5cm}
  \subfloat[\label{fig:x-regions-long}The $x_{\sL}$ plane with preimages of long cuts.]{\includegraphics{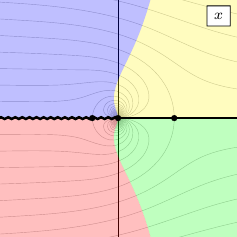}}
  
  \caption{\label{fig:x-u-regions-long}The $u_{\sL}$ plane with long cuts.}
\end{figure}
\begin{figure}
  \centering
  \includegraphics{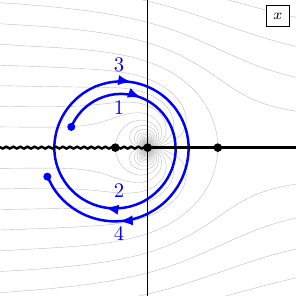}

  \caption{\label{fig:x-long-circle}A path wrapping twice around the origin of $x_{\sL}$.}
\end{figure}
\begin{figure}
  \centering
  \subfloat[]{\includegraphics{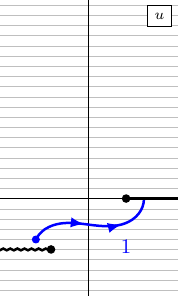}}
  \hspace{0.5cm}
  \subfloat[]{\includegraphics{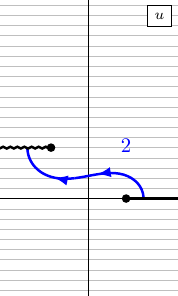}}
  \hspace{0.5cm}
  \subfloat[]{\includegraphics{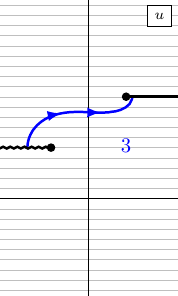}}
  \hspace{0.5cm}
  \subfloat[]{\includegraphics{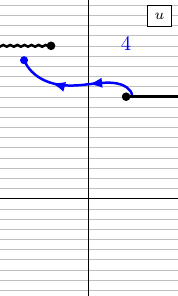}}

  \caption{\label{fig:u-long-circle}Four sheets of $u_{\sL}$ with long cuts.}
\end{figure}
Let us first see what it takes to cover a single sheet of the $x_{\sL}$ plane. A simple choice is to use ``\textit{long}'' cuts, where the upper and lower half of the $x_{\sL}$ planes are completely covered by one sheet of the $u_{\sL}$ plane each.\footnote{We will refer to cuts as long, and later as short, because in the $k=0$ limit they reduce to the well-known short and long cuts of the R-R theory.} Figure~\ref{fig:u-regions-long-upper} and~\ref{fig:u-regions-long-lower} show the two sheets of $u_{\sL}$ with long cuts, which cover the single  $x_{\sL}$ plane shown in Figure~\ref{fig:x-regions-long}. We have shown how the regions of the $x_{\sL}$ plane are mapped into the $u_{\sL}$ planes by using matching colours in the figures. In Figure~\ref{fig:x-regions-long} we have also explicitly drawn in black the \textit{pre-image} of the two long cuts of $u_{\sL}(x)$ shown in Figures~\ref{fig:u-regions-long-upper} and~\ref{fig:u-regions-long-lower}. We emphasize that these are not cuts in the $x_{\sL}$ plane and are drawn merely to help visualise the gluing of the two $u_{\sL}$ planes. The only cut of the $x_{\sL}$ plane is the log cut, whose location coincides with the wavy black line in Figure~\ref{fig:x-regions-long}. As is clear from the picture, the two $u_{\sL}$ sheets are connected by the solid black line cut on the right. If we instead go through the left cut, we simultaneously go through the log cut in the $x_{\sL}$ plane and thus come out on a new sheet of the $x_{\sL}$ plane, which comes with two new $u_{\sL}$ sheet. We stress that this is a new feature of the mixed-flux variables compared to the R-R ones. To see this in more detail, figure~\ref{fig:x-long-circle} shows a path that wraps twice around the origin in the $x_{\sL}$ plane. In the $u_{\sL}$ plane this corresponds to the path in figure~\ref{fig:u-long-circle}.

\begin{figure}
  \centering
  
  \subfloat[\label{fig:x-region-outside}]{\includegraphics{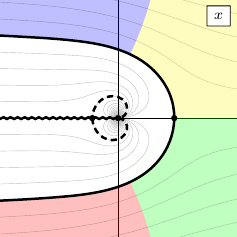}}
  \hspace{0.5cm}
  \subfloat[\label{fig:x-region-between}]{\includegraphics{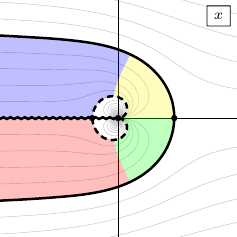}}
  \hspace{0.5cm}
  \subfloat[\label{fig:x-region-inside}]{\includegraphics{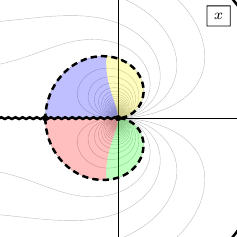}}
  \vspace{0.5cm}
  \subfloat[\label{fig:u-region-outside-small}]{\includegraphics{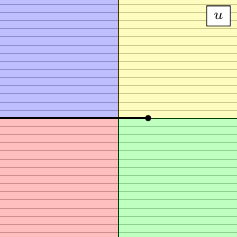}}
  \hspace{0.5cm}
  \subfloat[\label{fig:u-region-between-small}]{\includegraphics{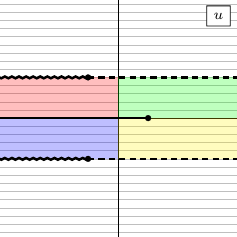}}
  \hspace{0.5cm}
  \subfloat[\label{fig:u-region-inside-small}]{\includegraphics{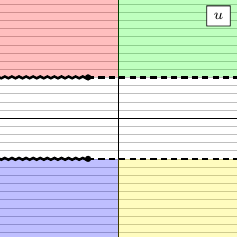}}
  \vspace{0.5cm}
  \subfloat[\label{fig:u-region-outside}]{\includegraphics{./figures/u-regions-outside}}
  \hspace{0.5cm}
  \subfloat[\label{fig:u-region-between}]{\includegraphics{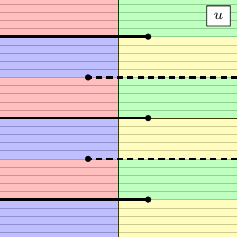}}
  \hspace{0.5cm}
  \subfloat[\label{fig:u-region-inside}]{\includegraphics{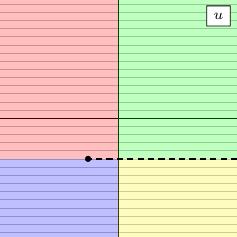}}
  
  \caption{The $x$ and $u$ planes with ``short'' cuts split into regions \emph{outside} the scallion (figures \protect\subref{fig:x-region-outside},~\protect\subref{fig:u-region-outside-small}  and~\protect\subref{fig:u-region-outside}), \emph{between} the scallion and the kidney (figures \protect\subref{fig:x-region-between},~\protect\subref{fig:u-region-between-small} and \protect\subref{fig:u-region-between}) and \emph{inside} the kidney (figures \protect\subref{fig:x-region-inside},~\protect\subref{fig:u-region-inside-small} and \protect\subref{fig:u-region-inside}). When entering a cut in the $u$ plane we come out on the next sheet in a region with the same colour. The middle figures~\protect\subref{fig:u-region-outside-small},~\protect\subref{fig:u-region-between-small} and~\protect\subref{fig:u-region-inside-small} show the parts of the $u_{\sL}$ plane with short cuts that cover a single sheet of the $x_{\sL}$ plane. The last row, with figures~\protect\subref{fig:u-region-outside},~\protect\subref{fig:u-region-between} and~\protect\subref{fig:u-region-inside}, shows three sheets of the $u_{\sL}$ plane after extending them through the log cuts. In the $x$ plane, the solid and dashed black lines correspond to the pre-images of the $u$ short cuts, while the wavy black line is the location of the $\log x$ cut. }
\end{figure}

A benefit of the long cuts is that each sheet of the $x_{\sL}$ plane is completely covered by two full sheets of the $u_{\sL}$ plane. However, in what follows we will find it more convenient to work with ``short'' cuts. These can be thought of as the complement of the long cuts. We cut each sheet of the $u_{\sL}$ plane into three pieces along horizontal lines coincident with the long cuts, and then glue them back together along the opposite cuts. This partitions the $x_{\sL}$ plane into a region outside the scallion, and region between the scallion and the kidney, and a region inside the kidney, as shown in figures~\ref{fig:x-region-outside},~\ref{fig:x-region-between} and~\ref{fig:x-region-inside}, respectively. In these figures, the solid and dashed black lines correspond to the pre-images of the  short cuts of $u_{\sL}(x)$. The wavy black line is the location of the $\log x$ cut, which further splits the inside region into an upper and a lower part.

If we only consider a single sheet of the $x_{\sL}$ plane, the short cuts split the $u_{\sL}$ plane into four parts. The outside region is covered by one sheet of $u_{\sL}$ with a single cut, see figure~\ref{fig:u-region-outside-small}. The between region corresponds to a horizontal strip of height $2k/h$, see figure~\ref{fig:u-region-between-small} and the inside region corresponds to two disconnected and offset half-planes, see figure~\ref{fig:u-region-inside-small}.

\begin{figure}
  \centering
  \subfloat[]{\includegraphics{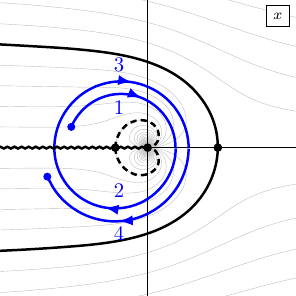}}
  \hspace{1cm}
  \subfloat[]{\includegraphics{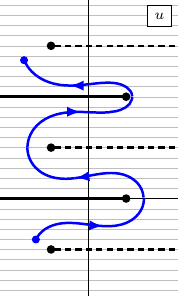}}
  
  \caption{\label{fig:x-short-circle}A path encircling the origin of the $x_{\sL}$ plane. The short cuts resolve the log cut.}
\end{figure}

However, we can now easily extend the $u_{\sL}$ plane through the log cuts in the between and inside regions. We then find that the between region gives rise to a periodic $u_{\sL}$ sheet as in figure~\ref{fig:u-region-between} while the region inside the kidney corresponds to a sheet with a single cut as in figure~\ref{fig:u-region-inside}.\footnote{Here we draw the sheets that corresponds to the upper half of the kidney region of the original $x_{\sL}$ plane. There is a second sheet corresponding to the extension of the lower half of the kidney, which looks identical except the cut is shifted up by $2ik/h$.} With  short cuts the log cut is thus completely resolved in the $u_{\sL}$ plane. As an example of this figure~\ref{fig:x-short-circle} shows the same path as in figures~\ref{fig:x-long-circle} and~\ref{fig:u-long-circle}. With short cuts the whole path sits on a single sheet.

\subsubsection{The full \texorpdfstring{$u_{\sL}$}{uL} plane}

\begin{figure}
  \centering
  \subfloat[\label{fig:p-simple-path}]{\includegraphics{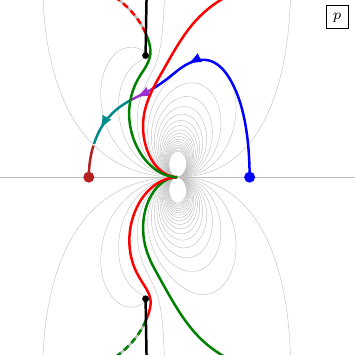}}
  \hspace{1cm}
  \subfloat[\label{fig:x-simple-path}]{\includegraphics{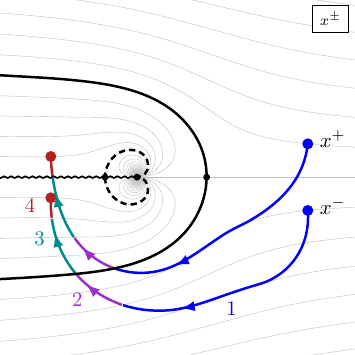}}
  
  \subfloat[\label{fig:u-simple-path-1}]{\includegraphics{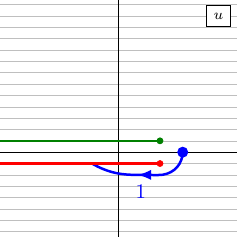}}
  \hspace{0.5cm}
  \subfloat[\label{fig:u-simple-path-2}]{\includegraphics{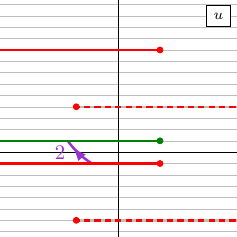}}
  \hspace{0.5cm}
  \subfloat[\label{fig:u-simple-path-34}]{\includegraphics{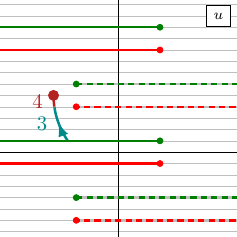}}
  
  \caption{\label{fig:simple-path}A simple path that takes $x^{\pm}$ from the outside of the scallion to the region between the scallion and the kidney.}
\end{figure}

In the previous section we considered the analytical structure of the function $u_{\sL}(x_{\sL})$ and its inverse. However, to describe the full kinematics of the mixed-flux string theory we need both Zhukovsky variables $x_{\sL}^{\pm}$. The full $u_{\sL}$ plane therefore has twice the number of cuts to what was discussed above. Let us illustrate this by considering a path in the momentum plane that goes from $(0,2\pi)$ to $(-2\pi,0)$ as shown in figure~\ref{fig:p-simple-path} on page~\pageref{fig:p-simple-path}. Its images in the $x^+$ and $x^-$ planes, drawn for compactness together in figure~\ref{fig:x-simple-path}, split into four parts:
\begin{enumerate}
\item $x^{\pm}$ start out with real momentum $p \in [0,2\pi]$ and $x^+$ is brought inside the scallion from below,
\item $x^-$ is brought inside the scallion,
\item $x^+$ goes through the log cut,
\item $x^{\pm}$ end up with real momentum $p \in [-2\pi,0]$.
\end{enumerate}
Figures~\ref{fig:u-simple-path-1}--\subref*{fig:u-simple-path-34} shows the same path in the $u$ plane. For real momentum $p \in [0,2\pi]$ there are two cuts corresponding to $x^+$ and $x^-$ going through the scallion. In the first step (figure~\ref{fig:u-simple-path-1}) we go through the $x^+$ scallion cut. We then come out on a new sheet (figure~\ref{fig:u-simple-path-2}) where we have a periodic set of $x^+$ cuts (in red) but a single $x^-$ cut (in green). We go through the $x^-$ cut. Finally, we come out (figure~\ref{fig:u-simple-path-34}) on a third sheet where both the red $x^+$ cuts and the green $x^-$ cuts are periodic. Since the $u$ plane resolves the log cut we can just go up to the real momentum line. Note that for real momentum $p \in [0,2\pi]$ we have chosen $u$ to be real. As we have just seen, this means that for $p \in [-2\pi,0]$ $u$ will not be real.

\begin{figure}
  \centering
  \subfloat[\label{fig:p-large-circle}]{\includegraphics{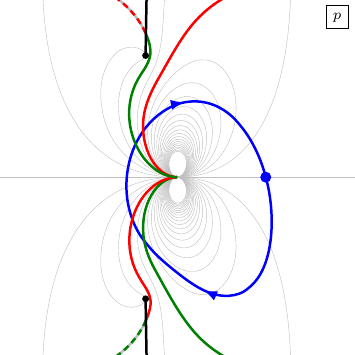}}
  \hspace{1cm}
  \subfloat[\label{fig:x-large-circle}]{\includegraphics{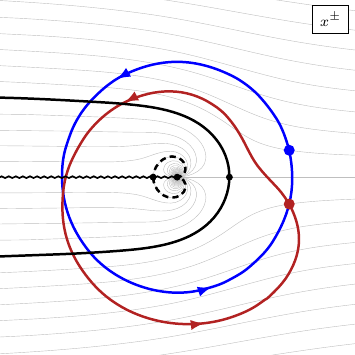}}
  
  \subfloat[\label{fig:u-large-circle-1}]{\includegraphics{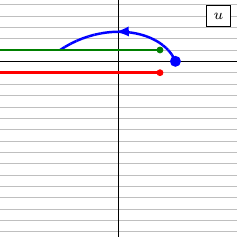}}
  \hspace{0.5cm}
  \subfloat[\label{fig:u-large-circle-2}]{\includegraphics{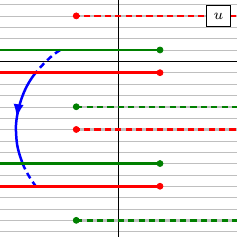}}
  \hspace{0.5cm}
  \subfloat[\label{fig:u-large-circle-3}]{\includegraphics{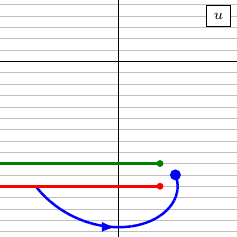}}
  
  \caption{\label{fig:large-circle}A path describing a large circle in the $x^+$ plane.}
\end{figure}

 Consider next a path in the $p$ plane that begins and ends on $(0,2\pi)$ and goes clockwise around $p=0$ as shown in figure~\ref{fig:p-large-circle}. The image of this path in the $x^+$ and $x^-$ planes, again drawn for compactness together in figure~\ref{fig:x-large-circle} where both $x^+$ and $x^-$ encircle the origin as well as the branch points at $s$ and $-1/s$ once in a anti-clockwise direction. The corresponding path in the $u$ plane is shown in figure~\ref{fig:u-large-circle-1}--\subref*{fig:u-large-circle-3}. From the point of view of $x^{\pm}$, the path takes us back to the exact same point we started at, but that is not true in the $u$ plane. Instead, the end point sits on a sheet which looks identical to the one we started on except everything is shifted \textit{down} by $2ik/h$. If instead we had gone around $p=0$ in an anti-clockwise direction, we would have ended up on a $u$ sheet where everything is shifted \textit{up} by $2ik/h$. This argument shows that it is only the \textit{difference} of $u(x^+)$ and $u(x^-)$ that is physical, rather than the value of each.

\begin{figure}
  \centering
  \subfloat[]{\label{fig:full-xp-region-0}\includegraphics{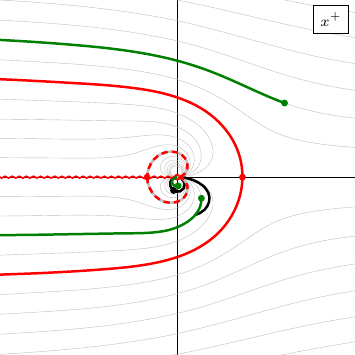}}
  \hspace{1cm}
  \subfloat[]{\label{fig:full-xm-region-0}\includegraphics{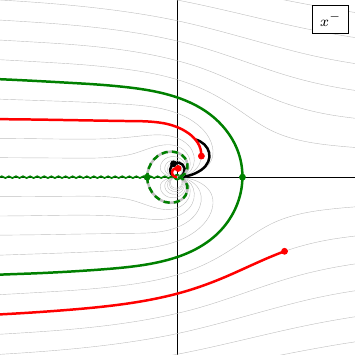}}
  \caption{\label{fig:full-xp-xm-region-0}The full $x^+$ and $x^-$ planes as seen by an excitation with $p \in (0,2\pi)$.}
\end{figure}

We can also think about separate $x^+$ and $x^-$ planes, where in, \eg, the $x^+$ plane we include not only the normal scallion and kidney, but also additional cuts where $x^-$ goes through the scallion and the kidney. See figure~\ref{fig:full-xp-xm-region-0} where we have also included the image of the $E(p)$ cut of the $p$ plane as a black line.

\subsection{The \texorpdfstring{$p$, $x^{\pm}$ and $u$}{p, x and u} planes}
\label{sec:p-x-u-planes}

We are now finally ready to discuss the full analytical structure of the $p$, $x^+$, $x^-$ and $u$ planes. Let us start by summarising the types of cuts we encounter in the various planes:
\begin{enumerate}
\item The \textbf{$\boldsymbol{p}$ plane} has two sheets connected by an infinite number of $E(p)$ cuts.
\item The \textbf{$\boldsymbol{x^{\pm}}$ planes} have an infinite number of sheets. The sheets are connected both through log cuts, along the negative real axis, and through the $x^{\mp}$ scallion and kidney cuts.
\item The \textbf{$\boldsymbol{u}$ plane} has an infinite number of sheets connected through the $x^{\pm}$ scallion and kidney cuts.
\end{enumerate}
A physical excitation with real momentum $p$ (by definition) sits along the real line on the first sheet of the $p$ plane. It is natural to divide this real line into regions $(2\pi n,2\pi (n+1))$. This corresponds to $x_{\sL}^- = (x_{\sL}^+)^*$ with $\Im x_{\sL}^+ > 0$ and $u_{\sL}$ along a line with $\Im u_{\sL} = kl/h$ for some $l \in \Integers$, where $l$ is an even (odd) integer when $n$ is even (odd).

\begin{figure}
  \centering
  \includegraphics{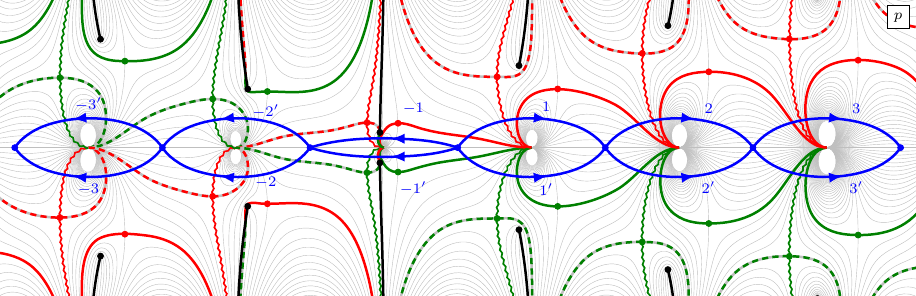}
  
  \caption{\label{fig:p-plane-path-between-region}To go from real momentum $p \in (2\pi n,2\pi (n+1))$ to momentum $p + 2\pi$ or $p - 2\pi$ we need to choose if we go above or below the point $2\pi(n+1)$ or $2\pi n$, respectively. Equivalently we need to choose if we go through the log cut in the $x_{\sL}^+$ or $x_{\sL}^-$ plane. The paths $-3,-2,\dotsc,3$ all cross the $x_{\sL}^+$ log cut while the paths $-3',-2',\dotsc,3'$ cross the $x_{\sL}^-$ log cut.}
\end{figure}
When $p$ approaches a multiple of $2\pi$ $x_{\sL}^{\pm}$ and $u_{\sL}$ all diverge. In order to continuously go between the different real momentum regions we need to go into the complex plane. At each point $2\pi n$ we need to choose if we go above or below the point. Some such paths are illustrated in figure~\ref{fig:p-plane-path-between-region}. Let us first consider going from the region 0 (with $p \in (0,2\pi)$) to the region 1 (with $p \in (2\pi,4\pi)$). The figure shows two paths:
\begin{enumerate}
\item[$1$:] Take $x_{\sL}^+$ into the scallion, through the log cut and then out of the scallion again.
\item[$1'$:] Take $x_{\sL}^-$ into the scallion, through the log cut and then out of the scallion again.
\end{enumerate}
In either case the other parameter ($x_{\sL}^{\mp}$) stays outside the scallion. Once we go down to the real momentum line again we find that $x_{\sL}^{\pm}$ are now located on the
\begin{equation}\label{eq:zhukovsky-parameters-n-1-region}
  x_{\sL}^{\pm} = \Xi_{\sL}^{\pm}(p-2\pi,m=k+1)
\end{equation}
contours, where we write the momentum as $p-2\pi$ to emphasise that $x_{\sL}^{\pm}$ has momentum in the $(2\pi,4\pi)$ range while the functions $\Xi_{\sL}^{\pm}(p,m)$ were define for $p \in (0,2\pi)$. Going to higher momentum regions along the paths $2,3,\dotsc$ or $2',3',\dotsc$ is just repeating a similar path.

To go to the $-1$ region (where $p \in (-2\pi,0)$) we take both $x_{\sL}^+$ and $x_{\sL}^-$ through both scallions, see path $-1$ and $-1'$ in the figure. The region $-2$ (where $p \in (-4\pi,-2\pi)$) takes us into the kidney, and to go to yet lower regions we just take one of $x_{\sL}^{\pm}$ out of the kidney, through the log cut and then back in to the kidney again.

\begin{figure}
  \centering
  \subfloat[$n=-3$]{\label{fig:u-physical-regions-min-3}\includegraphics{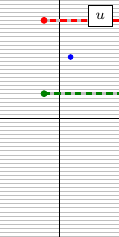}}
  \hspace{0.5cm}
  \subfloat[$n=-2$]{\label{fig:u-physical-regions-min-2}\includegraphics{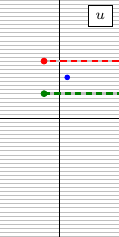}}
  \hspace{0.5cm}
  \subfloat[$n=-1$]{\label{fig:u-physical-regions-min-1}\includegraphics{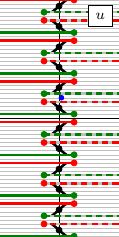}}
  \hspace{0.5cm}
  \subfloat[$n=0$]{\label{fig:u-physical-regions-0}\includegraphics{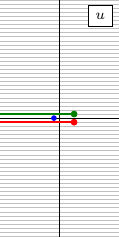}}
  \hspace{0.5cm}
  \subfloat[$n=1$]{\label{fig:u-physical-regions-1}\includegraphics{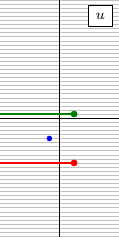}}
  \hspace{0.5cm}
  \subfloat[$n=2$]{\label{fig:u-physical-regions-2}\includegraphics{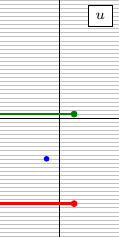}}
  
  \caption{\label{fig:u-physical-regions}The physical region in the $u$ plane for various momentum regions $2\pi n < p < 2\pi(n+1)$. The blue dots correspond to excitations with real momenta $p = 2\pi n + \pi$. The real momentum line in the middle of the cuts. The $u$ plane for $n=-1$ is periodic and shows an infinite number of both scallion and kidney cuts. The other regions each have two cuts.}
\end{figure}
It seems natural to pick a convention where we always take say $x_{\sL}^+$ through the log cut. This means that we go around the point $p = 2\pi n$ in the upper half-plane for $n \ge 0$ and in the lower half-plane for $n < 0$. In the $u_{\sL}$ plane real momenta in region $n$ then correspond to $\Im u_{\sL} = -\frac{nk}{h}$. Moreover, for $n \ge 0$ the $x_{\sL}^-$ scallion cut in the $u_{\sL}$ plane will be in a fixed position with imaginary part $+1/h$, with the $x_{\sL}^+$ scallion cut at $-(2nk+1)/h$, and for $n < -1$ the $x_{\sL}^-$ kidney cut will always have imaginary part $(k+1)/h$, with the $x_{\sL}^+$ kidney cut at $-((2n+1)k+1)/h$, see figure~\ref{fig:u-physical-regions}.

\begin{figure}
  \centering
  \subfloat[The sheet $\Re E > 0$]{\includegraphics{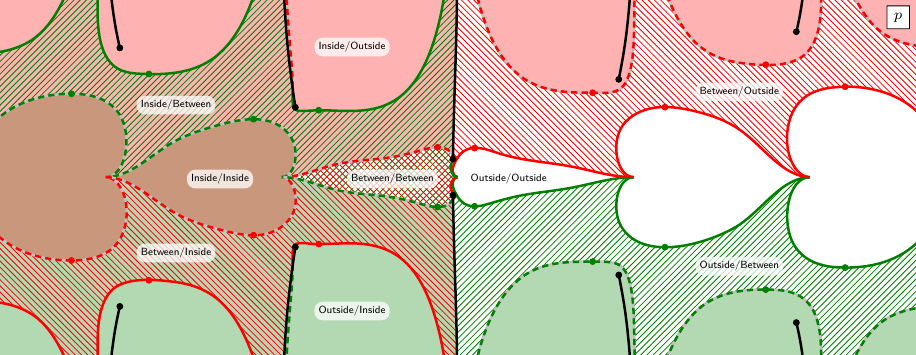}}

  \subfloat[The sheet $\Re E < 0$]{\includegraphics{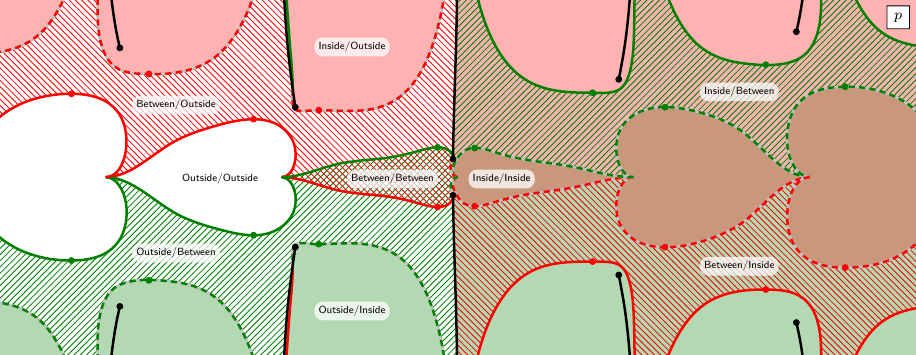}}
  
  \caption{\label{fig:p-short-cut-regions}The $p$ plane can be naturally subdivided into regions depending on the location of $x_{\sL}^{\pm}$. If $x_{\sL}^+$ ($x_{\sL}^-$) is inside the kidney the region is filled with a solid red (green) colour, if $x_{\sL}^+$ ($x_{\sL}^-$) is between the scallion and the kidney the region is filled with a diagonal red (green) pattern, and if $x_{\sL}^+$ ($x_{\sL}^-$) is outside the scallion we leave the region unfilled. The colours mix, so that, \eg, a solid brownish colour means that $x_{\sL}^{\pm}$ are both inside the kidney and a solid green colour means that $x_{\sL}^+$ is outside the scallion and $x_{\sL}^-$ is inside the kidney.}
\end{figure}
The scallion and kidney cuts naturally partition the $p$ planes into regions depending on where $x_{\sL}^+$ and $x_{\sL}^-$ sit: \emph{outside} the scallion, \emph{between} the scallion and kidney, or \emph{inside} the kidney, as illustrated in figure~\ref{fig:p-short-cut-regions}.

For an excitation with real momentum, $x_{\sL}^{\pm}$ sit in the same region. In particular, if the momentum $p$ is in the range $(2\pi n,2\pi (n+1))$, a physical excitation, which has positive energy and sits on the real line of the first sheet of the $p$ plane, has
\begin{equation}\label{eq:location-of-fundamental-physical-xpm}
  x_{\sL}^{\pm}(p,1) = \Xi_{\sL}^{\pm}(p - 2\pi n, 1 + nk)
\end{equation}
while a crossed excitation, with negative energy sitting on the real line of the second sheet of the $p$ plane, has
\begin{equation}\label{eq:location-of-fundamental-crossed-xpm}
  x_{\sL}^{\pm}(p,1) = \Xi_{\sL}^{\mp}\bigl(2\pi (n+1) - p, -1-(n+1)k\bigr) .
\end{equation}
As a result, a crossed real momentum excitation has $x_{\sL}^-$ in the upper half-plane and $x_{\sL}^+$ in the lower half-plane.

\pagebreak

\subsection{Bound states}
\label{sec:bound-states}

In addition to fundamental excitations the world-sheet spectrum contains bound states of such excitations. The basic construction of such bound states follows from $\algPSU(1|1)^4_{\ce}$ representation theory and is the same as in the pure R-R theory~\cite{Borsato:2012ud,Borsato:2013hoa,OhlssonSax:2019nlj}. 
An $m$-particle boundstate has $m$ constituent Zhukovsky parameters $x_j^{\pm}$, $j = 1,\dotsc,m$ satisfying\footnote{Here we will discuss bound states from the point of view of kinematics and representation theory. For a particular bound state to actually appear in the physical spectrum the (bound state) S matrix needs to have a pole corresponding to the formation of that bound state. We will discuss the poles of the S matrix in section~\protect\ref{sec:mixed-flux-scalar-factors}.} 
\begin{equation}\label{eq:bound-state-equations}
  x_{j}^- = x_{j+1}^+ , \quad j = 1,\dotsc,m-1 .
\end{equation}
The bound state condition~\eqref{eq:bound-state-equations} ensures that the momentum, energy and charge of the bound state only depends on the ``outer-most'' Zhukovsky parameters $X^+ = x_1^+$ and $X^- = x_1^-$. Using~\eqref{eq:energy-momentum-zhukovksy} we have
\begin{equation}
  E_{\text{tot}} = \sum_{j=1}^m E_j
  = -\frac{ih}{2} \sum_{j=1}^m x_j^+ - \frac{1}{x_j^+} - x_j^- + \frac{1}{x_j^-}
  =
  -\frac{ih}{2} \Bigl( X^+ - \frac{1}{X^+} - X^- + \frac{1}{X^-} \Bigr) ,
\end{equation}
and
\begin{equation}
  e^{ip_{\text{tot}}} = \frac{X^+}{X^-} = \prod_{j=1}^m \frac{x_j^+}{x_j^-} 
  \equiv \prod_{j=1}^m e^{ip_j} ,
\end{equation}
where $p_j$ is the momentum of the $j$-th constituent of the bound state. From the shortening condition~\eqref{eq:mixed-shortening-cond} we find the charge
\begin{equation}
  m + \kbar p_{\text{tot}}
  = -\frac{ih}{2} \sum_{j=1}^m x_j^+ + \frac{1}{x_j^+} - x_j^- - \frac{1}{x_j^-}
  = -\frac{ih}{2} \Bigl( X^+ + \frac{1}{X^+} - X^- - \frac{1}{X^-} \Bigr) .
\end{equation}
$X^{\pm}$ thus satisfy the L shortening condition~\eqref{eq:mixed-shortening-cond}.

In the $u_{\sL}$ plane the bound state condition~\eqref{eq:bound-state-equations} translates to a simple Bethe string configuration of the form\footnote{As we saw in the previous section we can shift $u$ by a multiple of $2ik/h$ by taking $x^{\pm}$ through log cuts, so for~\eqref{eq:bethe-string-condition-u} to hold we need to pick the correct log branch for each $u_j$.}
\begin{equation}\label{eq:bethe-string-condition-u}
  u_{j+1} = u_j - \frac{2i}{h} ,
\end{equation}
where
\begin{equation}
  u_j\pm \frac{i}{h}\equiv u(x^\pm_j),
\end{equation}
like in the pure R-R case.

\begin{figure}
  \centering

  \includegraphics{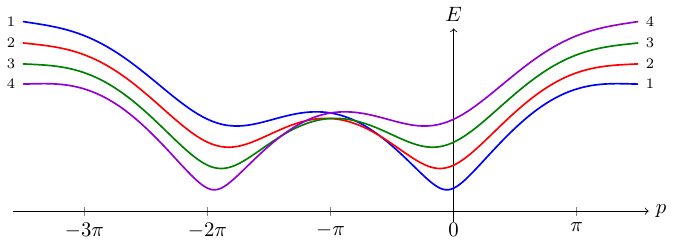}
  
  \caption{\label{fig:bs_disp_rel}Bound state dispersion relations for $h=2$, $k=5$ and $m = 1,\dotsc,k-1$. Note that for negative momentum $p$ and small bound state numbers $m$ the energy is \emph{decreasing} as $m$ increases.}
\end{figure}%
It is worth stressing a feature of the bound state dispersion relation not present in  R-R theories
\begin{equation}
E_{\sL}(p)=  \sqrt{(m + \kbar p)^2 + 4h^2\sin^2\frac{p}{2}} .
\end{equation}
In figure~\ref{fig:bs_disp_rel} we plotted the dispersion relation of bound states with $m=1,\dotsc,k-1$ for $k=5$. We note that for $p > 0$ the energy increases with increasing $m$, as happens in conventional relativistic theories and in R-R integrable backgrounds. However, as $p$ is lowered from $0$ to $-2\pi$ the energy levels cross each other, and for $p < -2\pi$ the energy \emph{decreases} as $m$ increases from $1$ to $k-1$. Hence, the lightest massive excitation for momentum on sheets with $p>0$,  has $m=1$, while on sheets with $p<-2\pi$ the lightest massive excitation has $m=k-1$. Therefore, bound states on $p>0$ sheets will be most naturally made up of $m=1$ excitations, while bound states on $p<-2\pi$ will be made up of $m=k-1$ excitations. As we will show in section~\ref{sec:m-is-k-1-L-rep-same-as-m-is-1-R-rep} below, an L excitation with $m=k-1$ and momentum $p$ is equivalent to an R $m=1$ excitation with momentum $p+2\pi$. As a result, on the $p<-2\pi$ sheets we can think of L fundamentals and L bound states as being made up of $R$ fundamentals.

From figure~\ref{fig:bs_disp_rel} we also see that the $-2\pi<p<0$ sheet does not have a massive excitation that is lightest for the full range of momenta. For momenta slightly less than 0, the $m=1$ L excitation is lightest and bound states can be most naturally made from it. For momenta slightly larger than $-2\pi$, the $m=k-1$ L excitation is lightest so bound states are naturally made from them or equivalently from the $m=1$ R excitations. The key feature of the $-2\pi<p<0$ sheet is then the fact that the notion of fundamental and bound state excitations does not make sense across the whole sheet, as there are momentum regions where so-called bound states become less energetic than the fundamentals they are made up from. We now discuss these observations more fully.

\paragraph{Bound states with \texorpdfstring{$p \in (0,2\pi)$}{p in (0,2pi)}.}

\begin{figure}
  \centering
  \subfloat[\label{fig:x-boundstates-0}$0<p<2\pi$]{\includegraphics{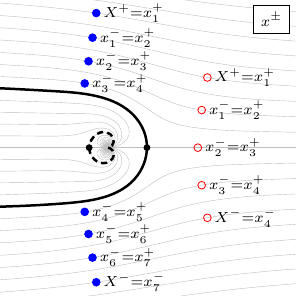}}
  \hspace{1cm}
  \subfloat[\label{fig:x-boundstates-1}$2\pi<p<4\pi$]{\includegraphics{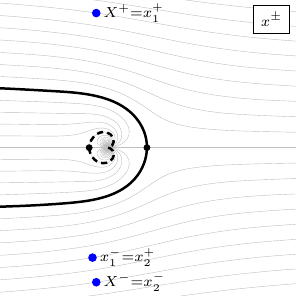}}
  
  \caption{\label{fig:x-boundstates}Typical configurations corresponding to bound states. Figure~\protect\subref{fig:x-boundstates-0} shows two bound states with momentum $p \in (0,2\pi)$. The filled dots on the right show the locations of the five parameters $x_1^{\pm},\dotsc,x_5^{\pm}$ making up a bound state with $m=7$. Similarly, the hollow dots on the rights show the locations of the four parameters describing a bound state with $m=4$. Figure~\protect\subref{fig:x-boundstates-1} shows a bound state with $m=2$ and $p \in (2\pi,4\pi)$.}
\end{figure}
Figure~\ref{fig:x-boundstates-0} shows two typical bound states with $m=4$ and $m=7$ and total momentum $p$ in the $(0,2\pi)$ range.\footnote{As in most figures in this paper, we set $k=5$ here.} The highest component sits at $x_1^+ = X^+ = \Xi_{\sL}^+(p,m)$. The next component sits two steps below that, at $x_1^- = x_2^+ = \Xi^+(q_1,m-2)$, where $q_1$ is a real parameter which can be determined through the shortening condition for $x_1^{\pm}$. The subsequent excitations sit along curves below this, with $m$ decreasing in steps of two, until we reach the lowest component $x_m^- = X^- = \Xi_{\sL}^-(p,m)$. In this way we can construct  bound states with total momentum in the $(0,2\pi)$ range, and with all constituents having (complex) momenta in the same range.

\begin{figure}
  \centering
  \includegraphics{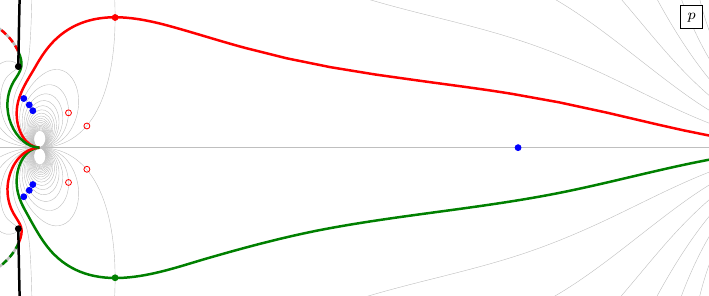}
  
  \caption{\label{fig:p-typical-bound-states}$p$ plane plot of two bound states with momentum in $(0,2\pi)$.}
\end{figure}
Figure~\ref{fig:p-typical-bound-states} shows the same two states as in figure~\ref{fig:x-boundstates-0} but in the $p$ plane. From conventional relativistic intuition of bound states one might expect that energy and momentum should be more or less evenly distributed among the constituents. This is indeed true for the $m=4$ states represented by the red hollow circles, because this state has relatively small total momentum. However, as we increase total momentum, most of the energy and momentum is carried by a single excitation as in the $m=7$ state represented by the blue dots in figure~\ref{fig:p-typical-bound-states}. Indeed, we can see that any excitation sitting on a contour $\Xi_{\sL}^{\pm}(p,m)$ with $m > 2$ is confined to a small region close to the origin in the $p$ plane. In the $u$-plane,  the $p>0$  bound states  sit at the same real value of $u$ and are separated by $\frac{2}{h}$ in the imaginary direction. In other words, they take the familiar ``Bethe string'' form~\eqref{eq:bethe-string-condition-u}, much like in higher-dimensional integrable holographic models~\cite{Roiban:2006gs,Dorey:2006dq}.

\paragraph{Bound states with \texorpdfstring{$p > 2\pi$}{p > 2pi}.}

\begin{figure}
  \centering
  \includegraphics{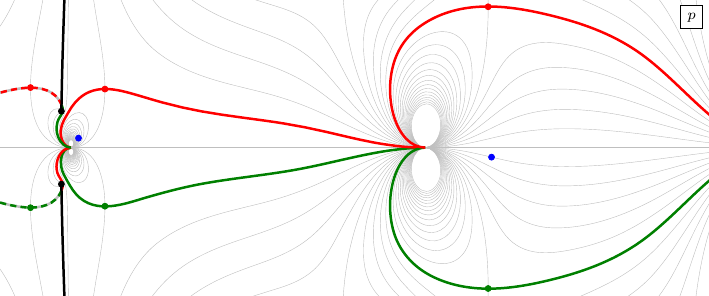}
  
  \caption{\label{fig:p-bound-state-region-1}A two particle bound state with total momentum in $(2\pi,4\pi)$ in the $p$ plane. Note that the constituent on the left has momentum which is almost purely imaginary.}
\end{figure}
Let us now consider a bound state with total momentum in $(2\pi,4\pi)$. We could try starting with two or more excitations each having momentum in $(2\pi,4\pi)$. However, the resulting state would have momentum larger than $4\pi$: the real part of the momentum of each excitation is at least $2\pi$ so a two-particle state would have total momentum in the $(4\pi,6\pi)$ range. Instead, starting with an bound state in the $(0,2\pi)$ range with bound state number $m<k$ , we can analytically continue the momentum of one of the constituents to the $(2\pi,4\pi)$ range, while imposing the bound state condition along the path.\footnote{The interested reader may find it useful to experiment themselves with such analytic continuations in the visualisation programme~\cite{pxu-gui}.} As a result, at the end of the analytic continuation, the remaining $m-1$ constituents will have momenta close to the origin of the $p$ plane. 

%However, if we want to avoid any excitation crossing the $x_{\sL}^-$ log cut the only option is to analytically continue the top component $x_1^{\pm}$ by taking $X^+ = x_1^+$ through the log cut.\footnote{This is probably easiest to see in the $u$ plane. If we take some $u_j$ with $j>1$ through the $x_{\sL}^+$ scallion cut, then $u_{j-1}$ will at the same time go through the $x_{\sL}^-$ cut which is located $2i/h$ above the $x_{\sL}^+$ cut.}

Consider the $m=2$ case in more detail: we begin with two excitations $x_1^{\pm}$ and $x_2^{\pm}$ which satisfy
\begin{equation}
  x_1^+ = X^+, \qquad
  x_1^- = x_2^+, \qquad
  x_2^- = X^- .
\end{equation}
For real momentum $p \in (0,2\pi)$, the middle roots $x_1^- = x_2^+$ are real\footnote{To be precise the middle root is real for momentum up to some critical value. Beyond that the middle root sit on one side of the scallion: $x_1^- = x_2^+ = \Xi^+(p',m=0)$ or $x_1^- = x_2^+ = \Xi^-(p',m=0)$.} and $X^{\pm} = \Xi_{\sL}^{\pm}(p,m=2)$. We now continue the top excitation to the next momentum region by taking $X^+$ through a log cut. Once we get back to a real momentum configuration we have
\begin{equation}
  X^+ = \Xi_{\sL}^+(p-2\pi,k+2) , \qquad
  x_1^- = x_2^+ = \Xi_{\sL}^-(p_1, k) , \qquad
  X^- = \Xi_{\sL}^-(p-2\pi,k+2) .
\end{equation}
Because $x_1^{\pm}$ sits in a higher momentum region there is a ``gap'' of size $2k$ in the Zhukovsky plane, as shown in figure~\ref{fig:x-boundstates-1}. Figure~\ref{fig:p-bound-state-region-1} shows the same state in the $p$ plane. We could instead have analytically continued the bottom excitation by taking $X^-$ through a log cut. This would result in a similar configuration, with $x^+$ and $x^-$ switched. 

Both these configurations are closely related to $m=k+2$ bound states considered in the $p\in(0,2\pi)$ paragraph above. For example, comparing the $m=k+2=7$ bound state shown in blue in figure~\ref{fig:x-boundstates-0} (recall that $k=5$ in the figure) to the $m=2$ configuration in figure~\ref{fig:x-boundstates-1}, we see that the bound state Zhukovsky variables $X^\pm$ are the same for both configurations, and the value of $x_6^-=x_7^+$ of the $m=7$ bound state agrees with $x_1^-=x_2^+$ of the $m=2$ bound state. As we will show in section~\ref{sec:singlet-states} below, the remaining $2k=10$  Zhukovsky variables of the $m=7$ bound state, not present in the $m=2$ bound state form a singlet state~\cite{Beisert:2005tm,Beisert:2006qh}, which has trivial scattering with all physical excitations. As a result, the $m=7$ bound state in figure~\ref{fig:x-boundstates-0} and the $m=2$ bound state in figure~\ref{fig:x-boundstates-1} are in fact the same state. 

If we had instead analytically continued the $X^-$ variable of the $m=2$ bound state to the higher momentum region, \textit{i.e.} through a $\log x$ cut, we would obtain a configuration similar to the one depicted in figure~\ref{fig:p-bound-state-region-1}, but mirrored around the horizontal axis. This state too is equivalent to 
the $m=7$ bound state in figure~\ref{fig:x-boundstates-0}, by an analogous argument involving another singlet state.

In a similar way we can obtain bound states in momentum regions $(2\pi n, 2\pi (n+1))$ with $n > 1$ by analytic continuation of one of its components. The singlets that imply equivalence of the different possible analytic continuations will in general involve $2k l$ Zhukovsky variables, with $l\in\Naturals^+$. For any such higher-momentum bound state almost all of the energy and momentum is carried by the component that is analytically continued to highest momentum sheet. The other constituents sit close to the origin of the $p$ plane. As we will show in section~\ref{sec:singlet-states}, these higher momentum bound states with momentum $p=2\pi l+p_0$ and bound state number $m_0$ are equivalent to bound states with momentum $p_0\in [0,2\pi]$ and  bound state number $m=m_0+kl$,

\paragraph{Bound states with \texorpdfstring{$p\in(-2\pi,0)$}{p in (-2pi,0)}.}

\begin{figure}
  \centering
  \subfloat[\label{fig:x-boundstates-min-1}$-2\pi<p<0$]{\includegraphics{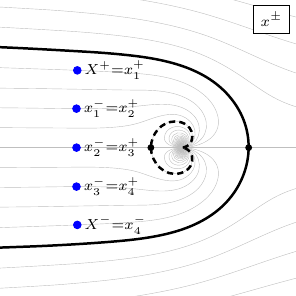}}
  \hspace{1cm}
  \subfloat[\label{fig:x-boundstates-min-2}$-4\pi<p<-2\pi$]{\includegraphics{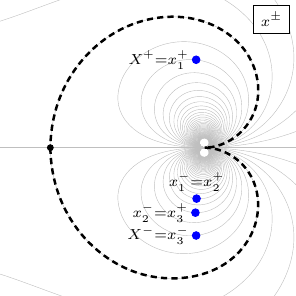}}
  
  \caption{\label{fig:x-boundstates-min}Typical configurations corresponding to bound states with $p < 0$. Figure~\protect\subref{fig:x-boundstates-min-1} shows a bound state with $m=4$ and $p \in (-2\pi,0)$ and figure~\protect\subref{fig:x-boundstates-min-2} shows a bound state with $m=2$ and $p \in (-4\pi,-2\pi)$. In figure~\protect\subref{fig:x-boundstates-min-2}, note $x_j^-$ gets closer to the kidney contour with increasing $j$.}
\end{figure}
\begin{figure}
  \centering
  \includegraphics{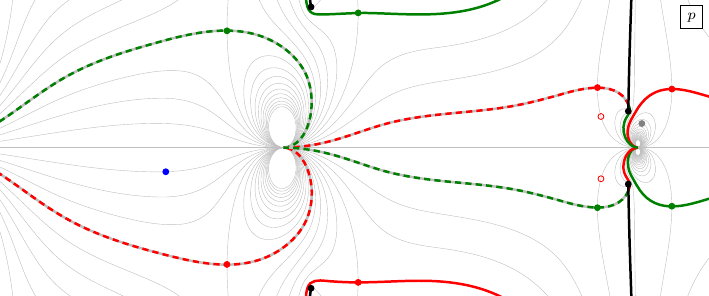}
  
  \caption{\label{fig:p-bound-state-regions-min-1-min-2}Two two-particle bound states. The red circles show a bound state with momentum in the $(-2\pi,0)$ range, while the dots show a bound state in the $(-4\pi,-2\pi)$ range. Note that the dot on the right, close to the origin of the $p$ plane, is drawn in grey to represent that it is located on the other sheet of the $p$ plane, and thus carries energy with a negative real part.}
\end{figure}
To construct a bound state with momentum in the range $(-2\pi,0)$, we can start with a bound state in $(0,2\pi)$ and analytically continue all Zhukovsky parameters through the scallion, essentially following the same path we would use for a single fundamental excitation, crossing $\log x$ cuts in the process.\footnote{If we were mainly interested in states with $p < 0$ it would be more natural to put the log cut along the positive real line. Another choice could be to put it along the interval from $x=0$ to $x=+s$ and from there along the upper half of the scallion contour. However, we find it easiest to leave the log cut along the conventional negative part of the Zhukovsky real line.} An example of such a bound state is shown in figure~\ref{fig:x-boundstates-min-1}.

\begin{figure}
  \centering
  \subfloat[\label{fig:p-bs-3-region-min-1}]{\includegraphics{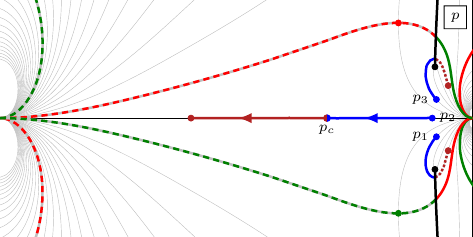}}
  \hspace{1cm}
  \subfloat[\label{fig:u-bs-3-region-min-1}]{\includegraphics{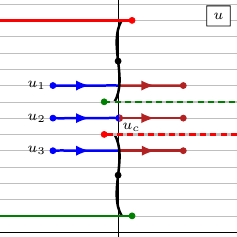}}
  
  \caption{\label{fig:bs-3-region-min-1}Three-particle bound state. Here $h=1$ and $k=7$.}
\end{figure}
In this momentum region bound state constituents exhibit a novel property we aluded to above when discussing figure~\ref{fig:bs_disp_rel}. To see this in detail, consider a three-particle state with momentum in the range $(-2\pi,0)$ and close to 0, as shown by the blue dots in figure~\ref{fig:bs-3-region-min-1}. The middle excitation has real momentum $p_2$ and thus real energy $E_2$, while the outer excitations have complex momenta $p_1$ and $p_3$ and complex energies $E_1$ and $E_3$. Since the total momentum and energy of the bound state are real, we have $p_1 = p_3^*$ and $E_1 = E_3^*$. Let us now make the total momentum more negative. This also decreases $p_2$. As $p_2$ hits some critical value $p_{\text{c}}$, the outer excitations hit the black line representing the cut of $E(p)$. This means that the real parts of $E_1$ and $E_3$ vanish. If we further lower the momentum, $\Re E_1$ and $\Re E_3$ will become negative. Hence, for $p_2 < p_{\text{c}}$ we find that a single ``fundamental'' $m=1$ L excitation with momentum $p_2$ has higher energy than the $m=3$ L three-particle bound state we built from it. We saw precisely this behaviour earlier in the plot of the bound state dispersion relations in figure~\ref{fig:bs_disp_rel}: at low momentum the energy decreases as the bound state number increases. This unusual non-relativistic behaviour does not lead to a pathology. If we think about the bound state dispersion relation as a function of $m$ for a fixed bound state momentum $p$, that function has a minimum at $m = -\kbar p$, which means that the spectrum still is bounded from below. As we have argued above, in this intermediate region of momenta, or in other words between the kidney and scallion, there is no single excitation which has lowest energy for all momenta. The analytic continuation we have just described starts in a sub-region where the $m=1$ excitation is least energetic and so it is natural to think of the $m=3$ excitation as a bound state of it. But, as we continue to lower momenta, the $m=3$ excitation becomes less energetic than its $m=1$ constituents. We can continue thinking of it as a bound state, if we allow the constituents to live on the  crossed momentum sheet. Alternatively, we can declare that here the $m=3$ excitation is ``fundamental''. 

Finally, we note that there are only $2k-1$ contours between the scallion and the kidney. Thus, only bound states with $m \le k$ can have all their constituents in the same region. If we want to go beyond that we need to include excitations that have, \eg, $x^-$ inside (or on) the scallion and $x^+$ outside the scallion. Allowing for that we can however build bound states for any $m$. These will be related to lower bound state number states  with more momentum, using the singlet argument from section~\ref{sec:singlet-states}.

\paragraph{Bound states with \texorpdfstring{$p < -2\pi$}{p < 2pi}.}

To construct a bound state with $p \in (-4\pi,-2\pi)$ we start with a bound state in the $(-2\pi,0)$ region and continue the top component to the $(-4\pi,-2\pi)$ region. In doing so the other constituents of our starting bound state will unavoidably go into the crossed $(0,2\pi)$ region. This results in a state where all the excitations sit inside the kidney, see figure~\ref{fig:x-boundstates-min-2}. The same construction can be used to build bound states with $p < -4\pi$. We see again the consequences of the discussion around figure~\ref{fig:bs_disp_rel}: insisting on viewing these states as bound states of $m=1$ fundamental excitations necessarily implies that some of the $m=1$ constituents must be on crossed momentum sheets. 

To obtain a more conventional picture of bound states for $p<-2\pi$, recall that in this region the $m=k-1$ L excitations are lightest. We can therefore form bound states of the $m=k-1$ excitations, whose constituents will remain on the physical $m=k-1$ momentum sheet, as long as the bound state itself is on the physical sheet. In fact, this can further be simplified, because, as we will show in section~\ref{sec:m-is-k-1-L-rep-same-as-m-is-1-R-rep} below, $m=k-1$ L excitations are equivalent to $m=1$ R excitations. Therefore, the bound states we are discussing here are equivalent to bound states of R fundamentals in the region \textit{outside} the R-scallion. But such bound states are constructed in essentially an identical way to the construction of boundstates from $m=1$ L fundamentals that we discussed in the $p>0$ and $p>2\pi$ paragraphs above.\footnote{For example, such R bound states are easily deduced from the L ones shown in figure~\ref{fig:x-boundstates-0} by reflecting in the imaginary axis.}

%Just as for bound states in the $(-2\pi,0)$, for a given momentum region $(2\pi n, 2\pi(n+1))$ with $n < -1$ there is only a finite number of states ($|n|k-1$) with all excitations inside the kidney.

\subsection{``Physical'' regions}
\label{sec:physical-regions}

In the pure R-R theory, the dispersion relation~\eqref{eq:R-R-disp-rel} is periodic. This is an unusual property for a theory on a decompactified worldsheet and allows for a periodic identification of momentum. For example, upon this identification, analytically continuing past the $p=2\pi$ point, following a path analogous to the blue line labelled 1 in figure~\ref{fig:p-plane-path-between-region}, brings one back to the $p=0$ region.\footnote{The interested reader may find it instructive to set $k=0$ in the accompanying visualisation programme~\cite{pxu-gui} to compare and contrast the R-R and mixed-flux complex momentum planes.} 
%In Zhukovsky variables, this translates into the physical region being outside the unit circle, with the convention that $x^+$ is in the UHP and $x^-$ in the LHP. 

As we have discussed at the begining of Section~\ref{sec:mixed-flux-kinematics}, for mixed-flux backgrounds the dispersion relation~\eqref{eq:disp-rels-mf} is no longer periodic and no corresponding identification of the momentum complex plane can be made. Since analytic continuation past the points $p=2\pi n$, for $n\in \Integers$, as shown by the blue paths in figure~\ref{fig:p-plane-path-between-region}, does not cross the black $E(p)$ cuts, it leaves us on the physical momentum plane. Its effect is simply to increase the physical excitation's momentum. Unlike in the periodic R-R theory, this is conventional behaviour for magnons on a decompactified worldsheet and implies that at least a neighbourhood of the whole real momentum line needs to be included in the physical region. In turn, analyticity of the S matrix requires that one cannot arbitrarily restrict to some sub-region  of the complex momentum plane which does not include the whole real momentum line. 

%Therefore, the \textit{physical region} of the mixed flux theory is the whole complex momentum plane $p$ as shown in figure~\ref{fig:Ep-branch-cuts} and the \textit{crossed region} is a second complex momentum plane reached by analytic continuation through one of the black cuts, as we discuss more fully in section~\ref{sec:cross-transformation} below.

Based on experiences with R-R theories~\cite{Arutyunov:2007tc}, it may seem useful to identify the physical region in the $x^\pm$ and $u$  variables. In R-R theories the physical region is the minimal region such that all excitations used to build up physical states are located in it. Conventionally, it is taken to be the region outside the unit disc $\left|x^\pm\right|>1$ or equivalently the full ``top'' $u$ sheet with two cuts. As we have seen in section~\ref{sec:p-x-u-planes}, in the mixed-flux setting, there are an infinite number of $u$ planes labelled by $n\in\Integers$, each one containing the image of a real momentum interval $p\in [2\pi n,2\pi (n+1)]$.
Analytically continuing past the points $p=2\pi n$ for $n\in \Integers$, as shown by the blue paths in figure~\ref{fig:p-plane-path-between-region}, takes one through the (pre-image) of a log cut to \textit{new} $u$-rapidity planes. Apart from the $n=-1$ plane, each of these $u$-planes contains a pair of ``short'' cuts, see figure~\ref{fig:u-physical-regions}, with the image of the corresponding real momentum interval a horizontal line marked by a blue dot. The short cuts are images of scallions for $n> -1$ and kidneys for $n<-1$.

For excitations with momentum $p\in[0,2\pi]$, we have seen in the preceding section that fundamentals and bound state constituents are all located outside the scallions (\textit{cf.} figure~\ref{fig:x-boundstates-0}). Hence, the physical region is the whole $n=0$ $u$ plane with ``short''  red and green cuts shown in figure~\ref{fig:u-physical-regions-0}. These cuts are the images of the $x^+$ and $x^-$ scallions and the physical $u$ plane corresponds to  the region \textit{outside} the (red and green) scallions depicted in figure~\ref{fig:full-xp-xm-region-0}.\footnote{As in the R-R case, we take $x^+$ to be in the UHP and $x^-$ in the LHP in order for the excitation to have positive energy.} 
%As we will see below,  the physical Zhukovsky regions to be outside the scallions, because, as we will see, the dressing phases are most naturally expressed in terms of integrals along the scallion and kidney contours.\footnote{Just as in the R-R case, where dressing phases' integration contours lie along the unit circle of the Zhukovsky plane, it is of course possible to deform the contours of integration slightly away from those arbitrary boundaries and correspondingly extend or reduce the physical region.} This choice of physical region is consistent with  bound states which have momenta $p>0$ constructed in the preceding sub-section since the $m=1$ constitutents of these bound states were always located outside the scallions, as illustrated in figure~\ref{fig:x-boundstates-0}. 
As mentioned in the preceding section and discussed more fully in section~\ref{sec:singlet-states} below,  bound states of momentum $p_0$ with bound state number $m=m_0+kl$, with $l,m_0\in\Integers^+$, are equivalent to bound states with $m=m_0$ and $p=p_0+2\pi l$. This equivalence means that we can also include all the  $u_{n>0}$ sheets with pairs of scallion cuts in our definition of the $p>0$ physical region. The inclusion of these sheets introduces a redundancy in the physical region, by giving equivalent ways of describing the same excitaitons. At the same time, including them allows us to understand the consequences of analytic continuation across the \textit{full} momentum plane using the identifications of section~\ref{sec:singlet-states}, without overcounting the number of independent excitations.

  For bound states with momenta $p<-2\pi$, we saw that it is most natural to think of these as bound states made up of $m=k-1$ L bound states or equivalently as $m=1$ R fundamentals,\footnote{See section~\ref{sec:m-is-k-1-L-rep-same-as-m-is-1-R-rep} below, for more details.} since these are the lightest massive excitations in this momentum region.  When expressed in terms of $m=1$ R fundamentals, the physical region is outside the R scallions, in direct analogy with the $p>0$ physical region being outside the L scallions of the preceding paragraph.\footnote{In terms of the $m=k-1$ L variables, this is the region \textit{inside} the L kidney.} Similarly, the physical region corresponds to the full $u_{R, -2\pi<p_R<0}$ sheet with two ``short'' cuts where the R scallions (a.k.a. L kidneys) sit.  As in the case of the  $u_{n>0}$ sheets from the previous paragraph, the equivalence of bound states whose bound state numbers differ by integer multiples of $k$ and momenta differ by corresponding multiples of $2\pi$ implies that all $u_{R, p_R<-2\pi}$ sheets with single R scallions (equivalently L kidney cuts) can also be included as part of the physical region. 
  
 In summary, ignoring the $u_{n=-1}$ sheet for now, we find that for $n>-1$, the physical region is always \textit{outside} the two L scallions, while for $n<-1$, the physical region is always \textit{inside} the two L kidneys. In terms of the $u$ variable, the physical region is the union of all the $u_L$ sheets with two L scallion cuts and $u_R$ sheets with two R scallion cuts, the latter being equivalent to $u_L$ sheets with two L kidney cuts. This ``outside/outside''  and ``inside/inside''  mnemonic for the physical Zhukovsky regions is shown in figure~\ref{fig:p-short-cut-regions}. All these regions are therefore quite similar to those in the R-R cases and reduce to regions outside the unit disc in the formal $k\rightarrow 0$ limit.

%in the corresponding Zhukovsky planes is the region \textit{outside} the (red and green) scallions shown in figure~\ref{fig:full-xp-xm-region-0}, and is the full $u$-plane in figure~\ref{fig:u-simple-path-1}, with ``short'' cuts in red and green.\footnote{As in the R-R case, we take $x^+$ to be in the UHP and $x^-$ in the LHP in order for the excitation to have positive energy.} The real momentum interval $p\in[0,2\pi]$ is mapped to the contour $\Xi_{\sL}^{+}(p,m=1)$, which is the light-grey line directly above the red scallion in the $x^+$ plane in figure~\ref{fig:full-xp-region-0}; similarly it is mapped to the contour $\Xi_{\sL}^{-}(p,m=1)$, which is the light-grey line directly below the green scallion in the $x^-$ plane in figure~\ref{fig:full-xm-region-0}. 

The $n=-1$ $u$ plane in figure~\ref{fig:u-physical-regions-min-1} is rather different for two reasons. Firstly, with ``short'' cuts, it has a novel $\frac{2k}{h}$ periodicity along the imaginary $u$ direction. Secondly, the black $E(p)$ cuts are now visible on the same $u$ sheet as the image of the real momentum interval $p\in[-2\pi,0]$, as shown in figure~\ref{fig:u-physical-regions-min-1}. As we have discussed in the previous sub-section, for certain values of momenta the ``fundamental'' $m=1$ excitations that make up an $m>1$ bound state in this region inevitably end up crossing from the physical momentum sheet to the crossed one by going through a black cut, \textit{cf.}  figure~\ref{fig:bs-3-region-min-1}. This is an unavoidable consequence of the mixed flux dispersion relation and shows that on the $u_{n=-1}$ sheet there is no intrinsic notion of a physical $u$ sheet beyond the one inherited from the momentum plane, when viewing bound states. In fact, when establishing the periodicity of the $u$ plane one crosses the $E(p)$ cut twice, going from physical to crossed and back to physical momentum sheets. This point will be further emphasized when we consider the singlet states in section~\ref{sec:singlet-states}. As we will see in detail there, singlet states consist of \textit{physical} $m=k-1$ L bound states and \textit{crossed} R fundamentals, whose Zhukovsky and $u$ variables sit at the \textit{same} locations. Therefore, in this setting the only notion of ``physical'' or ``crossed'', is inherited from the momentum plane itself.\footnote{Note that these conclusions remain unchanged if one considered ``long'' cuts instead of ``short'' ones, since these cuts are unrelated to the cuts of $E(p)$.}

\subsection{The R representation}

\begin{figure}
  \centering
  \subfloat[\label{fig:pR-plane}The $p_{\sR}$ plane]{\includegraphics{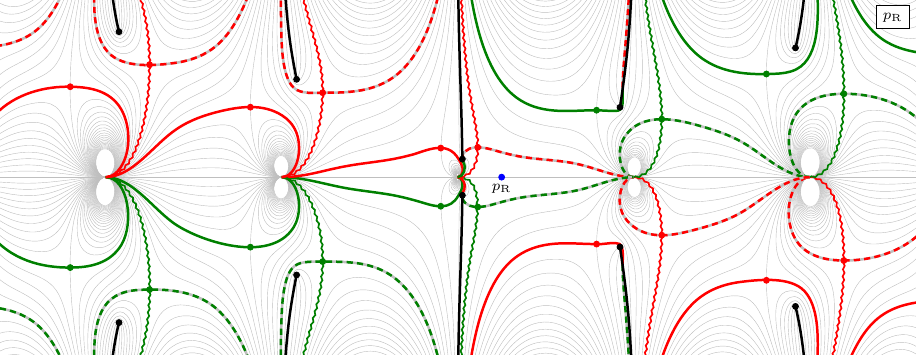}}

  \subfloat[\label{fig:xR-plane}The $x_{\sR}$ plane]{\includegraphics{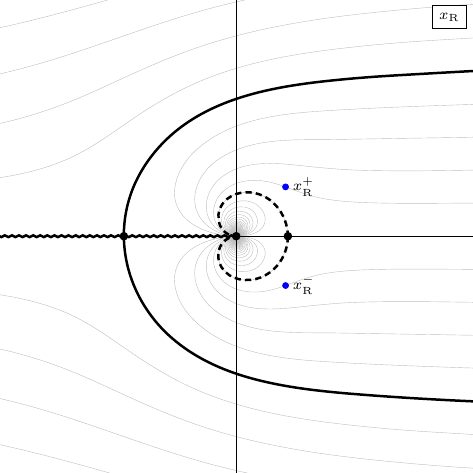}}
  \hspace{2cm}
  \subfloat[\label{fig:uR-plane}The $u_{\sR}$ plane]{\includegraphics{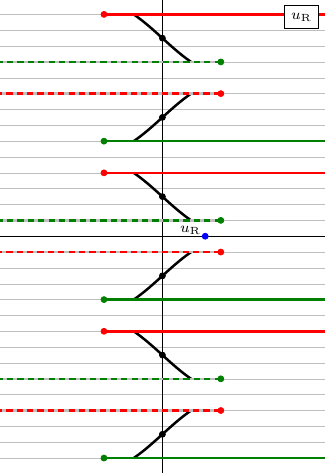}}
  
  \caption{\label{fig:R-p-x-u-planes}The $p_{\sR}$, $x_{\sR}$ and $u_{\sR}$ planes. The blue dots show the location of a fundamental excitation with momentum $p \in (0,2\pi)$. Red (green) solid lines represent the $x_{\sR}^+$ ($x_{\sR}^-$) scallion and red (green) dashed lines represent the $x_{\sR}^+$ ($x_{\sR}^-$) kidney. The solid black lines in the $p_{\sR}$ and $u_{\sR}$ planes show the cut of the R dispersion relation. These cuts are not shown in the $x_{\sR}$ plane since they would be located at different spots for $x_{\sR}^+$ and $x_{\sR}^-$.}
\end{figure}
So far we have mainly discussed the L representation. As can be seen from the shortening condition~\eqref{eq:mixed-shortening-cond} the R representation is essentially the parity conjugate of the L representation. Figure~\ref{fig:pR-plane} shows the $p_{\sR}$ plane, which looks like the $p_{\sL}$ plane but reflected along the imaginary axis. In figure~\ref{fig:xR-plane} we show the $x_{\sR}$ plane, with the same scallion and kidney cuts as in the L case, but again reflected along the imaginary axis. Note that we still put the log cut along the negative real line. The $u_{\sR}$ rapidity is defined through the function
\begin{equation}
  u_{\sR}(x)  = x + \frac{1}{x} + \frac{2\kbar}{h} \log x , \qquad
  u_{\sR}(x_{\sR}^{\pm}) = u_{\sR} \pm \frac{i}{h} .
\end{equation}
The $u_{\sR}$ plane is shown in figure~\ref{fig:uR-plane}. We note that the scallion and kidney cuts  in the $u_{\sR}$ plane go in the opposite direction to those of the $u_{\sL}$ plane, and that the whole plane is shifted by $ik/h$, so that the real line represents the momentum range $(0,2\pi)$, which for R excitations corresponds to the region between the scallion and the kidney in the $x_{\sR}$ plane.

\vspace{1cm}

\textbf{Note added:} Preliminary discussions of mixed flux kinematics and bound states were presented by OOS at the 
\href{https://www.youtube.com/watch?v=Zsj9IJ789_E&list=PLPEZqJwA6SghjkCqzuqRLBddyMVpM71EZ&index=11}
{London Integrability Journal Club} on the 4th of June 2020 and by BS during the \href{https://sites.google.com/view/low-d-integrability-hmi/homepage}{``Integrability in Lower Dimensional AdS/CFT''} virtual workshop held between the 17th and 20th of August 2021, where the scallion, kidney and constant $m$ contours were introduced.\footnote{We would like to thank Sergey Frolov and Alessandro Sfondrini for their questions and discussions at the time of those presentations.} The recent paper~\cite{Frolov:2023lwd} has also independently analysed the dispersion relation, $x^\pm$ and $u$  variables and bound states introduced for the mixed-flux backgrounds in~\cite{Hoare:2013lja,Lloyd:2014bsa}. As we discussed in detail above, different choices can be made for the location of cuts in the $u$ plane, much as long or short cuts can be used in R-R backgrounds. We analysed explicitly cut configurations where \textit{both} kidney and scallion cuts are ``long'' and where \textit{both} are ``short''.  In ~\cite{Frolov:2023lwd}, a combination of a ``short'' cut for the L-scallion and a ``long'' cut for the L-kidney was studied in detail (similarly for the R excitations). While physical quantities do not depend on the choice of cuts, the advantage of selecting all ``short'' cuts is that they exhibit more clearly the novel u-plane periodicity for excitations whose physical momenta lie in the region between the scallion and kidney.

\section{The crossing transformation}
\label{sec:cross-transformation}

In this section we discuss the crossing transformation in mixed-flux kinematics. Analogously to the R-R case, the mixed-flux crossing transformation sends\footnote{The bar over $p$, $E$ and $x^\pm_I$ is often used in the literature to denote the values of variables at the crossed point and should not be confused with complex conjugation.}
\begin{equation}
  p \to \bar{p} = -p , \qquad
  E(p) \to \bar{E}(\bar{p}) = -E(-p) .
\end{equation}
In contrast to the R-R case, the mixed-flux model is not parity invariant. In particular, the dispersion relations are not even functions of the momentum $p$, which means that the negation of the momentum in the last expression above is important. To understand how crossing acts in the Zhukovsky variables let us start at a position $x_{\sL}^{\pm}(p) = \Xi_{\sL}^{\pm}(p,1)$. Crossing takes us through the cut in the dispersion relation so after crossing we should arrive at $\bar{x}_{\sL}^{\pm}(\bar{p}) = \tilde{\Xi}_{\sL}^{\pm}(\bar{p},1)$. Now we can use the relations
\begin{equation}
  \tilde{\Xi}_{\sL}^{\pm}(p,m) = -\frac{1}{\Xi_{\sL}^{\mp}(p,m)} , \qquad
  \Xi_{\sR}^{\pm}(p,m) = -\Xi_{\sL}^{\mp}(-p,m) ,
\end{equation}
to find that $\bar{x}_{\sL}^{\pm}(\bar{p}) = -1/\Xi_{\sL}^{\mp}(\bar{p},1) = 1/\Xi_{\sR}^{\pm}(p,1)$ so that the crossing transformation can be expressed as
\begin{equation}\label{eq:zhukovsky-vars-crossing}
  x_{\sL}^{\pm}(p) \to \bar{x}_{\sL}^{\pm}(\bar{p}) = \frac{1}{x_{\sR}^{\pm}(p)} .
\end{equation}
The fact that crossing an L excitation gives an R excitation can be anticipated from~\eqref{eq:mixed-shortening-cond}:  $1/x_{\sR}^{\pm}(p)$ satisfy the L shortening conditions while $1/x_{\sL}^{\pm}(p)$ do not.

\begin{figure}
  \centering    
  \subfloat[]{\includegraphics{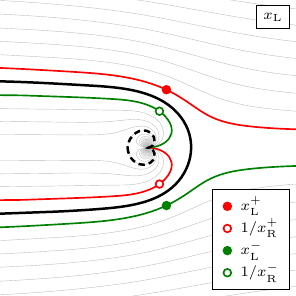}}
  \hspace{2cm}
  \subfloat[]{\includegraphics{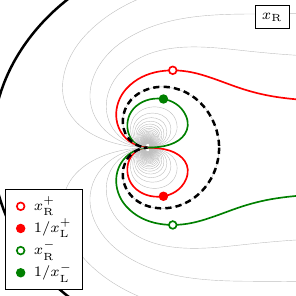}}
  \caption{\label{fig:crossing-xL-xR-0-2pi}Location of $x_{\sL}^{\pm}(p)$, $x_{\sR}^{\pm}(p)$, $1/x_{\sL}^{\pm}(p)$, and $1/x_{\sR}^{\pm}(p)$ for real momentum in the range $0 \le p < 2\pi$. Crossing of $x_{\sL}^{\pm}(p)$ is associated with going across the outer black scallion contour and crossing of $x_{\sR}^{\pm}(p)$ is associated with going across the inner black dashed kidney contour.}
\end{figure}
\begin{figure}
  \centering
  \subfloat[]{\includegraphics{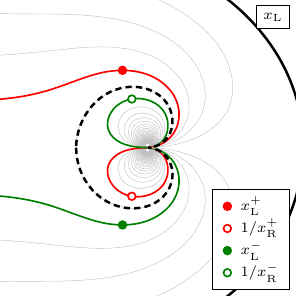}}
  \hspace{2cm}
  \subfloat[]{\includegraphics{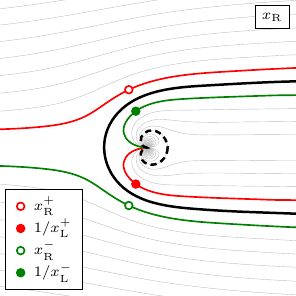}}
  \caption{\label{fig:crossing-xL-xR--2pi-0}Location of $x_{\sL}^{\pm}(p)$, $x_{\sR}^{\pm}(p)$, $1/x_{\sL}^{\pm}(p)$, and $1/x_{\sR}^{\pm}(p)$ for real momentum in the range $-2\pi \le p < 0$. Crossing of $x_{\sL}^{\pm}(p)$ is associated with going across the inner black dashed kidney contour and crossing of $x_{\sR}^{\pm}(p)$ is associated with going across the outer black scallion contour.}
\end{figure}
Figures~\ref{fig:crossing-xL-xR-0-2pi} and~\ref{fig:crossing-xL-xR--2pi-0} show the points $x_{\sL}^{\pm}$ and $x_{\sR}^{\pm}$ of two physical excitations with real momentum $p$, as well as the corresponding crossed points $1/x_{\sR}^{\pm}$ and $1/x_{\sL}^{\pm}$, for momenta in the ranges $0 \le p < 2\pi$ and $-2\pi \le p < 0$, respectively. These figures show that the scallion and kidney contours make up a natural border between the physical and crossed regions, for these ranges of momenta: the crossing transformation is an analytical continuation of the S matrix  from \textit{outside} to \textit{inside} these contours.

\begin{figure}
  \centering
  \includegraphics{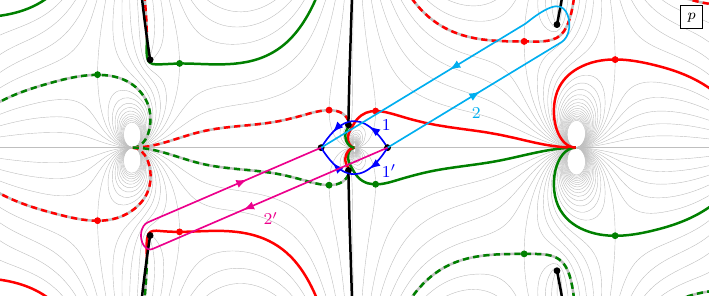}
   \caption{\label{fig:p-plane-short-cuts-crossing-choices}A few of the possible choices for crossing paths in the $p$ plane with short cuts.}	 
  \end{figure}

Since the Zhukovsky and $u$ planes are made up of many sheets we need to have a convention for how to perform the analytic continuation from a physical to a crossed point. From the point of view of the $p$ plane it is clear that we have to take a path from $p$ to $-p$ and cross one of the black $E_I(p)$ cuts depicted in figure~\ref{fig:Ep-branch-cuts}. \textit{A priori} we can cross any one of these cuts and some choices are depicted in figure~\ref{fig:p-plane-short-cuts-crossing-choices}. Depicted there are two dark blue paths labelled $1$ and $1^\prime$, which take a  direct path between $p$ and $\bar p$. 
\begin{figure}
  \centering
  \subfloat[$x^+_{\sL}$]{\includegraphics{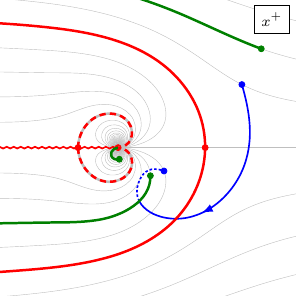}}
  \hspace{1cm}
  \subfloat[$x^-_{\sL}$]{\includegraphics{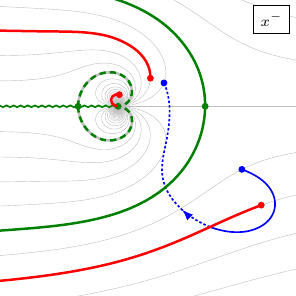}}
  \caption{\label{fig:xp-plane-short-cuts-main-crossing-path}Our chosen crossing path in the $x^\pm_{\sL}$ planes, for starting momenta $p\in [0,2\pi]$.}	
\end{figure}
The image of  path $1$ in Zhukovsky variables is shown in figure~\ref{fig:xp-plane-short-cuts-main-crossing-path}. The convention we adopt is that both $x^+_{\sL}$ and $x^-_{\sL}$ scallions are entered from the \textit{lower}-half-plane.\footnote{This convention is chosen in analogy with the R-R case and higher-dimensional integrable backgrounds, where the crossing paths enter the unit discs from the LHP of both $x^+$ and $x^-$.}. We will adopt this path as our chosen crossing path when solving the crossing equations below. Other crossing paths such as $1'$, or the light-blue and pink ones depicted in figure~\ref{fig:p-plane-short-cuts-crossing-choices} as well as many others can also be chosen and they lead to the same required transformations of the Zhukovsky variables~\eqref{eq:zhukovsky-vars-crossing}. However, their paths through the various cuts is often more complicated than our minimal choice and would lead to more complicated expressions for dressing phase solutions.\footnote{The path $1^\prime$ is equivalent in simplicity to the path $1$. It does not go over naturally to the usual definition of crossing for R-R theories and it is for this reason we do not adopt it. As in the R-R cases, it would be straightforward to find dressing phases for this definition of crossing, whose form would be very similar to the conventional ones.}

\begin{figure}
  \centering
  \includegraphics{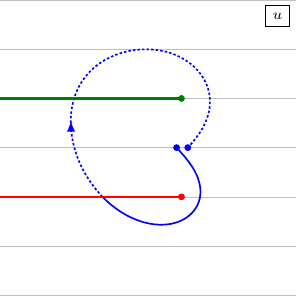}
  \caption{\label{fig:u-crossing-0}Our chosen crossing path in the $u_{\sL}$ plane, for starting momenta $p\in [0,2\pi]$.}	 
\end{figure}
Our chosen crossing path takes a simple form in the $u$ plane as depicted in figure~\ref{fig:u-crossing-0}. This is entirely analogous to the familiar R-R case where crossing involves going through both the $x^+$ and $x^-$  cuts a single time in the same direction. Note that, unlike the R-R case, the end-point of crossing in the $u$ plane is not the same value of $u$, because of the modified definition of the Zhukovsky map~\eqref{eq:mixed-u-of-x}. Instead, it takes us from $u = u_{\sL}(x_{\sL}^{\pm}(p)) \mp \frac{i}{h}$ to $\bar{u} = u_{\sR}(x_{\sR}^{\pm}(p)) \mp \frac{i}{h}$.

\section{The two-particle S matrix}
\label{sec:two-particle-s-matrix}

As discussed in detail in~\cite{Lloyd:2014bsa}, given two short representations\footnote{All (non-trivial) short representations of $\algPSU(1|1)^4_{\ce}$ consist of two bosons and two fermions and are parameterised by the momentum $p$ and the charge $M$, which in the mixed-flux case takes the form $M = m \pm \kbar p$ with $m \in \Integers$.} of $\algPSU(1|1)^4_{\ce}$ the two-particle S matrix is fully determined by the $\algPSU(1|1)^4_{\ce}$ symmetry up to an overall scalar factor. If we consider only massive fundamental excitations that leaves us with four undetermined functions, which we can take as the matrix element for scattering two highest weight states
\begin{equation}
  A_{\sLL}^2(p,q) , \qquad
  A_{\sLR}^2(p,q) , \qquad
  A_{\sRL}^2(p,q) , \qquad
  A_{\sRR}^2(p,q) .
\end{equation}
These functions are further restricted by additional symmetries of the model
\paragraph{Physical unitarity.} The S matrix of real momentum excitation should be  pure phase
  \begin{equation}
    \begin{aligned}
    A_{\sLL}^2(p,q)^{\dag}A_{\sLL}^2(q,p)
    &= 1, \quad &
    A_{\sLR}^2(p,q)^{\dag}A_{\sLR}^2(q,p)
    &= 1, \\
    A_{\sRL}^2(p,q)^{\dag}A_{\sRL}^2(q,p)
    &=1, &
    A_{\sRR}^2(p,q)^{\dag}A_{\sRR}^2(q,p)
    &=
    1 .
  \end{aligned}
  \end{equation}
\paragraph{Braiding unitarity.} For the Zamolodchikov-Faddeev algebra to be consistent we need to impose the conditions~\cite{Arutyunov:2009ga}
  \begin{equation}
    A_{\sLL}^2(p,q)A_{\sLL}^2(q,p)
    =
    A_{\sLR}^2(p,q)A_{\sRL}^2(q,p)
    =
    A_{\sRR}^2(p,q)A_{\sRR}^2(q,p)
    =
    1
  \end{equation}
\paragraph{Parity.} The mixed-flux world-sheet theory is not parity invariant. However, a parity transformation simply swaps the L and R representations.\footnote{Compared to the R-R theory this symmetry combines parity symmetry theory with the so called ``LR symmetry''. These are separate symmetries in the R-R theory but in the mixed-flux model only the combination of the two operations gives an actual symmetry.} The parity transformation acts on the Zhukovsky variables by sending $x_{\sL}^{\pm} \to -x_{\sR}^{\mp}$. From the matrix part of the S matrix we find
  \begin{equation}
    \begin{aligned}
      \mathbb{S}_{\sRR}^t(-p,-q) \mathbb{S}_{\sLL}(p,q) &= A_{\sLL}^2(p,q) F_{\sRR}^2(-p,-q) \mathds{1}, \\
      \mathbb{S}_{\sRL}^t(-p,-q) \mathbb{S}_{\sLR}(p,q) &= A_{\sLR}^2(p,q) F_{\sRL}^2(-p,-q) \mathds{1} .
    \end{aligned}
  \end{equation}
  From this we find the following constraints on the scalar factors
  \begin{equation}
    \begin{aligned}
      A_{\sRR}^2(-p,-q) A_{\sLL}^2(p,q) &= \frac{x^-}{x^+} \frac{y^+}{y^-} \bigl( \frac{x^+ - y^-}{x^- - y^+} \bigr)^2 , \\
      A_{\sRL}^2(-p,-q) A_{\sLR}^2(p,q) &= \frac{x^+}{x^-} \frac{y^+}{y^-} \bigl( \frac{1 - \frac{1}{x^+y^+}}{1 - \frac{1}{x^-y^-}} \bigr)^2 , \\
    \end{aligned}
  \end{equation}
\paragraph{Crossing symmetry.} 
The crossing equations in the mixed-flux background were found in~\cite{Lloyd:2014bsa}. It is useful to write them directly in terms of S matrix elements since those equations are independent of the normalisation of the S matrix. Thus, let us define the matrix elements
\begin{equation}
  \begin{aligned}
    \Upsilon^{\sLL}_{pq} & = \braket{ Y^{\sL}_q Y^{\sL}_p | \Smat | Y^{\sL}_p Y^{\sL}_q } = A_{\sLL}^2(p,q) , \quad &
    \Upsilon^{\sRL}_{pq} & = \braket{ Y^{\sL}_q Y^{\sR}_p | \Smat | Y^{\sR}_p Y^{\sL}_q } = D_{\sRL}^2(p,q) ,         \\
    % \Upsilon^{\sRR}_{pq} & = \braket{ Y^{\sR}_q Y^{\sR}_p | \Smat | Y^{\sR}_p Y^{\sR}_q } = F_{\sRR}^2(p,q) , \quad &
    % \Upsilon^{\sLR}_{pq} & = \braket{ Y^{\sR}_q Y^{\sL}_p | \Smat | Y^{\sL}_p Y^{\sR}_q } = B_{\sLR}^2(p,q) , \\
    \Lambda^{\sRL}_{pq}  & = \braket{ Z^{\sL}_q Y^{\sR}_p | \Smat | Y^{\sR}_p Z^{\sL}_q } = F_{\sRL}^2(p,q) , \quad &
    \Lambda^{\sLL}_{pq}  & = \braket{ Z^{\sL}_q Y^{\sL}_p | \Smat | Y^{\sL}_p Z^{\sL}_q } = B_{\sLL}^2(p,q).
    % \Lambda^{\sLR}_{pq}  & = \braket{ Z^{\sR}_q Y^{\sL}_p | \Smat | Y^{\sL}_p Z^{\sR}_q } = A_{\sLR}^2(p,q) , \quad &
    % \Lambda^{\sRR}_{pq}  & = \braket{ Z^{\sR}_q Y^{\sR}_p | \Smat | Y^{\sR}_p Z^{\sR}_q } = D_{\sRR}^2(p,q) .
  \end{aligned}
\end{equation}
The crossing equations can then be written as
\begin{equation}\label{eq:crossing-eq-A}
  \begin{aligned}
    1 &= \Upsilon^{\sLL}_{pq} \Upsilon^{\sRL}_{\bar{p}q}  = A_{\sLL}^2(p,q) D_{\sRL}^2(\bar{p},q)
    = \frac{y^+}{y^-} \Bigl( \frac{x^+-y^-}{x^+-y^+} \Bigr)^2 A_{\sLL}^2(p,q) A_{\sRL}^2(\bar{p},q), \\
    1 &= \Lambda^{\sRL}_{pq} \Lambda^{\sLL}_{\bar{p}q} = F_{\sRL}^2(p,q) B_{\sLL}^2(\bar{p},q)
    = \frac{y^+}{y^-} \Bigl( \frac{\frac{1}{x^-}-y^-}{\frac{1}{x^-}-y^+} \Bigr)^2 A_{\sRL}^2(p,q) A_{\sLL}^2(\bar{p},q) .
    % 1 &= \Upsilon^{\sRR}_{pq} \Upsilon^{\sLR}_{\bar{p}q} = F_{\sRR}^2(p,q) B_{\sLR}^2(\bar{p},q)
    % = \frac{y^-}{y^+} \Bigl( \frac{x^--y^+}{x^--y^-} \Bigr)^2 A_{\sRR}^2(p,q) A_{\sLR}^2(\bar{p},q), \\
    % 1 &= \Lambda^{\sLR}_{pq} \Lambda^{\sRR}_{\bar{p}q} = A_{\sLR}^2(p,q) D_{\sRR}^2(\bar{p},q)
    % = \frac{y^-}{y^+} \Bigl( \frac{\frac{1}{x^+}-y^+}{\frac{1}{x^+}-y^-} \Bigr)^2 A_{\sLR}^2(p,q) A_{\sRR}^2(\bar{p},q) ,
  \end{aligned}
\end{equation}
If we perform a second crossing transformation we find the \emph{double} crossing relations
\begin{equation}\label{eq:double-crossing-equations}
  \begin{aligned}
    A_{\sLL}^2(\bar{\bar{p}},q) &= \Bigl( \frac{x^+-y^-}{x^+-y^+} \frac{x^--y^+}{x^--y^-} \Bigr)^2 A_{\sLL}^2(p,q) , \\
    A_{\sRL}^2(\bar{\bar{p}},q) &= \Bigl( \frac{\frac{1}{x^+} - y^+}{\frac{1}{x^+} - y^-} \frac{\frac{1}{x^-} - y^-}{\frac{1}{x^-} - y^+} \Bigr)^2 A_{\sRL}^2(p,q) .
  \end{aligned}
\end{equation}
Since $x_{\sI}^{\pm}(\bar{\bar{p}}) = x_{\sI}^{\pm}(p)$ the double crossing equations describe non-trivial monodromies of the scalar factors. 

\section{The R-R scalar factors}
\label{sec:r-r-scalar-factors}

In the R-R theory the scalar factors were chosen as~\cite{Borsato:2014hja,Borsato:2016kbm,Borsato:2016xns}
\begin{equation}
  \begin{aligned}
    A_{\sLL}^2(p,q) &=
    \frac{x^+}{x^-} \frac{y^-}{y^+} \frac{x^- - y^+}{x^+ - y^-} \frac{1-\frac{1}{x^-y^+}}{1-\frac{1}{x^+y^-}} \sigma_{\sLL}^{-2}(p,q) ,\\
    A_{\sRL}^2(p,q) &=
    \frac{x^+}{x^-} \frac{1-\frac{1}{x^+y^+}}{1-\frac{1}{x^-y^-}} \frac{1-\frac{1}{x^-y^+}}{1-\frac{1}{x^+y^-}} \sigma_{\sRL}^{-2}(p,q) .
  \end{aligned}
\end{equation}
Inserting this into~\eqref{eq:crossing-eq-A} we find the crossing equations for the dressing phases $\sigma_{\sIJ}$~\cite{Borsato:2013hoa}
\begin{equation}\label{eq:old-crossing}
  \begin{aligned}
    \sigma_{\sLL}^2(p,q) \sigma_{\sRL}^2(\bar{p},q) & = \frac{(x^+-y^-)(x^--y^+)}{(x^+-y^+)(x^--y^-)}
    \biggl( \frac{y^-}{y^+} \biggr)^{\!\!2} \frac{x^--y^+}{x^+-y^-}  \frac{1-\frac{1}{x^-y^+}}{1-\frac{1}{x^+y^-}} , \\
    \sigma_{\sRL}^2(p,q) \sigma_{\sLL}^2(\bar{p},q) & =
    \frac{\bigl(\frac{1}{x^+} - y^+\bigr)\bigl(\frac{1}{x^-} - y^-\bigr)}{\bigl(\frac{1}{x^+} - y^-\bigr) \bigl(\frac{1}{x^-} - y^+\bigr)}
    \biggl( \frac{y^-}{y^+} \biggr)^{\!\!2}
    \frac{x^--y^+}{x^+-y^-}
    \frac{1-\frac{1}{x^-y^+}}{1-\frac{1}{x^+y^-}} .
  \end{aligned}
\end{equation}
In reference~\cite{Borsato:2013hoa} the dressing phases were written as $\sigma_{\sIJ}^2(p,q) = \exp\left[2i\theta_{\sIJ}(x^{\pm},y^{\pm})\right]$ and then expanded in a chi decomposition
\begin{equation}\label{eq:chi-decomp}
  \theta_{\sIJ}(x^{\pm},y^{\pm}) =
  \chi_{\sIJ}(x^+,y^+) - \chi_{\sIJ}(x^+,y^-) - \chi_{\sIJ}(x^-,y^+) + \chi_{\sIJ}(x^-,y^-) .
\end{equation}
The crossing equations were then solved by splitting the dressing phases into two parts
\begin{equation}
  \chi_{\sLL}(x,y) = \chi_{\BES}(x,y) + \chi^{\text{extra}}_{\sLL}(x,y) , \qquad
  \chi_{\sRL}(x,y) = \chi_{\BES}(x,y) + \chi^{\text{extra}}_{\sRL}(x,y) ,
\end{equation}
where $\chi_{\BES}$ is the BHL/BES phase~\cite{Beisert:2006ib,Beisert:2006ez} and the extra terms can be written in terms of the HL phase $\chi_{\HL}$ and a new phase $\chi_-$ as
\begin{equation}
  \begin{aligned}
    \chi^{\text{extra}}_{\sLL}(x,y) &= -\frac{1}{2} \chi_{\HL}(x,y) + \frac{1}{2} \chi_-(x,y) , \\
    \chi^{\text{extra}}_{\sRL}(x,y) &= -\frac{1}{2} \chi_{\HL}(x,y) - \frac{1}{2} \chi_-(x,y) .
  \end{aligned}
\end{equation}

Here we instead find it convenient to split the dressing factors into even and odd parts, following~\cite{Beisert:2006ib}
\begin{equation}
\sigma_{\sIJ}=\sigma_{\sIJ,\text{even}}\sigma_{\sIJ,\text{odd}}\,.
\end{equation}
The even dressing factors are  defined to be trivial/homogeneous under double crossing, with the odd part containing the remainder of the dressing factors. Analogously,  we may split the $\chi_{\sIJ}$ into even and odd parts
\begin{equation}\label{eq:chi-odd-even-split}
  \chi_{\sLL}(x,y) = \chi_{\sLL}^{\text{even}}(x,y) + \chi_{\sLL}^{\text{odd}}(x,y) ,
  \qquad
  \chi_{\sRL}(x,y) = \chi_{\sRL}^{\text{even}}(x,y) + \chi_{\sRL}^{\text{odd}}(x,y) .
\end{equation}
The even parts are  given by
\begin{equation}
  \chi_{\sLL}^{\text{even}}(x,y) = \chi_{\sRL}^{\text{even}}(x,y) = \chi_{\BES}(x,y) - \chi_{\HL}(x,y) \,,
\end{equation}
and satisfy the 
\begin{equation}\label{eq:R-R-even-crossing-sigma}
  \begin{aligned}
    \sigma_{\sLL,\text{even}}^2(\bar{p},q)
    \sigma_{\sRL,\text{even}}^2(p,q)
    &=
    \bigl(\frac{y^-}{y^+}\bigr)^2 \frac{x^--y^+}{x^+-y^-} \frac{1-\frac{1}{x^-y^+}}{1-\frac{1}{x^+y^-}} ,
    \\
    \sigma_{\sLL,\text{even}}^2(p,q)
    \sigma_{\sRL,\text{even}}^2(\bar{p},q)
    &=
    \bigl(\frac{y^-}{y^+}\bigr)^2 \frac{x^--y^+}{x^+-y^-} \frac{1-\frac{1}{x^-y^+}}{1-\frac{1}{x^+y^-}} .
  \end{aligned}
\end{equation}
Since the right hand sides of~\eqref{eq:R-R-even-crossing-sigma} are invariant under $x^{\pm} \to 1/x^{\pm}$, we see explicitly that the even phases are homogeneous under double crossing. Note that the right hand sides of~\eqref{eq:R-R-even-crossing-sigma} corresponds to the first term on the right-hand side of the full crossing equations~\eqref{eq:old-crossing}. Using these phases we can build the even matrix elements
\begin{equation}
  \begin{aligned}\label{eq:R-R-even-matrix-elements}
    A_{\sLL,\text{even}}^2(p,q) &=
    \frac{x^+}{x^-} \frac{y^-}{y^+} \frac{x^--y^+}{x^+-y^-} \frac{1-\frac{1}{x^-y^+}}{1-\frac{1}{x^+y^-}} \sigma_{\sLL,\text{even}}^{-2}(p,q) ,\\
    A_{\sRL,\text{even}}^2(p,q) &=
    \frac{x^+}{x^-} \frac{y^-}{y^+} \sigma_{\sRL,\text{even}}^{-2}(p,q) ,
  \end{aligned}
\end{equation}
which satisfy the \emph{homogenous} crossing equations\footnote{Note that even though the solutions~\eqref{eq:R-R-even-matrix-elements} satisfy the homogenous crossing equation, they have a highly non-trivial analytic structure associated to the BES factors.}
\begin{equation}\label{eq:R-R-A-even}
  A_{\sLL,\text{even}}^2(\bar{p},q) A_{\sRL,\text{even}}^2(p,q) = 1 ,
  \qquad
  A_{\sLL,\text{even}}^2(p,q) A_{\sRL,\text{even}}^2(\bar{p},q) = 1 .
\end{equation}

Similarly, the odd matrix elements are given by
\begin{equation}\label{eq:R-R-A-odd}
  \begin{aligned}
    A_{\sLL,\text{odd}}^2(p,q) =
    \sigma_{\sLL,\text{odd}}^{-2}(p,q) ,\quad
    A_{\sRL,\text{odd}}^2(p,q) =
    \frac{y^+}{y^-} \frac{1-\frac{1}{x^+y^+}}{1-\frac{1}{x^-y^-}} \frac{1-\frac{1}{x^-y^+}}{1-\frac{1}{x^+y^-}} \sigma_{\sRL,\text{odd}}^{-2}(p,q) .
  \end{aligned}
\end{equation}
Inserting these into the crossing equations we find that the odd phases should satisfy the odd crossing equation
\begin{equation}\label{eq:R-R-odd-crossing}
  \begin{aligned}
    \sigma_{\sLL,\text{odd}}^2(\bar{p},q) \sigma_{\sRL,\text{odd}}^2(p,q) &= \frac{\frac{1}{x^+} - y^+}{\frac{1}{x^+} - y^-} \frac{\frac{1}{x^-} - y^-}{\frac{1}{x^-} - y^+} ,
    \\
    \sigma_{\sLL,\text{odd}}^2(p,q) \sigma_{\sRL,\text{odd}}^2(\bar{p},q) &= \frac{x^+ - y^-}{x^+ - y^+} \frac{x^- - y^+}{x^- - y^-} .
  \end{aligned}
\end{equation}
Note that the right hand side of these equations corresponds to the second and third terms on the right-hand side of the full crossing equations~\eqref{eq:old-crossing}, with the first terms already accounted for by the even crossing equations~\eqref{eq:R-R-even-crossing-sigma}. As expected, the square of the right hand sides of~\eqref{eq:R-R-odd-crossing} are precisely the factors appearing in the double crossing equations~\eqref{eq:double-crossing-equations}.

\begin{figure}
  \centering
  \subfloat[\label{fig:x-integration-contour-RR-circle}]{\includegraphics{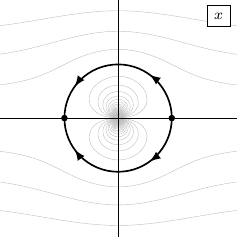}}
  \hspace{1cm}
  \subfloat[\label{fig:x-integration-contour-RR-line}]{\includegraphics{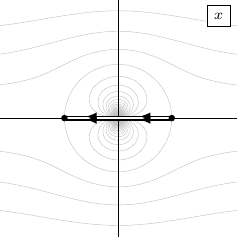}}
  \caption{\label{fig:x-integration-contours-RR}Integration contours for the odd R-R phases.}
\end{figure}

To find a solution to these equations let us start with the expression found in~\cite{Borsato:2013hoa} and call the two odd phases obtained from there $\phi_{\sLL}$ and $\phi_{\sRL}$. We have
\begin{equation}
  \begin{aligned}
    \phi_{\sLL}(x,y) &= +\intudrl \frac{dz}{16\pi} \biggl( \frac{\log\frac{(y-z)^2}{y}}{x-z} - \frac{\log\frac{(x-z)^2}{x}}{y-z} \biggr) , \\
    \phi_{\sRL}(x,y) &= -\intudrl \frac{dz}{16\pi} \biggl( \frac{\log\frac{(y-z)^2}{y}}{\frac{1}{x}-z} - \frac{\log\frac{(x-\frac{1}{z})^2}{x}}{y-z} \biggr) ,
  \end{aligned}
\end{equation}
where the integration is taken over the contour shown in figure~\ref{fig:x-integration-contour-RR-circle}.
If we want to understand the cuts seen by $x$ it is useful to integrate by parts
\begin{equation}
  \begin{aligned}
    \phi_{\sLL}(x,y) &= +\intudrl \frac{dz}{8\pi} \frac{\log\frac{(y-z)^2}{y}}{x-z}
    \\ &\qquad
    - \frac{1}{16\pi} \Bigl( \log\frac{(x-1)^2}{x} \log\frac{(y-1)^2}{y} + \log\frac{(x+1)^2}{x} \log\frac{(y+1)^2}{y} \Bigr)
    , \\
    \phi_{\sRL}(x,y) &= -\intudrl \frac{dz}{8\pi} \frac{\log\frac{(y-z)^2}{y}}{\frac{1}{x}-z}
    \\ &\qquad
    + \frac{1}{16\pi} \Bigl( \log\frac{(x-1)^2}{x} \log\frac{(y-1)^2}{y} - \log\frac{(x+1)^2}{x} \log\frac{(y+1)^2}{y} \Bigr)
  \end{aligned}
\end{equation}
We expect the phases to have cuts along the unit circle, which we indeed do get from the above expressions. However, the boundary terms have an additional unwanted log cut along the negative real lines in both the $x$ and $y$ planes and the integral term also has such a log cut in the $y$ plane. To get of rid of these we can add an extra anti-symmetric term
\begin{equation}
  \delta\phi(x,y) =
  \frac{1}{16\pi} \log\frac{(x+1)^2}{x} \log\frac{(y-1)^2}{y}
  -
  \frac{1}{16\pi} \log\frac{(x-1)^2}{x} \log\frac{(y+1)^2}{y} .
\end{equation}
The full phases $\chi_{\sLL}(x,y) = \phi_{\sLL}(x,y) + \delta\phi(x,y)$ and $\chi_{\sRL}(x,y) = \phi_{\sRL}(x,y) - \delta\phi(x,y)$ then take the form
\begin{equation}
  \begin{aligned}
    \chi_{\sLL}(x,y) 
    &=
    +\intudrl \frac{dz}{8\pi} \frac{\log\frac{(y-z)^2}{y}}{x-z}
    +\frac{1}{16\pi} \log\frac{(x+1)^2}{(x-1)^2} \Bigl( \log\frac{(y+1)^2}{y} + \log\frac{(y-1)^2}{y} \Bigr) ,
    \\
    \chi_{\sRL}(x,y) 
    &=
    -\intudrl \frac{dz}{8\pi} \frac{\log\frac{(y-z)^2}{y}}{\frac{1}{x}-z}
    -\frac{1}{16\pi}\log\frac{(x+1)^2}{(x-1)^2} \Bigl( \log\frac{(y+1)^2}{y} + \log\frac{(y-1)^2}{y} \Bigr) .
  \end{aligned}
\end{equation}
This modification of the phase of~\cite{Borsato:2013hoa} was originally proposed by Andrea Cavagli\`a and Simon Ekhammar~\cite{CavagliaEkhammar} and leaves the crossing equation satisfied by the full phase $\theta_{\sLL}$ unchanged. It is possible to show that~\cite{CavagliaEkhammarMajumderStefanskiTorrielli} this simple modification of~\cite{Borsato:2013hoa} reproduces the more involved expressions given by~\cite{Frolov:2021fmj}, which were in turn found by generalising the expressions for massless dressing factors given in~\cite{Bombardelli:2018jkj}. It is easy to verify that the branch points of the S matrix are \textit{not} of square-root type, in line with the key finding of~\cite{Borsato:2013hoa}. The analysis~\cite{OhlssonSax:2019nlj} of Dorey-Hofman-Maldacena (DHM) double-poles also remains unchanged by the addition of $\delta\phi$.

We find it useful to bring the cuts down from the unit circle to the interval $[-1,1]$. If $x$ and $y$ are outside the unit circle the LL phase remains unchanged
\begin{equation}
  \chi_{\sLL}(x,y) =
  +\int_{1}^{-1} \frac{dz}{2\pi} \frac{\log(y-z)}{x-z}
  + \frac{1}{4\pi} \log\frac{x+1}{x-1} \Bigl( \log(y+1) + \log(y-1) \Bigr) .
\end{equation}
However, when we deform the integration contour in the RL phase from the two half circles in figure~\ref{fig:x-integration-contour-RR-circle} to the interval in figure~\ref{fig:x-integration-contour-RR-line} we pick up a pole at $1/x$ whose residue depends on which half-plane $x$ sits in
\begin{equation}\label{eq:R-R-chi-RL}
  \begin{aligned}
    \chi_{\sRL}(x,y) &=
    -\int_{1}^{-1} \frac{dz}{2\pi} \frac{\log(y-z)}{\frac{1}{x}-z}
    -\frac{1}{4\pi} \log\frac{x+1}{x-1} \Bigl( \log(y+1) + \log(y-1) \Bigr)
    \\ &\qquad
    -\frac{i}{2} \sign (\Im x) \log\bigl( y - \frac{1}{x} \bigr).
  \end{aligned}
\end{equation}
Note that the integral in the RL phase gives long cuts along $(-\infty,-1]\cup[1,\infty)$, but the expression in the second line exactly cancels these discontinuities so that the full expression has a cut only along the short interval $[-1,1]$.

Using the above expression we can check the crossing equations. Going through the cut from below the phases pick up additional terms
\begin{equation}
  \begin{aligned}
    \delta_{\uparrow}\phi_{\sLL}(x,y) &= -i\log(y-x) + \frac{i}{2} \bigl( \log(y+1) + \log(y-1) \bigr),
    \\
    \delta_{\uparrow}\phi_{\sRL}(x,y) &= +i\log\bigl(y-\frac{1}{x}\bigr) - \frac{i}{2} \bigl( \log(y+1) + \log(y-1) \bigr) .
  \end{aligned}
\end{equation}
Combining this with the results of~\cite{Borsato:2013hoa} we find
\begin{equation}
  \begin{aligned}\label{eq:R-R-chi-cross}
    \chi_{\sLL}\bigl(\frac{1}{x},y\bigr) + \chi_{\sRL}(x,y) &=
    -\frac{i}{2} \log\bigl(y-\frac{1}{x}\bigr) + \frac{i}{4} \bigl( \log(y+1) + \log(y-1) \bigr) ,
    \\
    \chi_{\sLL}(x,y) + \chi_{\sRL}\bigl(\frac{1}{x},y\bigr) &=
    +\frac{i}{2} \log(y-x) - \frac{i}{4} \bigl( \log(y+1) + \log(y-1) \bigr) .
  \end{aligned}
\end{equation}
The last term on each line cancels out in the full dressing phase~\eqref{eq:chi-decomp}.

Before ending this section let us consider the poles of the R-R S matrix. We have split the R-R matrix elements into two parts. The odd part transforms non-trivially under crossing, but does not have any singularities in the physical or crossed regions. This is easy to see for the LL phase: neither the rational pre-factor nor the phase has any relevant poles. For the RL case we note that for excitations with real momenta (so that $x^+$ and $x^-$ is in the upper and lower half-planes respectively) the term on the last line in equation~\eqref{eq:R-R-chi-RL} exactly cancels the rational factor in $A_{\sRL,\text{odd}}^2$. The even parts of the matrix elements, on the other hand, are trivial under crossing but responsible for all physically relevant poles. We could thus think of it as a CDD factor~\cite{Castillejo:1955ed}, though in general we expect it to have an intricate analytic structure. In $A_{\sLL,\text{even}}^2$ in~\eqref{eq:R-R-A-even} we directly find an S channel zero\footnote{A bound state corresponds to a pole on the right hand side of the Bethe equations, but the Bethe equations contain the inverse matrix element $A_{\sLL,\text{even}}^{-2}$, so here we find a zero instead of a pole.} at $x^- = y^+$ which corresponds to the formation of a bound state. The even phase has a pole at $x^- = 1/y^+$ which cancels the zero in $A_{\sLL,\text{even}}^2$ and provides us with the expected T channel pole in $A_{\sRL,\text{even}}^2$.\footnote{The BES phase does not have any simple poles, but the HL phase does~\cite{Borsato:2013hoa} which means that the even phase has a corresponding zero.} Finally the BES part of the phase gives rise to DHM double poles corresponding to the formation of an on-shell bound state in an intermediate channel~\cite{Dorey:2007xn,OhlssonSax:2019nlj}.

\section{The mixed-flux scalar factors}
\label{sec:mixed-flux-scalar-factors}

In this section we will construct the scalar factors in the mixed-flux theory. As in the pure R-R theory, we split the scalar factors into odd and even parts~\eqref{eq:chi-odd-even-split}.

\subsection{The odd scalar factor}
\label{sec:odd-phase-mixed}

\begin{figure}
  \centering
  \subfloat[\label{fig:x-integration-contour-scallion-kidney}]{\includegraphics{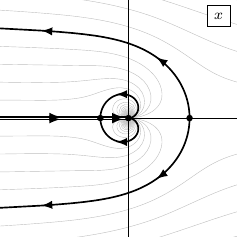}}
  \hspace{1cm}
  \subfloat[\label{fig:x-integration-contour-line}]{\includegraphics{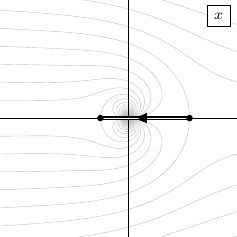}}
  
  \caption{\label{fig:x-integration-contours}Mixed-flux integration contours}
\end{figure}
For the odd scalar factor we start with the odd matrix elements which are similar to the odd R-R matrix elements
\begin{equation}\label{eq:mixed-A-odd}
  \begin{aligned}
    A_{\sLL,\text{odd}}^2(p,q) &=
    \Bigl(
    \frac{\alpha(x^-)\alpha(y^+)}{\alpha(x^+)\alpha(y^-)}
    \Bigr)^{\!\!1/k}
    \sigma_{\sLL,\text{odd}}^{-2}(p,q) ,\\
    A_{\sRL,\text{odd}}^2(p,q) &=
    \Bigl(
    \frac{\alpha(1/x^+)\alpha(y^-)}{\alpha(1/x^-)\alpha(y^+)}
    \Bigr)^{\!\!1/k}
    \frac{y^+}{y^-} \frac{1-\frac{1}{x^+y^+}}{1-\frac{1}{x^-y^-}} \frac{1-\frac{1}{x^-y^+}}{1-\frac{1}{x^+y^-}} \sigma_{\sRL,\text{odd}}^{-2}(p,q) ,
  \end{aligned}
\end{equation}
where
\begin{equation}
  \alpha(x) = (x-s)\Bigl(\frac{1}{x}+s\Bigr) .
\end{equation}
The additional factors in~\eqref{eq:mixed-A-odd} compared to the R-R case~\eqref{eq:R-R-A-odd} trivially cancel in the crossing equations.\footnote{Ignoring the discrete nature of $k$, in the formal $k\rightarrow 0$ limit these factors are not well behaved. This is consistent with the expectation that the $k=0$ and $k\neq 0$ theories are very different: the latter being a modulus deformation of the WZW theory, while the former being a pure R-R flux background. While the $k\rightarrow 0$ limit in classical and semi-classical near-BMN worldsheet theories is smooth, it is known that 1-loop worldsheet dressing phases do not match with exact ones even in the R-R background, see for example point 3 on page 35 
of~\cite{Frolov:2021fmj}.  The above factors have 1-loop (or $1/k$) scaling, and based on the experience with the R-R theory, we do not expect the formal divergence in the $k\rightarrow 0$ limit of the exact S matrix to be visible in perturbation theory. Nevertheless, it will be important to revisit this issue once the full dressing phases are known.} As we discuss in section~\ref{sec:constraints-on-s-mat} below, these terms play an important role in fusion. The generalisation of the R-R odd phases presented in the previous section to the mixed-flux case is
\begin{equation}\label{eq:chis-in-nice-x-form}
  \begin{aligned}
    \chi_{\sLL}(x,y) &=
    +\!\!\int_{s}^{-s^{-1}} \frac{dz}{2\pi} \frac{\log(y-z)}{x-z}
    + \frac{1}{4\pi} \log\frac{x+s^{-1}}{\!x-s} \Bigl( \log(y+s^{-1}) + \log(y-s) \Bigr) ,
    \\
    \chi_{\sRL}(x,y) &=
    -\!\!\int_{s}^{-s^{-1}} \frac{dz}{2\pi} \frac{\log(y-z)}{\frac{1}{x}-z}
    -\frac{1}{4\pi} \log\frac{\!x+s}{x-s^{-1}} \Bigl( \log(y+s^{-1}) + \log(y-s) \Bigr)
    \\ &\qquad
    -\frac{i}{2} \sign (\Im x) \log\bigl( y - \frac{1}{x} \bigr).
  \end{aligned}
\end{equation}
The integrals are taken over the interval $[-1/s,s]$ as shown in figure~\ref{fig:x-integration-contour-line}. It is simple to check that these phases satisfy the same crossing equations as in the pure R-R case\footnote{In the mixed-flux case there are extra terms proportional to $\log s$ on the right hand side of the analogue of~\eqref{eq:R-R-chi-cross}, which cancel out in the full dressing phase.}
\begin{equation}
  \begin{aligned}\label{eq:odd-mixed-phases}
    \sigma_{\sLL,\text{odd}}^2(\bar{p},q) \sigma_{\sRL,\text{odd}}^2(p,q) &= \frac{\frac{1}{x^+} - y^+}{\frac{1}{x^+} - y^-} \frac{\frac{1}{x^-} - y^-}{\frac{1}{x^-} - y^+} ,
    \\
    \sigma_{\sLL,\text{odd}}^2(p,q) \sigma_{\sRL,\text{odd}}^2(\bar{p},q) &= \frac{x^+ - y^-}{x^+ - y^+} \frac{x^- - y^+}{x^- - y^-} .
  \end{aligned}
\end{equation}
The above integrals over $[-1/s,s]$ are useful in checking the crossing equation. It is sometimes also useful to do an integration by parts on half of the integrands above to find\footnote{Here we have dropped some terms that do no contribute to the full phase $\theta_{\sRL}$.}
\begin{equation}
  \begin{aligned}\label{eq:mixed-L-chis-short-contour}
    \chi_{\sLL}(x,y) &=
    +\!\!\int_{s}^{-s^{-1}} \frac{dz}{4\pi} \left[\frac{\log(y-z)}{x-z}-\frac{\log(x-z)}{y-z}\right]
    \\ & \qquad
    + \frac{1}{4\pi} \left[\log(x+s^{-1}) \log(y-s) - \log(x-s) \log(y+s^{-1})\right]  ,
    \\
    \chi_{\sRL}(x,y) &=
    -\!\!\int_{s}^{-s^{-1}} \frac{dz}{4\pi} \frac{\log(y-z)}{\frac{1}{x}-z}
    +\!\!\int_{s^{-1}}^{-s} \frac{dz}{4\pi} \frac{\log(x-z)}{\frac{1}{y}-z}
    \\ &\qquad
    - \frac{1}{4\pi} \left[\log(x+s) \log(y-s) - \log(x-s^{-1}) \log(y+s^{-1})\right]  ,
    \\ &\qquad
    -\frac{i}{4} \sign (\Im x) \log\bigl( y - \frac{1}{x} \bigr)
    +\frac{i}{4} \sign (\Im y) \log\bigl( x - \frac{1}{y} \bigr)
  \end{aligned}
\end{equation}
As we discussed in section~\ref{sec:cross-transformation}, the scallion and kidney contours are natural boundaries when considering crossing. It is therefore natural to deform $[-1/s,s]$  into scallion and kidney contours, analogously to what is depicted in figure~\ref{fig:x-integration-contours-RR} for the R-R case. Notice that the integral over $[-1/s,s]$ has no monodromy around infinity. To preserve this property after deforming the integration contour to a scallion and kidney (for non-quadratic cuts), it is necessary to introduce additional integrals over $(-\infty,0]$. The deformed contour is shown in figure~\ref{fig:x-integration-contour-scallion-kidney} and the mixed-flux expressions for $\chi_{\sLL}$ and $\chi_{\sRL}$ take the same form as in~\eqref{eq:mixed-L-chis-short-contour}, but with the following replacement of integrals
\begin{equation}
  \int_{s}^{-s^{-1}}\longrightarrow \int_{\mathcal{I}_L}+2\!\!\int_{-\infty}^0 ,
  \qquad
  \int_{s^{-1}}^{-s}\longrightarrow \int_{\mathcal{I}_R}+2\!\int_0^{\infty}
\end{equation}
where the contours $\mathcal{I}_L$ and $\mathcal{I}_R$ are defined as
\begin{equation}
  \mathcal{I}_L=S_L^+-S_L^-+K_L^+-K_L^- ,
  \qquad
  \mathcal{I}_R=S_R^+-S_R^-+K_R^+-K_R^- ,
\end{equation}
with $K_L^+$ and $K_L^-$ the upper and lower halves of the L kidney contour traversed in an anti-clockwise direction, and similarly $S_L^+$ and $S_L^-$ for the L scallion, $K_R^+$ and $K_R^-$ for the R kidney and $S_R^+$ and $S_R^-$ for the R scallion.

It will sometimes be useful to do a ``complete'' integration by parts for $\chi_{\sRL}(x,y)$ to make its analytic structure in the $y$-plane clear
\begin{equation}
  \begin{aligned}\label{eq:chis-in-nice-y-form}
    \chi_{\sRL}(x,y) &=
    +\!\!\int_{s^{-1}}^{-s} \frac{dz}{2\pi} \frac{\log(x-z)}{\frac{1}{y}-z}
    +\frac{1}{4\pi} \log\frac{\!y+s^{-1}}{y-s} \Bigl( \log(x+s) + \log(x-s^{-1}) \Bigr)
    \\ &\qquad
    +\frac{i}{2} \sign (\Im y) \log\bigl( x - \frac{1}{y} \bigr).
  \end{aligned}
\end{equation}
The analogous expression for the LL phase follows trivially from the anti-symmetry of $\chi_{\sLL}$, $\chi_{\sLL}(x,y)=-\chi_{\sLL}(y,x)$.

The expressions presented in this sub-section for $\chi_{\sLL}$ and $\chi_{\sRL}$ are valid for momenta in the region outside the left and right kidneys, \textit{i.e.} for $p_{\sL}\in(0,2\pi)$ and $p_{\sR}\in(-2\pi,0)$. These regions shares many similarities with the physical region of the R-R theory, which is outside the unit disc. In particular, it is useful to expand the dressing phases at large values of the Zhukovsky parameters. Expanding the integrand in the expression for $\chi_{\sLL}$ in~\eqref{eq:odd-mixed-phases} and then integrating gives
\begin{equation}
\chi_{\sLL}=\frac{1}{4\pi}\sum_{m,n=1}^\infty\frac{x^{-m}y^{-n}}{mn}\biggl[\frac{m-n}{m+n}\bigl(s^{n+m}-(-s^{-1})^{n+m}\bigr)+(-1)^m s^{n-m}-(-1)^ns^{m-n}\biggr].
\end{equation}
In the case of $\chi_{\sRL}$, it is helpful to deform the integration contour in~\eqref{eq:odd-mixed-phases} to two semi-circles, similar to those in the R-R case shown in figure~\ref{fig:x-integration-contour-RR-circle}, but now going between $z=s$ and $z=-s^{-1}$. These semi-circles appeared in~\cite{Babichenko:2014yaa} and are useful here, because they remove the term proportional to $\sign (\Im x)$\footnote{On the $p_{\sR}\in(-2\pi,0)$ sheet $1/x^\pm_R$ sit inside the left kidney, while the circle contours are outside it.} After this contour deformation, it is again straightforward to expand the integrand at large values of the Zhukovsky parameters and the integrate to get
\begin{align}
\chi_{\sRL}&=\frac{\log s}{\pi}\sum_{n=1}^\infty \frac{1}{n}y^{-n}x^{-n} +\frac{1}{4\pi}\sum_{n=1}^\infty \frac{(-1)^n\bigl(s^{-2n}-s^{2n}\bigr)}{n^2}y^{-n}x^{-n}
\\ &\quad \nonumber
+\frac{1}{4\pi}\sum_{\substack{
    m,n=1 \\
    m\neq n}}^\infty \frac{x^{-m}y^{-n}}{mn}\biggl[ \frac{n+m}{n-m}\bigl(s^{n-m}-(-s)^{m-n}\bigr)+(-1)^ns^{-n-m}-(-1)^ms^{n+m}\biggr]
\\ &\quad \nonumber
-\frac{\log s}{\pi}\log y+\frac{1}{2\pi}\sum_{n=1}^\infty \frac{s^n-(-s)^{-n}}{n^2}y^{-n}    \, ,
\end{align}
where the terms on the last line can be dropped, since they are $x$-independent and so cancel out in the expression for $\theta_{\sRL}$.

\textbf{Note added:} The odd dressing phase found here is the same as the odd part of the phase that has since been proposed in~\cite{Frolov:2024pkz}.

\subsection{The even scalar factor}

As in the R-R case the even scalar factors satisfy the homogeneous crossing equations but provide us with physical poles and zeros. We expect the LL S matrix to have a simple pole corresponding to the formation of a bound state as discussed in section~\ref{sec:bound-states}. This can be accommodated by a simple rational factor as in the pure R-R case. However, after a proper analytical continuation we also expect to find double poles corresponding to the exchange of an on-shell bound state~\cite{Dorey:2007xn,OhlssonSax:2019nlj}. In the R-R case the double poles were provided by the BES phase, and in the mixed-flux case it seems natural to expect a generalisation of this phase to appear.

A new feature of the mixed-flux theory is the appearance of an infinite number of momentum regions $(2\pi n,2\pi(n+1))$. Here we will discuss what poles the fundamental S matrix is expected to have in different regions, and what constraints that gives for the even dressing phases.

\subsubsection{Scalar factors in the \texorpdfstring{$(0,2\pi)$}{(0,2pi)} region}

Let us start by considering the case of two fundamental excitations with momenta in $(0,2\pi)$.
Here we take the even matrix element to be the same as in the R-R model (see equation~\eqref{eq:R-R-even-matrix-elements}), when expressed in terms of the rapidity\footnote{In the R-R case $u_{\sL}(x^-) - u_{\sL}(y^+) = (x^- - y^+)(1 - \frac{1}{x^-y^+})$, but in the mixed-flux theory these expressions are not equivalent.} $u_{\sL}$
\begin{equation}\label{eq:A-even-0-2pi}
  \begin{gathered}
    A_{\sLL,\text{even}}^2(p,q) =
    \frac{x^+}{x^-} \frac{y^-}{y^+} \frac{u_{\sL}(x^-)-u_{\sL}(y^+)}{u_{\sL}(x^+)-u_{\sL}(y^-)} \sigma_{\sLL,\text{even}}^{-2}(p,q) ,
    \\
    A_{\sRL,\text{even}}^2(p,q) = \frac{x^+}{x^-} \frac{y^-}{y^+} \sigma_{\sRL,\text{even}}^{-2}(p,q) .
  \end{gathered}
\end{equation}
The even dressing phases satisfy the crossing equations (\cf~\eqref{eq:R-R-even-crossing-sigma})
\begin{equation}\label{eq:even-crossing-sigma}
  \begin{aligned}
    \sigma_{\sLL,\text{even}}^2(\bar{p},q)
    \sigma_{\sRL,\text{even}}^2(p,q)
    &=
    \bigl(\frac{y^-}{y^+}\bigr)^2 \frac{u_{\sL}(x^-)-u_{\sL}(y^+)}{u_{\sL}(x^+)-u_{\sL}(y^-)},
    \\
    \sigma_{\sLL,\text{even}}^2(p,q)
    \sigma_{\sRL,\text{even}}^2(\bar{p},q)
    &=
    \bigl(\frac{y^-}{y^+}\bigr)^2 \frac{u_{\sR}(x^-)-u_{\sL}(y^+)}{u_{\sR}(x^+)-u_{\sL}(y^-)} .
  \end{aligned}
\end{equation}
The LL factor has the expected S channel zero at $x^+ = y^-$, and as in the R-R case we expect the even dressing phase to cancel any other relevant zero or pole in $A_{\sLL,\text{even}}^2(p,q)$ and to provide us with the T channel pole of $A_{\sRL,\text{even}}^2(p,q)$.

\subsubsection{Scalar factors in the \texorpdfstring{$(2\pi,4\pi)$}{(2pi,4pi)} region}

Consider bringing $x^{\pm}$ into the $(2\pi,4\pi)$ region. The Beisert-Dippel-Staudacher (BDS) factor~\cite{Beisert:2004hm} in~\eqref{eq:A-even-0-2pi} is clearly analytic as a function of $u_{\sL}$ and thus does not change when we go to a different momentum region. However, let us consider the poles and zeros we expect in the LL S matrix. As above we expect a zero when $x^- = y^+$ and a pole when $x^+ = y^-$, but the analytic continuation has taken us to a different sheet of the $u_{\sL}$ plane. If we follow a path where $x^+$ goes around the origin we then find that the LL matrix element should be given by
\begin{equation}
  A_{\sLL,\text{even}}^2(p,q) =
  \frac{x^+}{x^-} \frac{y^-}{y^+} \frac{u_{\sL}(x^-)-u_{\sL}(y^+)}{u_{\sL}(x^+)-u_{\sL}(y^-) + \frac{2ik}{h}} \, \sigma_{\sLL,\text{even}}^{-2}(p,q) .
\end{equation}
If we instead take both $x^{\pm}$ and $y^{\pm}$ to be in the $(2\pi,4\pi)$ region we instead expect
\begin{equation}
  A_{\sLL,\text{even}}^2(p,q)
  =
  \frac{x^+}{x^-} \frac{y^-}{y^+} \frac{u_{\sL}(x^-)-u_{\sL}(y^+) - \frac{2ik}{h}}{u_{\sL}(x^+)-u_{\sL}(y^-) + \frac{2ik}{h}} \,
  \sigma_{\sLL,\text{even}}^{-2}(p,q) .
\end{equation}
Both of these expressions indicate that we should pick up a factor from the dressing phase when we change momentum region. In particular when we take $p$ to $p+2\pi$ with $p$ and $q$ in $(0,2\pi)$ we should pick up a factor
\begin{equation}
  \frac{u_{\sL}(x^+)-u_{\sL}(y^-)}{u_{\sL}(x^+)-u_{\sL}(y^-) + \frac{2ik}{h}}
\end{equation}
when we go into the scallion contour from below and exit it on the top.

\subsubsection{Scalar factor for \texorpdfstring{$p_{\sL}$}{pL} in \texorpdfstring{$(-2\pi,0)$}{(-2pi,0)}}

In the $(-2\pi,0)$ region we have seen that the $u_{\sL}$ plane is periodic with period $2ik/h$, and we expect the S matrix to respect this. To accommodate this we propose that in this region the scalar factor takes the form
\begin{equation}\label{eq:mixed-ALL-even-min-2pi-0}
  A_{\sLL,\text{even}}^2(p,q) =
  \frac
  {\sinh\bigl(\frac{\pi h}{2k}\bigl(u_{\sL}(x^-)-u_{\sL}(y^+)\bigr)\bigr)}
  {\sinh\bigl(\frac{\pi h}{2k}\bigl(u_{\sL}(x^+)-u_{\sL}(y^-)\bigr)\bigr)}
  \sigma_{\sLL,\text{even}}^{-2}(p,q) .
\end{equation}
Note that up to the dressing factor this is the S matrix for the XXZ model at a root of unity, \ie, with anisotropy of the form $\Delta = \cos\frac{\pi}{k}$ with $k$ an integer~\cite{Fabricius:2000yx,Baxter:2001sx,Braak:2001}. To see this more explicitly we introduce the rapidities $u$ and $v$ such that
\begin{equation}
  u_{\sL}(x^{\pm}) = u \pm \frac{i}{h} , \qquad
  u_{\sL}(y^{\pm}) = v \pm \frac{i}{h} ,
\end{equation}
which gives
\begin{equation}
  A_{\sLL,\text{even}}^2(p,q) =
  \frac
  {\sinh\bigl(\frac{\pi h}{2k}\bigl(u-v-\frac{2i}{h}\bigr)\bigr)}
  {\sinh\bigl(\frac{\pi h}{2k}\bigl(u-v+\frac{2i}{h}\bigr)\bigr)}
  \sigma_{\sLL,\text{even}}^{-2}(p,q) .
\end{equation}
The above factor provides us with the expected poles and zeros at $x^{\pm} = y^{\mp}$. For the even RL matrix element with the L excitation in $(-2\pi,0)$ and the R excitation in $(0,2\pi)$, \ie, with both excitation between respective scallion and kidney, we take
\begin{equation}\label{eq:mixed-ARL-even-min-2pi-0}
  A_{\sRL,\text{even}}^2(p,q) =
  \frac
  {\sinh\bigl(\frac{\pi h}{2k}\bigl(u_{\sR}(x^+)-u_{\sL}(y^-)\bigr)\bigr)}
  {\sinh\bigl(\frac{\pi h}{2k}\bigl(u_{\sR}(x^-)-u_{\sL}(y^+)\bigr)\bigr)}
  \sigma_{\sRL,\text{even}}^{-2}(p,q) .
\end{equation}
With the above choice of normalisation the dressing phases in this region satisfy the homogeneous crossing equations
\begin{equation}
  \sigma_{\sLL,\text{even}}^2(\bar{p},q) \sigma_{\sRL,\text{even}}^2(p,q)
  =
  1
  =
  \sigma_{\sLL,\text{even}}^2(p,q) \sigma_{\sRL,\text{even}}^2(\bar{p},q) .
\end{equation}

We could also ask what happens to the LL matrix element when $p$ is in $(-2\pi,0)$ and $q$ is in $(0,2\pi)$. In principle we could have an S channel pole when, \eg, $x^- = y^+$ is exactly on the scallion, but as we will see soon we expect that pole to not be present which means that we should take the matrix element to be
\begin{equation}
  A_{\sLL,\text{even}}^2(p,q) = \frac{x^+}{x^-} \frac{y^-}{y^+}
  \sigma_{\sLL,\text{even}}^{-2}(p,q) .
\end{equation}

\subsection{Relation to relativistic limit}

\begin{figure}
  \centering
  \subfloat[\label{fig:p-plane-h01-k5}]{\includegraphics{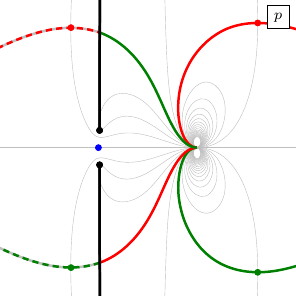}}
  \hspace{0.5cm}
  \subfloat[\label{fig:u-plane-h01-k5}]{\includegraphics{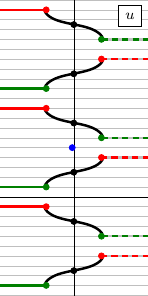}}
  \hspace{0.5cm}
  \subfloat[\label{fig:p-plane-h0-k5}]{\includegraphics{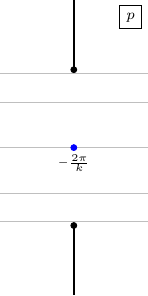}}
  \hspace{0.5cm}
  \subfloat[\label{fig:u-plane-h0-k5}]{\includegraphics{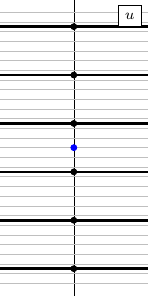}}
  
  \caption{\label{fig:p-and-u-in-rel-limit}The $p$ and $u$ planes with $k=5$ in the relativistic limit. Figures~\protect\subref{fig:p-plane-h01-k5} and~\protect\subref{fig:u-plane-h01-k5} show the $p$ and $u$ planes for $h=1/10$. As we lower $h$ even more the scallion and kidney cuts in the $u$ plane move further away from the imaginary axis and the black cuts become flatter. The relativistic limit zooms in at a narrow strip around the imaginary axis in the $u$ plane, which corresponds to the a region close to $p=-2\pi/k$ in the $p$ plane, as shown in figures~\protect\subref{fig:p-plane-h0-k5} and~\protect\subref{fig:u-plane-h0-k5}. The $p$ plane now looks like that of a standard massive relativistic excitation, and the $u$ plane goes to the corresponding rapidity plane. The blue dots in each plot show the location of an excitation with $p=-2\pi/k$.}

\end{figure}

In this sub-section we consider the relativistic limit of the mixed-flux odd dressing phases found above. Relativistic limits have been used in integrable $\AdS_3$ models in R-R backgrounds~\cite{Bombardelli:2018jkj} to investigate massless modes, where a difference form of the S matrix and dressing phases was observed~\cite{Fontanella:2019baq,Fontanella:2019ury}, leading to novel expressions for massless dressing factors. In mixed-flux backgrounds such limits helped identify a close relationship to relativistic integrable models~\cite{Fendley:1991ve,Fendley:1992dm} as well as q-deformed holographic ones~\cite{Arutyunov:2012zt,Arutyunov:2012ai,Hoare:2013ysa}. In~\cite{Frolov:2023lwd} a modified version of such relativistic limit was proposed, in which the Hopf algebra and hence S matrix and dressing phases were the same as in~\cite{Fontanella:2019ury}, but a different scaling of momentum was introduced. In this limit, $h\rightarrow 0$ and momentum is  expanded around
\begin{equation}\label{eq:rel-limit-min}
p_{\sL}=-\frac{2\pi m}{k},\qquad p_{\sR}=\frac{2\pi m}{k},
\end{equation}
with the corrections scaling with $h$
\begin{equation}\label{eq:rel-limit-scale}
\begin{aligned}
p_{\sL}&=-\frac{2\pi m}{k}+\frac{4\pi h}{k}\left| \sin \left(\frac{m \pi }{k}\right)\right|\sinh\vartheta,
\\
p_{\sR}&=+\frac{2\pi m}{k}+\frac{4\pi h}{k}\left| \sin \left(\frac{m \pi }{k}\right)\right|\sinh\vartheta.
\end{aligned}
\end{equation}
Above, $\vartheta$ is a direct analogue of the rapidity variable present in relativistic integrable models. Notice that in taking this limit we are zooming in on momenta \textit{between} the scallions and kidneys, in other words on
\begin{equation}\label{eq:re-limit-mom}
p_{\sL}\in (-2\pi,0),\qquad\qquad p_{\sR}\in (0,2\pi).
\end{equation}
Furthermore, in the limit the $E_{\sI}$ branch-points  inside those regions move closer to the real axis and the effective dynamics, to leading order, matches that of a massive relativistic theory. Similarly, the $u$ variable becomes
\begin{equation}
u(x) \rightarrow -\frac{k}{\pi h}\left(\vartheta - i \pi\right),
\end{equation}
in other words, up to some simple shifts and rescalings we can identify $u$ and $\vartheta$. Indeed, as shown in figure~\ref{fig:p-and-u-in-rel-limit}, in this limit the $(-2\pi,0)$ sheet of the $u$-plane matches the usual $\theta$ plane of relativistic theories.  From~\eqref{eq:explicit-s-and-sinv}  we note that
\begin{equation}
s\rightarrow\infty,\qquad \qquad s^{-1}\rightarrow 0.
\end{equation}
It is worth pointing out that the expansion around the momenta in~\eqref{eq:rel-limit-min}, which restricts the momenta to lie on the sheets~\eqref{eq:re-limit-mom} \textit{de facto} puts an upper bound on the bound state number to be no more than $k$. This is because bound states with $m>k$ and $p$ in the range~\eqref{eq:rel-limit-min} are related to states with bound state number $m-k$ but momentum $p+2\pi$, as we discuss in section~\ref{sec:bound-state-and-mom-shift}. The latter states necessarily decouple in the relativistic limit, since their momenta are not near the minimal values~\eqref{eq:rel-limit-min}.

In this relativistic limit the crossing equations~\cite{Fontanella:2019ury} can be solved~\cite{Fontanella:2019ury,Frolov:2023lwd}, with an explicit expression in terms of Barnes G-function, or equivelently an infinite product of ratios of $\Gamma$ functions for both dressing phases given in~\cite{Frolov:2023lwd}. While those expressions are complicated, the $\vartheta$ derivative of the dressing phases is simple
\begin{equation}\label{eq:der-fps-phases}
  \begin{aligned}
    \partial_{\vartheta}\theta_{\sLL,\,\mbox{\tiny FPS}}&=
    \frac{\sin \frac{\pi}{k}}{\cosh \vartheta -\cos \frac{2\pi}{k}} \Bigl(
    \frac{2}{k} \cos\frac{\pi}{k}
    -
    \frac{\vartheta}{\pi} \coth\frac{\vartheta}{2}\sin\frac{\pi}{k}
    \Bigr)
    \\
    \partial_{\vartheta}\theta_{\sRL,\,\mbox{\tiny FPS}}&=
    \frac{\sin \frac{\pi}{k}}{\cosh \vartheta + \cos \frac{2\pi}{k}} \Bigl(
    \bigl(1-\frac{2}{k}\bigr)\cos\frac{\pi}{k}
    +
    \frac{\vartheta}{\pi} \tanh\frac{\vartheta}{2}\sin\frac{\pi}{k}
    \Bigr)
  \end{aligned}
\end{equation}
where $\vartheta\equiv\vartheta_1-\vartheta_2$.

We would like to compare this with relativistic limit of the exact  odd phases and scalar factors found in the previous sub-sections. The expressions for $\chi_{\sLL}$ and $\chi_{\sRL}$ given in section~\ref{sec:odd-phase-mixed} are valid on the $(0,2\pi)$ Zhukovsky sheets with the logarithms appearing in these expressions being on the principal branch. To compare to the relativistic limit we need to analytically continue all left momenta to the $(-2\pi,0)$ sheet as discussed in section~\ref{sec:p-x-u-planes}. When analytically continuing $x^\pm_{\sL}$ and $y^\pm_{\sL}$ to the $(-2\pi,0)$ region via path $-1$ in figure~\ref{fig:p-plane-path-between-region}, $x^+$ and $y^+$ cross the $(-\infty,-1/s]$ interval from below. As a result, the logarithms with arguments involving $x^+$ and $y^+$ continue to a different branch, for example
\begin{equation}
\log\left(x^++s^{-1}\right) \longrightarrow  \log\left(x^++s^{-1}\right) -2\pi i.
\end{equation}
However, the complete expression for $\chi_{\sLL}$ is  analytic across $(-\infty,-1/s]$ and, as one can check explicitly, the change in branches cancels between the integral and non-integral terms in~\eqref{eq:chis-in-nice-x-form}. For analytic continuation of $\chi_{\sRL}$ in $y^\pm$ to the $(-2\pi,0)$ sheet, it is useful to use~\eqref{eq:chis-in-nice-y-form}. While the integral part of the expression explicitly has cuts along path $-1$, the non-integral terms do not. However, it is straightforward to check that $e^{-2i\theta_{\sRL}}$ has no cuts. As a result, we can also use the principal branch of the logarithm in~\eqref{eq:chis-in-nice-x-form} when finding the relativistic limit of $\chi_{\sLL}$ and $\chi_{\sRL}$. 

It is then straightforward to check that in the relativistic limit the non-integral part of $\chi_{\sLL}$ in~\eqref{eq:chis-in-nice-x-form} cancels out in $\theta_{\sLL}$. Similarly, the non-integral part in the first line of $\chi_{\sRL}$ in~\eqref{eq:chis-in-nice-x-form} cancels out in $\theta_{\sRL}$, while the non-integral term on the second line cancels the contribution of the rational term in $A_{\sRL,\text{odd}}^2$ in~\eqref{eq:mixed-A-odd}.

Turning to the integral parts of $\chi_{\sLL}$ and $\chi_{\sRL}$, we note that in the relativistic limit 
\begin{equation}
y^\pm\partial_{y^\pm}\rightarrow \partial_\vartheta.
\end{equation}
Using this, we find
\begin{equation}
y^\pm\partial_{y^\pm}\chi_{\sLL}(x^{\pm'},y^{\pm})\longrightarrow \frac{1}{2\pi}
\frac{e^{\theta_2\mp\frac{i \pi }{k}}}{e^{\theta_1\mp'\frac{i \pi }{k}}-e^{\theta_2\mp\frac{i \pi }{k}}}
\left(\log\left(e^{\theta_1\mp' \frac{i \pi }{k}}\right)-\log\left(e^{\theta_2\mp \frac{i \pi }{k}}\right)\right),
\end{equation}
from which one can check that the relativistic limit of our $\partial_\vartheta\theta_{\sLL}$ reduces to the expression in~\eqref{eq:der-fps-phases}. Similarly,  the \textit{integral part} of $\chi_{\sRL}$ gives in the relativistic limit
\begin{equation}
y^\pm\partial_{y^\pm}\chi_{\sRL}(x^{\pm'},y^{\pm})\longrightarrow 
-\frac{1}{2\pi}
\frac{e^{\theta_2\mp\frac{i \pi }{k}}}{e^{\theta_1\mp'\frac{i \pi }{k}}+e^{\theta_2\mp\frac{i \pi }{k}}}
\left(\log\left(e^{\theta_1\mp' \frac{i \pi }{k}}\right)-\log\left(e^{\theta_2\mp \frac{i \pi }{k}}\right)\right),
\end{equation}
from which one can check that the relativistic limit of our $\partial_\vartheta\theta_{\sRL}$ reduces to the expression in~\eqref{eq:der-fps-phases}.\footnote{In verifying this, we take into account the cancellation between the rational term in $A_{\sRL,\text{odd}}^2$ in~\eqref{eq:mixed-A-odd} and the non-integral term on the second line of $\chi_{\sRL}$ in~\eqref{eq:chis-in-nice-x-form} discussed in the preceding paragraph.} 

In addition to matching the dressing phase in the relativistic limit we can also consider the even matrix elements. If we assume that the even phase becomes trivial in the relativistic limit we find
\begin{equation}
  \begin{aligned}
    A_{\sLL,\text{even}}^2 &\approx
    \frac
    {\sinh\bigl(\frac{\pi h}{2k}\bigl(u_{\sL}(x^-)-u_{\sL}(y^+)\bigr)\bigr)}
    {\sinh\bigl(\frac{\pi h}{2k}\bigl(u_{\sL}(x^+)-u_{\sL}(y^-)\bigr)\bigr)}
    \sigma_{\sLL,\text{even}}^{-2}(p,q)
    \to
    \frac
    {\sinh\bigl(\frac{\vartheta}{2} +\frac{i\pi}{k}\bigr)}
    {\sinh\bigl(\frac{\vartheta}{2} - \frac{i\pi}{k}\bigr)} ,
    \\
    A_{\sRL,\text{even}}^2 &\approx
    \frac
    {\sinh\bigl(\frac{\pi h}{2k}\bigl(u_{\sR}(x^+)-u_{\sL}(y^-)\bigr)\bigr)}
    {\sinh\bigl(\frac{\pi h}{2k}\bigl(u_{\sR}(x^-)-u_{\sL}(y^+)\bigr)\bigr)}
    \sigma_{\sRL,\text{even}}^{-2}(p,q)
    \to
    \frac
    {\cosh\bigl(\frac{\vartheta}{2} - \frac{i\pi}{k}\bigr)}
    {\cosh\bigl(\frac{\vartheta}{2} + \frac{i\pi}{k}\bigr)} .
  \end{aligned}
\end{equation}
This perfectly agrees with the relativistic ``CDD'' factors of~\cite{Frolov:2023lwd}.

We have further checked that in a neighbourhood of physical values of momenta near~\eqref{eq:rel-limit-min} our odd phase matches numerically the one in~\cite{Frolov:2023lwd}. Together with the matching of the crossing equations, even phase and the derivatives above this completes the check that our S matrix reduces in the relativistic limit to the one in~\cite{Frolov:2023lwd}. In particular, it indicates that the even part of a putative mixed-flux BES phase will have to trivialise in the relativistic limit, if the proposed relativistic limit is to correctly capture some of the dynamics of the full mixed-flux theory. It will be interesting to verify this in the future.

\section{Constraints from the bound state S matrix}
\label{sec:bound-state-s-mat-and-phases}

In integrable models the bound-state S matrix, \ie, the S matrix describing scattering of a bound state with either a fundamental excitation or a second bound state, can be obtained from the S matrix for two fundamental excitations through the fusion procedure.~\cite{Roiban:2006gs,Dorey:2006dq,Arutyunov:2007tc,Arutyunov:2008zt} This procedure is simpler in the $\AdS_3\times\Sphere^3\times\Torus^4$ world-sheet theory than in a general integrable field theories, because all short representations of $\algPSU(1|1)^4_{\ce}$ are the same up to the values of the central charges, and it is straightforward to check that the matrix part of the S matrix fuses.

\subsection{Special bound state configurations}
\label{sec:special-bound-state}

As we have discussed in section~\ref{sec:physical-regions}, the non-periodicity of the mixed-flux dispersion relations implies that physical fundamental and bound state excitations can be analytically continued to any value of real total momentum. Because of the shift symmetry~\eqref{eq:mom-m-shift-sym}, L states with momentum $p=p_0+2\pi n$, for $n\in\Integers$ and bound state number $m=m_0$ have the same energy and charge $M\equiv m+\kbar P$, as states with momentum $p=p_0$ and bound state number $m=m_0+n$. In section~\ref{sec:bound-states} we already saw an example of this, when we compared a state with $p_{tot}\in[0,2\pi]$ and $m=7$ with a state with $p_{tot}\in[2\pi,4\pi]$ and $m=2$, \textit{cf.} figure~\ref{fig:x-boundstates}. Since we can continue to any real value of momentum, analytic continuation appears to lead to an \textit{infinite} degeneracy of identically-charged states. Fortunately, as we now discuss, this physically undesirable conclusion can be avoided if we demand that all states that are singlets under the symmetry algebra scatter trivially.\footnote{Trivial scattering of singlets in the context of R-R theories was first considered in~\cite{Beisert:2005tm,Beisert:2006qh}.} We focus on configurations with bound state numbers close to $k$ which do not have direct counterparts in the pure R-R model, begining with the $m=k$ singlet states. We then consider $m=k+1$ bound states and show how they are equivalent, up to a singlet, to fundamentals with higher momentum. Finally, we turn to $m=k-1$ bound states, which up to a singlet are equivalent to fundamental R excitations. In the next sub-section we discuss how fusion relations place additional constraints on the world-sheet S matrix that follow from these equivalences.

\subsubsection{Singlet states}
\label{sec:singlet-states}

\begin{figure}
  \centering
  \includegraphics{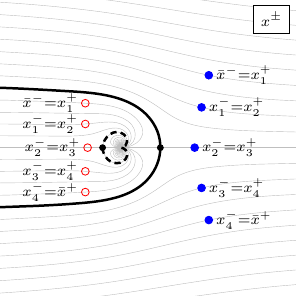}
  \caption{\label{fig:x-singlet-region-0}Two singlet configurations. The blue filled dots show a singlet made up of excitations outside the scallion, while the red hollow dots show a singlet made up of excitations sitting between the scallion and the kidney. Note that all such singlets carry exactly the same charges.}
\end{figure}
Consider a state consisting of a bound state with $m = k-1$ and momentum $p \in (0,2\pi)$, described by bound state Zhukovsky parameters $X^{\pm} = \Xi_{\sL}^{\pm}(p,k-1)$, plus a fundamental crossed excitation with momentum $-2\pi - p$. According to equation~\eqref{eq:location-of-fundamental-crossed-xpm}, the crossed excitation has parameters $\bar{x}^{\pm} = \Xi_{\sL}^{\mp}(p,k-1)$. The full state has total momentum $-2\pi$ and vanishing energy and charge $M\equiv m+\kbar P=0$, which makes it a singlet of the $\algPSU(1|1)^4_{\ce}$ algebra. Since the dynamics of the theory, \ie, the S matrix, is fully determined through $\algPSU(1|1)^4_{\ce}$ representation theory we expect this singlet state to completely decouple from the theory, which means that is should scatter trivially with any other state. 

This construction of a singlet as a $k-1$ physical bound state with momentum $p$ plus a crossed fundamental excitation with momentum $-2\pi - p$ works for $p$ in any range. Moreover, we can generalise it by starting with a bound state with $nk-1$ excitations and a crossed excitation with momentum $-2\pi(n+1) - p$, which leads to a singlet with total momentum $-2\pi(n+1)$. Figure~\ref{fig:x-singlet-region-0} shows two such singlet configurations.

We can also decompose the singlet in various other ways. The exact same state can be viewed as a physical L bound state with bound state number $m$ with $1 \le m < k-1$ and a crossed L bound state with bound state number $k - m$.

\subsubsection{The \texorpdfstring{$m=k+1$}{m=k+1} bound state as a fundamental excitation}
\label{sec:bound-state-and-mom-shift}

Let us now consider an $m=k+1$ bound state. It is described by the parameters
\begin{equation}
  \begin{aligned}
    x_1^+ &= \Xi_{\sL}^+(p,k+1), \\
    x_1^- = x_2^+ &= \Xi_{\sL}^+(q_1,k-1) , \\
    & \vdotswithin{=} \\
    x_k^- = x_{k+1}^+ &= \Xi_{\sL}^-(q_1,k-1) , \\
    x_{k+1}^- &= \Xi_{\sL}^-(p,k+1) .
  \end{aligned}
\end{equation}
We now rename the parameters
\begin{equation}
  \begin{gathered}
    x^+ = x_1^+ , \quad
    x^- = x_{k+1}^- , \quad
    \bar{y}^+ = x_{k+1}^+ , \quad
    \bar{y}^- = x_1^- , \\
    y_j^{\pm} = x_{j+1}^{\pm}, \quad \text{for $j=1,\dotsc,k-1$} .
  \end{gathered}
\end{equation}
The parameters $y_j^{\pm}$ and $\bar{y}^{\pm}$ describe a singlet of exactly the same form discussed above. It contributes $-2\pi$ to the momentum but has no energy or charge. We are left with $x^{\pm} = \Xi_{\sL}^{\pm}(p,k+1)$. This is exactly the parameters for a fundamental physical excitation with momentum $p+2\pi$.

This construction too works for any momentum range. Since we expect the singlet state to decouple from the theory we are left to conclude that a $m=k+1$ bound state with momentum $p$ should scatter in the same way as a fundamental excitation with momentum $p+2\pi$. We may also repeat the construction $n$ times and we are led to conclude that an $m = nk+1$ bound state with momentum $p$ should be identified with a fundamental excitation at momentum $p+2\pi n$. If we did not make such identification, we would have an infinite degeneracy of representations which would in turn lead to an unphysically large spectrum.

\subsubsection{The \texorpdfstring{$m=k-1$}{m=k-1} bound state as an R excitation}
\label{sec:m-is-k-1-L-rep-same-as-m-is-1-R-rep}

Let us now consider an $m=k-1$ bound state. As we saw above, we can construct a singlet from this bound state by adding a fundamental crossed excitation. Hence, the $m=k-1$ bound state acts as an anti-particle to the fundamental excitation. But from representation theory we expect the anti-particle of an L excitation to be an R excitation. Indeed if we let the bound state have momentum $p-2\pi$ we find that it has energy
\begin{equation}
  E_{\sL}(p-2\pi, m=k-1) = \sqrt{(1 - \kbar p)^2 + 4h^2\sin^2\frac{p}{2}} = E_{\sR}(p,m=1)
\end{equation}
which is exactly the same as that of an R excitation of momentum $p$. This means that from the perspective of representations, the $m=k-1$ bound state at momentum $p-2\pi$ is indistinguishable from a fundamental R excitation with momentum $p$. Since we demand that the singlet scatters trivially, we therefore conclude that also the S matrix for these two states is identical.

To make this identification more precise, we note that the Zhukovsky parameters $X^{\pm}$ for an $m=k-1$ bound state with momentum $p-2\pi$ and the parameters $x_{\sR}^{\pm}$ for a fundamental R excitation of momentum $p$ satisfy
\begin{equation}
  X^{\pm}(p-2\pi, k-1)
  = -\frac{1-\kbar p - \sqrt{(1-\kbar p)^2 + 4h^2\sin^2\frac{p}{2}}}{2h\sin\frac{p}{2}} e^{\pm\frac{ip}{2}}
  = \frac{1}{x_{\sR}^{\mp}(p)}
\end{equation}
It is straightforward to check for the \textit{matrix part} of the S matrix~\cite{Lloyd:2014bsa} that this replacement maps the LL S matrix to the S matrices for RL, LR and RR scattering. However, because of fusion, the identification does put constraints on the scalar factors of these S matrices, as we discuss in the next sub-section.

\subsubsection{The \texorpdfstring{$m=k$}{k=m} bound state as a massless excitation}

Finally, we will consider a physical bound state with $m=k$. Such a state has energy
\begin{equation}
  E_{\sL}(p,m=k) = \sqrt{\kbar^2 (p + 2\pi)^2 + 4h^2\sin^2\frac{p}{2}} = E_{\sL}(p+2\pi,m=0) .
\end{equation}
When such a bound state exists it will thus look just like a massless excitation in a higher momentum region.
Taking into account the identification of an $m=k-1$ L bound state with momentum $p$ and an R fundamental excitation of momentum $p+2\pi$, this massless mode can also be seen as a bound state between fundamental L and R excitations.

\vspace{0.5cm}

\bigskip\noindent
In summary, we have shown that the full massive sector of the mixed-flux world-sheet theory is captured by considering only L excitations with bound state numbers in the range $m \le k$ and any momenta.

\subsection{Constraints on the S matrix}
\label{sec:constraints-on-s-mat}

In this section we will check some of the above constraints on the S matrix, focusing on the LL S matrix in the $(-2\pi,0)$ region. As discussed in section~\ref{sec:mixed-flux-scalar-factors} the even part of the highest weight matrix element takes the form
\begin{equation}
  A_{\sLL,\text{even}}^2(x^{\pm},y^{\pm}) =
  \frac
  {S\bigl[u_{\sL}(x^-)-u_{\sL}(y^+)\bigr]}
  {S\bigl[u_{\sL}(x^+)-u_{\sL}(y^-)\bigr]}
  \sigma_{\sLL,\text{even}}^{-2}(x^{\pm},y^{\pm})
\end{equation}
where we have introduced the functions
\begin{equation}
  S[u] = \sinh\bigl(\tfrac{\pi h}{2k} u \bigr) ,
  \qquad
  C[u] = \cosh\bigl(\tfrac{\pi h}{2k} u \bigr) ,
\end{equation}
satisfying
\begin{equation}
  S\bigl[u \pm \tfrac{ik}{h}\bigr] = \pm i C[u] ,
  \qquad
  S\bigl[u \pm \tfrac{2ik}{h}\bigr] = -S[u] ,
  \qquad
  C\bigl[u \pm \tfrac{2ik}{h}\bigr] = -C[u] .
\end{equation}
It is straightforward to fuse this matrix element $m$ times\footnote{%
  All cancellations in the fusion procedure become manifest if we multiply the matrix element $A_{\sLL}^2$ by identity in the form
  \begin{equation*}
    1 = \frac
    {S\bigl[u_{\sL}(x^-)-u_{\sL}(y^-)\bigr]}
    {S\bigl[u_{\sL}(x^+)-u_{\sL}(y^+)\bigr]} .
  \end{equation*}
}%
\begin{equation}
  \begin{aligned}
    A_{\sL^m\sL,\text{even}}^2(X^{\pm},y^{\pm}) &=
    \frac
    {S\bigl[u_{\sL}(X^-)-u_{\sL}(y^+)\bigr]}
    {S\bigl[u_{\sL}(X^+)-u_{\sL}(y^-)\bigr]}
    \frac
    {S\bigl[u_{\sL}(X^-)-u_{\sL}(y^-)\bigr]}
    {S\bigl[u_{\sL}(X^+)-u_{\sL}(y^+)\bigr]}
    \sigma_{\sL^m\sL,\text{even}}^{-2}(X^{\pm},y^{\pm}) .
    \\
    &=
    \frac
    {S\bigl[U_{\sL}-v_{\sL}-\frac{i(m+1)}{h}\bigr]}
    {S\bigl[U_{\sL}-v_{\sL}+\frac{i(m+1)}{h}\bigr]}
    \frac
    {S\bigl[U_{\sL}-v_{\sL}-\frac{i(m-1)}{h}\bigr]}
    {S\bigl[U_{\sL}-v_{\sL}+\frac{i(m-1)}{h}\bigr]}
    \sigma_{\sL^m\sL,\text{even}}^{-2}(X^{\pm},y^{\pm}) ,
  \end{aligned}
\end{equation}
where we have introduced $U_{\sL}$ and $v_{\sL}$ defined through the relations
\begin{equation}
  u_{\sL}(X^{\pm}) = U_{\sL} \pm \frac{im}{h} ,
  \qquad
  u_{\sL}(y^{\pm}) = v_{\sL} \pm \frac{i}{h} .
\end{equation}
In particular for $m=k-1$ we get
\begin{align}
  A_{\sL^{(k-1)}\sL,\text{even}}^2(X^{\pm},y^{\pm})
  &=
      \frac
      {S\bigl[U_{\sL}-v_{\sL}-\frac{ik}{h}\bigr]}
      {S\bigl[U_{\sL}-v_{\sL}+\frac{ik}{h}\bigr]}
      \frac
      {S\bigl[U_{\sL}-v_{\sL}+\frac{2i}{h}-\frac{ik}{h}\bigr]}
      {S\bigl[U_{\sL}-v_{\sL}-\frac{2i}{h}+\frac{ik}{h}\bigr]}
      \sigma_{\sL^{(k-1)}\sL,\text{even}}^{-2}(X^{\pm},y^{\pm})
  \nonumber \\ \label{eq:ALL-even-fused}
  &=
      \frac
      {C\bigl[U_{\sL}-v_{\sL}+\frac{2i}{h}\bigr]}
      {C\bigl[U_{\sL}-v_{\sL}-\frac{2i}{h}\bigr]}
      \sigma_{\sL^{(k-1)}\sL,\text{even}}^{-2}(X^{\pm},y^{\pm})
\end{align}
To compare this with the RL matrix element $A_{\sRL}^2$ we should use the relations $X^{\pm} = 1/x_{\sR}^{\mp}$. In terms of the $u$-plane this translates to
\begin{equation}
  u_{\sL}(X^+) = u_{\sR}(x_{\sR}^-) + \frac{2ik}{h},
  \qquad
  u_{\sL}(X^-) = u_{\sR}(x_{\sR}^+) ,
\end{equation}
or
\begin{equation}
  u_{\sR}  = U_{\sL} - \frac{ik}{h} .
\end{equation}
Inserting this into the above matrix element we find
\begin{equation}
  \begin{aligned}
    A_{\sL^{(k-1)}\sL,\text{even}}^2(1/x_{\sR}^{\mp},y^{\pm})
    &=
    \frac
    {S\bigl[u_{\sR}(x_{\sR}^+)-u_{\sL}(y^-)\bigr]}
    {S\bigl[u_{\sR}(x_{\sR}^-)-u_{\sL}(y^+)\bigr]}
    \sigma_{\sL^{(k-1)}\sL,\text{even}}^{-2}(1/x_{\sR}^{\mp},y^{\pm})
    \\
    &=
    \frac
    {S\bigl[u_{\sR}-v_{\sL}+\frac{2i}{h}\bigr]}
    {S\bigl[u_{\sR}-v_{\sL}-\frac{2i}{h}\bigr]}
    \sigma_{\sL^{(k-1)}\sL,\text{even}}^{-2}(1/x_{\sR}^{\mp},y^{\pm}) .
  \end{aligned}
\end{equation}
This exactly matches the even RL matrix element~\eqref{eq:mixed-ARL-even-min-2pi-0} as long as the even dressing phases satisfy the fusion condition by themselves
\begin{equation}
  \sigma_{\sL^{(k-1)}\sL,\text{even}}^2(1/x_{\sR}^{\mp},y^{\pm})
  =
  \sigma_{\sRL,\text{even}}^2(x_{\sR}^{\pm},y^{\pm}) .
\end{equation}

Let us now turn to the odd S matrix element. Fusing the odd LL matrix element $k-1$ times we get
\begin{equation}
  A_{\sL^{(k-1)}\sL,\text{odd}}^2(1/x^{\mp},y^{\pm})
  =
  \Bigl( \frac{\alpha(1/x^+)}{\alpha(1/x^-)} \Bigr)^{\!\!1/k}
  \Bigl( \frac{\alpha(y^+)}{\alpha(y^-)} \Bigr)^{\!1-\frac{1}{k}}
  e^{-2i\theta_{\sLL}(1/x^{\mp},y^{\pm})} .
\end{equation}
To compare this with the RL matrix element we first note that the RL phase in~\eqref{eq:chis-in-nice-x-form} can be rewritten (up to terms that do not contribute to $\theta_{\sRL}$) as
\begin{equation}
  \begin{aligned}
        \chi_{\sRL}(x,y) &=
    -\!\!\int_{s}^{-s^{-1}} \frac{dz}{2\pi} \frac{\log(y-z)}{\frac{1}{x}-z}
    -\frac{1}{4\pi} \log\frac{\frac{1}{x}+s}{\frac{1}{x}-s^{-1}} \bigl( \log(y+s^{-1}) + \log(y-s) \bigr)
    \\ &\qquad
    -\frac{i}{2} \sign (\Im x) \Bigl( \log\bigl( y - \frac{1}{x} \bigr) - \frac{1}{2} \bigl( \log(y+s^{-1}) + \log(y-s) \bigr) \Bigr) ,
  \end{aligned}
\end{equation}
where the first line equals $-\chi_{\sLL}(1/x,y)$. For a physical real momentum excitation $\Im x^+ > 0$ and $\Im x^- < 0$ and we can use the expression in the second line above to simplify the rational terms in~\eqref{eq:mixed-A-odd} to
\begin{equation}\label{eq:A-RL-odd-nice}
  A_{\sRL,\text{odd}}^2(x^{\pm},y^{\pm})
  =
  \Bigl(
  \frac{\alpha(1/x^+)}{\alpha(1/x^-)}
  \Bigr)^{\!\!1/k}
  \Bigl( \frac{\alpha(y^+)}{\alpha(y^-)} \Bigr)^{\!1-\frac{1}{k}}
  e^{2i\theta_{\sLL}(1/x^{\pm},y^{\pm})} ,
\end{equation}
which perfectly matches what we got by fusing the odd LL matrix element.
Even though we found this expression by assuming that the momentum of the $x^{\pm}$ excitation is real it is straight forward to extend it to the whole complex plane by picking up the right branch of the dressing phase.

Note that the $k-1$ times fused S matrix element~\eqref{eq:ALL-even-fused} does not have an S channel pole. This means that we cannot extend it to an $m=k$ bound state. This has a direct parallel in the XXZ model~\cite{Fabricius:2000yx}. However, in~\cite{Braak:2001} it was pointed out that the XXZ model in fact does contain zero modes that correspond to $m=k$ bound states, even though they do not appear as solutions to the Bethe ansatz equations.
Since our proposed scalar factor has the same form as the XXZ chain, we expect similar results to hold for $\AdS_3$ also. In particular, this should clarify how massless states appear as a results of fusion. We hope to return to these questions in the near future.

Here we have focused our discussion about the relation between different bound states to the case where the bound state sits in the region between the scallion and the kidney. We expect similar results to hold in other momentum regions. For the odd part of the matrix elements the above calculation should be easy to analytically continue to a different region. However, it is clear that the even part of the matrix element requires more care. Consider for example a $k+1$ bound state with momentum in $(0,2\pi)$ scattering with a fundamental excitation in the same region. From~\eqref{eq:A-even-0-2pi} we find
\begin{multline}
  A_{\sL^{(k+1)}\sL,\text{even}}^2(X^{\pm},y^{\pm}) =
  \frac{u_{\sL}(X^-)-u_{\sL}(y^+)}{u_{\sL}(X^+)-u_{\sL}(y^-)}
  \frac{u_{\sL}(X^+)-u_{\sL}(y^+)}{u_{\sL}(X^-)-u_{\sL}(y^-)}
  \\ \times
  \frac{X^+}{X^-} \Bigl( \frac{y^-}{y^+} \Bigr)^{k+1}
  \sigma_{\sL^{(k+1)}\sL,\text{even}}^{-2}(X^{\pm},y^{\pm}) .
\end{multline}
Since the rapidity plane is not periodic we cannot drop the second factor here. This matrix element should equal $A_{\sLL,\text{even}}^2(p+2\pi,q)$ with two fundamental excitations. For this to be possible we need to pick up a compensating factor from the even dressing phase when we analytically continue from $p$ to $p+2\pi$. This should provide a test of the even dressing phase.

\section{Conclusions}
\label{sec:conclusions}

We have analysed the kinematical structure of worldsheet excitations in mixed-flux  $\AdS_3\times \Sphere^3\times \Torus^4$ geometries and found a structure far richer than in the case of R-R backgrounds. Momenta of these excitations are no longer periodic and the related Zhukovsky variables have infinitely many sheets, which we have shown can be reached through analytic continuations around the singular $p=2\pi n$ ($n\in \mathbb{Z}$) points. Similarly, the variable $u$ that is commonly used in integrable holography has a mixed-flux generalisation for each pair of Zhukovsky sheets. We have identified one set of Zhukovsky sheets and corresponding $u$ variable, which exhibit a novel type of periodicity. Alongside this paper we include a graphics programme that can be used to visualise this complicted analytic structure for both fundamental and bound state excitations~\cite{pxu-gui}.

Using these analyticity insights, we have shown that excitations which appear fundamental are equivalent to bound states upon suitable shifts of momenta. Similarly, R excitations can be constructed from L bound states. These constraints place restrictions on the S matrix of the theory, particularly on its scalar factors. We have solved the odd part of the crossing equations and proposed scalar factors for the S matrices that have the correct pole structure for the bound state spectrum of the theory. Focusing on the $(-2\pi,0)$ momentum region, we showed that these odd dressing phases and scalar factors have the required properties under fusion. Further, we proposed a scalar factor closely resembling\footnote{Up to simple rescalings of the rapidity variable.} the XXZ model at a root of unity, \ie, with anisotropy of the form $\Delta = \cos\frac{\pi}{k}$. This factor ensures consistency with $u$-periodicity in this region, the correct bound-state simple poles expected in the theory, as well as the required properties under fusion. 
Because of the known connections between XXZ models and quantum groups, it would be interesting to investigate whether there is an underlying quantum group structure in the mixed flux backgrounds. Such connections have been studied in $\eta$-like deformations of both R-R~\cite{Seibold:2019dvf,Seibold:2021lju} and mixed flux backgrounds~\cite{Hoare:2022asa}, as well as in more general ``elliptic'' deformations~\cite{Hoare:2023zti}. We leave to future work finding a more precise relation between these deformations and our results.

Unlike $\AdS_5$, integrable $\AdS_3$ theories' dressing factors remain non-trivial in the $h\rightarrow 0$ limit, both in R-R and mixed-flux cases, signalling the non-square-root nature of the branch points~\cite{Borsato:2013hoa}. This is an important difference in the weakly-coupled regime: the $\AdS_5$ case reduces to an XXX spin-chain in the $\alg{su}(2)$ sector, while in the $\AdS_3$ case, depending on the momentum region we would get XXX or XXZ spin-chains, \textit{dressed} by non-trivial phases. Intriguingly, this suggests that the planar spectrum of the WZW model deformed by the axion modulus should be governed by such ``dressed'' spin-chains.

Having gained a better understanding of the analyticity properties of the mixed-flux theory across the full range of $p$, we revisited its relativistic limit. Related limits were introduced in R-R backgrounds~\cite{Bombardelli:2018jkj} where they led to a difference-form for the massless S matrix and dressing factors~\cite{Fontanella:2019baq,Fontanella:2019ury}, fixing certain potential CDD ambiguities and leading to novel expressions for massless dressing factors. Relativistic limits were also studied in mixed-flux backgrounds~\cite{Fontanella:2019ury}, where a  close relationship to relativistic integrable models~\cite{Fendley:1991ve,Fendley:1992dm} as well as q-deformed holographic ones~\cite{Arutyunov:2012zt,Arutyunov:2012ai,Hoare:2013ysa} was identified. An improved version of this relativistic limit was proposed in~\cite{Frolov:2023lwd}, for which the Hopf algebra and hence S matrix and dressing phases were the same as in~\cite{Fontanella:2019ury}, but whose dispersion relation matched the bound-state structure more closely to the $\AdS_3$ case. We showed that the correct way to interpret the relativistic limit is as coming from the region between the scallion and kidney, zooming in on momenta near the $E_I$ branch points, as illustrated in figure~\ref{fig:p-and-u-in-rel-limit}. As we explained, from this perspective the truncation of the bound state spectrum is natural, with all but a finite number of bound states decoupling because they can equivalently be viewed as states whose momenta are not near the minimal values~\eqref{eq:re-limit-mom}. With these clarifications, we then showed that the odd dressing phase and scalar factors we have found in this paper reduce to the corresponding expressions given in~\cite{Frolov:2023lwd,Fontanella:2019ury}. 

In section~\ref{sec:bound-state-s-mat-and-phases}, we discussed the role of fusion and bound states in the mixed flux backgrounds. There, we showed that singlet states of the type first introduced in~\cite{Beisert:2005tm,Beisert:2006qh}  play an important role in mixed flux theories. As in those higher-dimensional examples, our crossing equations follow from the trivial scattering of such singlets. In turn, this implies the equivalence of, for example,  $m=k+1$ bound states with momentum $p$ and $m=1$ fundamentals with momentum $p+2\pi$, as well as the equivalence of $m=k-1$ L bound states with momentum $p-2\pi$ and $m=1$ R fundamentals with momentum $p$. As we discussed in section~\ref{sec:physical-regions}, these equivalences are important because they ensure that the spectrum of excitations does not include an infinite degeneracy of identically charged states when analytically continuing in the full $p$ plane. These equivalences of the excitations provide a \textit{unified} description for allowed values of excitation momenta and on bound state numbers.\footnote{Specifically, the kinematics at $k=1$ is similar to the cases of $k>1$ and so a shift of momentum by $2\pi$ is equivalent to a shift of $m$ by $k=1$. Hence,  from the point of view of kinematics, viewing the excitations as fundamental massive or massless in the case of $k=1$ should be equivalent.}

Further, consistency of these equivalences with fusion places constraints on the scalar factors of the S matrix that we discussed in section~\ref{sec:constraints-on-s-mat}.  For the odd dressing phases, it is this important consistency that has led us to include the $\alpha^{1/k}$ factors in the normalisations~\eqref{eq:mixed-A-odd}. From semi-classical near-BMN worldsheet intuition, one might have expected  a smooth $k\rightarrow 0$ limit. However, it is well known that in $\AdS_3$ integrable theories 1-loop worldsheet calculations often do not match with corresponding expansions of 1-loop results, see for example point 3 on page 35 of ~\cite{Frolov:2021fmj}. There is an expectation that the one-loop cancellation between massless bosons and fermions, analogous to what happens in flat space, does not fully capture the contributions to the S matrix or energies of states at that order and that naively sub-leading terms should be included.\footnote{We would like to thank Kolya Gromov and Arkady Tseytlin for discussions about this.}. Nevertheless, it will be important to revisit these fusion constraints and the $\alpha^{1/k}$ factors when including the even dressing phases, for example those proposed recently in~\cite{Frolov:2024pkz}.

The relations between massless modes and bound states with shifted momenta discussed in section~\ref{sec:special-bound-state} suggests that wrapping corrections coming from massless and mixed mass effects may be easier to analyse in the mixed-flux setting than in the R-R case, because one may be able to view all modes as being massive, under suitable momentum shift. In light of this, it may be worth revisiting the findings of~\cite{Abbott:2015pps}, as well as generalising the QSC computations of~\cite{Cavaglia:2022xld}.

To complete the determination of the S matrix of mixed-flux backgorunds it remains to find the even dressing factors, which should amount to finding the mixed-flux generalisation of the BES phase. We leave this for future work. It would be interesting to understand the mixed-flux TBA and QSC and use them to solve the spectral problem on $\AdS_3\times \Sphere^3\times \Torus^4$ backgrounds. This would provide much needed data for testing the $\AdS_3/\CFT_2$ duality and understanding the connection to the $\Sym^N(\Torus^4)$ dual CFT,  the $k=1$ results using the hybrid formalism~\cite{Eberhardt:2018ouy}, as well as the appearance of integrability on the Higgs branch of the infrared limit of the D1-D5-brane gauge theory~\cite{Sax:2014mea}.

\section*{Acknowledgements}
We would like to thank Andrea Cavagli\`a, Simon Ekhammar, Kolya Gromov, Suvajit Majumder, Arkady Tseytlin, Dima Volin and Konstantin Zarembo for many interesting discussions and collaborations. We have also enjoyed on several occasions discussing  $\AdS_3$ mixed-flux integrability with Sergey Frolov and Alessandro Sfondrini. We are especially grateful to Alessandro Torrielli for numerous enlightening conversations and sharing his insights with us, as well as comments on the manuscript.

D.R.\@ acknowledges studentship funding from The Science and Technology Facilities Council ST/W507398/1 and from the School of Science and Technology, City, University of London. B.S.\@ acknowledges funding support from The Science and Technology Facilities Council through Consolidated Grants ``Theoretical Particle Physics at City, University of London''  ST/T000716/1 and ST/X000729/1. B.S.\@ would lke to thank the organisers of the NORDITA workshop ``New Perspectives on Quantum Field Theory with Boundaries, Impurities, and Defects'' for hospitality and support where parts of this work were undertaken. The work of O.O.S\@ was supported by VR grant 2021-04578. Nordita is supported in part by NordForsk.

\appendix

\bibliographystyle{oos}
\bibliography{refs,pxu}

\makeatletter \@ifundefined{Sphere}{\newcommand{\Sphere}{{S}{}}}{}
  \@ifundefined{AdS}{\newcommand{\AdS}{{AdS}{}}}{}
  \@ifundefined{CFT}{\newcommand{\CFT}{{CFT}{}}}{}
  \@ifundefined{CP}{\newcommand{\CP}{CP}}{}
  \@ifundefined{Torus}{\newcommand{\Torus}{{T}{}}}{}
  \@ifundefined{superN}{\newcommand{\superN}{\mathcal{N}}}{}
  \@ifundefined{grpOSp}{\newcommand{\grpOSp}{\text{OSp}}}{}
  \@ifundefined{grpPSU}{\newcommand{\grpPSU}{\text{PSU}}}{}
  \@ifundefined{grpSU}{\newcommand{\grpSU}{\text{SU}}}{}
  \@ifundefined{grpU}{\newcommand{\grpU}{\text{U}}}{}
  \@ifundefined{grpD}{\newcommand{\grpD}{\text{D}}}{}
  \@ifundefined{grpSL}{\newcommand{\grpSL}{\text{SL}}}{}
  \@ifundefined{grpSp}{\newcommand{\grpSp}{\text{Sp}}}{}
  \@ifundefined{grpUSp}{\newcommand{\grpUSp}{\text{USp}}}{}
  \@ifundefined{grpSO}{\newcommand{\grpSO}{\text{SO}}}{}
  \@ifundefined{grpO}{\newcommand{\grpO}{\text{O}}}{}
  \@ifundefined{algOSp}{\newcommand{\algOSp}{\text{osp}}}{}
  \@ifundefined{algPSU}{\newcommand{\algPSU}{\text{psu}}}{}
  \@ifundefined{algSU}{\newcommand{\algSU}{\text{su}}}{}
  \@ifundefined{algSp}{\newcommand{\algSp}{\text{sp}}}{}
  \@ifundefined{algSL}{\newcommand{\algSL}{\text{sl}}}{}
  \@ifundefined{algGL}{\newcommand{\algGL}{\text{gl}}}{}
  \@ifundefined{algU}{\newcommand{\algU}{\text{u}}}{}
  \@ifundefined{algSO}{\newcommand{\algSO}{\text{so}}}{}
  \@ifundefined{algO}{\newcommand{\algO}{\text{o}}}{}
  \@ifundefined{Integers}{\newcommand{\Integers}{\text{Z}}}{}
  \@ifundefined{Reals}{\newcommand{\Reals}{\text{R}}}{}
  \@ifundefined{Complex}{\newcommand{\Complex}{\text{C}}}{} \makeatother
\begin{thebibliography}{10}
\ifx\href\asklfhas\newcommand{\href}[2]{#2}\fi
\ifx\arxivref\asklfhas\newcommand{\arxivref}[2]{\href{http://arxiv.org/abs/#1}{#2}}\fi
\raggedright
\small
\parskip 0pt

%%CITATION = 0912.1723;%%
\bibitem[Babichenko \textit{et~al.\@}(2010)Babichenko, Stefa{\'n}ski, and
  Zarembo]{Babichenko:2009dk}
A.~Babichenko, B.~Stefa{\'n}ski,~jr. and K.~Zarembo,
\textit{Integrability and the {AdS${}_{3}$/CFT${}_{2}$} correspondence},
\textsf{JHEP~1003,~058~(2010)},
\texttt{\arxivref{0912.1723}{arxiv:0912.1723}}.
%
%%CITATION = 1106.2558;%%
\bibitem[Ohlsson~Sax and Stefa{\'n}ski(2011)]{OhlssonSax:2011ms}
O.~Ohlsson~Sax and B.~Stefa{\'n}ski,~jr.,
\textit{Integrability, spin-chains and the {AdS${}_{3}$/CFT${}_{2}$}
  correspondence},
\textsf{JHEP~1108,~029~(2011)},
\texttt{\arxivref{1106.2558}{arxiv:1106.2558}}.
%
%%CITATION = ARXIV:1209.4049;%%
\bibitem[Cagnazzo and Zarembo(2012)]{Cagnazzo:2012se}
A.~Cagnazzo and K.~Zarembo,
\textit{{B}-field in {AdS${}_{3}$/CFT${}_{2}$} correspondence and
  integrability},
\textsf{JHEP~1211,~133~(2012)},
\texttt{\arxivref{1209.4049}{arxiv:1209.4049}}.
%
%%CITATION = 0804.3267;%%
\bibitem[David and Sahoo(2008)]{David:2008yk}
J.~R.~David and B.~Sahoo,
\textit{Giant magnons in the {D1}-{D5} system},
\textsf{JHEP~0807,~033~(2008)},
\texttt{\arxivref{0804.3267}{arxiv:0804.3267}}.
%
%%CITATION = 1005.0501;%%
\bibitem[David and Sahoo(2010)]{David:2010yg}
J.~R.~David and B.~Sahoo,
\textit{{S}-matrix for magnons in the {D1}-{D5} system},
\textsf{JHEP~1010,~112~(2010)},
\texttt{\arxivref{1005.0501}{arxiv:1005.0501}}.
%
%%CITATION = ARXIV:1804.02023;%%
\bibitem[Ohlsson~Sax and Stefa{\'n}ski(2018)]{OhlssonSax:2018hgc}
O.~Ohlsson~Sax and B.~Stefa{\'n}ski,~jr.,
\textit{Closed strings and moduli in {AdS${}_{3}$/CFT${}_{2}$}},
\textsf{JHEP~1805,~101~(2018)},
\texttt{\arxivref{1804.02023}{arxiv:1804.02023}}.
%
%%CITATION = ARXIV:1410.0866;%%
\bibitem[Lloyd \textit{et~al.\@}(2015)Lloyd, Ohlsson~Sax, Sfondrini, and
  Stefa{\'n}ski]{Lloyd:2014bsa}
T.~Lloyd, O.~Ohlsson~Sax, A.~Sfondrini and B.~Stefa{\'n}ski,~jr.,
\textit{The complete worldsheet {S} matrix of superstrings on {$\AdS_3 \times
  \Sphere^3 \times \Torus^4$} with mixed three-form flux},
\textsf{Nucl.~Phys.~B891,~570~(2015)},
\texttt{\arxivref{1410.0866}{arxiv:1410.0866}}.
%
%%CITATION = ARXIV:1303.5995;%%
\bibitem[Borsato \textit{et~al.\@}(2013)Borsato, Ohlsson~Sax, Sfondrini,
  Stefa{\'n}ski, and Torrielli]{Borsato:2013qpa}
R.~Borsato, O.~Ohlsson~Sax, A.~Sfondrini, B.~Stefa{\'n}ski,~jr. and
  A.~Torrielli,
\textit{The all-loop integrable spin-chain for strings on {$\AdS_3 \times
  \Sphere^3 \times \Torus^4$}: the massive sector},
\textsf{JHEP~1308,~043~(2013)},
\texttt{\arxivref{1303.5995}{arxiv:1303.5995}}.
%
%%CITATION = ARXIV:1403.4543;%%
\bibitem[Borsato \textit{et~al.\@}(2014)Borsato, Ohlsson~Sax, Sfondrini, and
  Stefa{\'n}ski]{Borsato:2014exa}
R.~Borsato, O.~Ohlsson~Sax, A.~Sfondrini and B.~Stefa{\'n}ski,~jr.,
\textit{Towards the all-loop worldsheet {S} matrix for {$\AdS_3 \times
  \Sphere^3 \times \Torus^4$}},
\textsf{Phys.~Rev.~Lett.~113,~131601~(2014)},
\texttt{\arxivref{1403.4543}{arxiv:1403.4543}}.
%
%%CITATION = ARXIV:1406.0453;%%
\bibitem[Borsato \textit{et~al.\@}(2014)Borsato, Ohlsson~Sax, Sfondrini, and
  Stefa{\'n}ski]{Borsato:2014hja}
R.~Borsato, O.~Ohlsson~Sax, A.~Sfondrini and B.~Stefa{\'n}ski,~jr,
\textit{The complete {$\AdS_3 \times \Sphere^3 \times \Torus^4$} worldsheet
  {S}-matrix},
\textsf{JHEP~1410,~66~(2014)},
\texttt{\arxivref{1406.0453}{arxiv:1406.0453}}.
%
%%CITATION = ARXIV:1311.1794;%%
\bibitem[Hoare \textit{et~al.\@}(2014)Hoare, Stepanchuk, and
  Tseytlin]{Hoare:2013lja}
B.~Hoare, A.~Stepanchuk and A.~Tseytlin,
\textit{Giant magnon solution and dispersion relation in string theory in
  {$\AdS_3 \times \Sphere^3 \times \Torus^4$} with mixed flux},
\textsf{Nucl.~Phys.~B879,~318~(2014)},
\texttt{\arxivref{1311.1794}{arxiv:1311.1794}}.
%
%%CITATION = HEP-TH/9905064;%%
\bibitem[Larsen and Martinec(1999)]{Larsen:1999uk}
F.~Larsen and E.~J.~Martinec,
\textit{{$\grpU(1)$} charges and moduli in the {D1}-{D5} system},
\textsf{JHEP~9906,~019~(1999)},
\texttt{\arxivref{hep-th/9905064}{hep-th/9905064}}.
%
%%CITATION = HEP-TH/0001053;%%
\bibitem[Maldacena and Ooguri(2001)]{Maldacena:2000hw}
J.~M.~Maldacena and H.~Ooguri,
\textit{Strings in {$\AdS_3$} and {$\grpSL(2,R)$} {WZW} model. {I}},
\textsf{J.~Math.~Phys.~42,~2929~(2001)},
\texttt{\arxivref{hep-th/0001053}{hep-th/0001053}}.
%
\bibitem[Ohlsson~Sax(2023)]{pxu-gui}
O.~Ohlsson~Sax,
\textit{{PXU}},
2023.
Available at \url{https://olofos.github.io/pxu-gui/}.
%
%%CITATION = ARXIV:1405.4857;%%
\bibitem[Gromov \textit{et~al.\@}(2015)Gromov, Kazakov, Leurent, and
  Volin]{Gromov:2014caa}
N.~Gromov, V.~Kazakov, S.~Leurent and D.~Volin,
\textit{Quantum spectral curve for arbitrary state/operator in
  {AdS${}_{5}$/CFT${}_{4}$}},
\textsf{JHEP~1509,~187~(2015)},
\texttt{\arxivref{1405.4857}{arxiv:1405.4857}}.
%
%%CITATION = ARXIV:1701.00473;%%
\bibitem[Bombardelli \textit{et~al.\@}(2017)Bombardelli, Cavagli{\`a},
  Fioravanti, Gromov, and Tateo]{Bombardelli:2017vhk}
D.~Bombardelli, A.~Cavagli{\`a}, D.~Fioravanti, N.~Gromov and R.~Tateo,
\textit{The full quantum spectral curve for {AdS${}_{4}$/CFT${}_{3}$}},
\texttt{\arxivref{1701.00473}{arxiv:1701.00473}}.
%
\bibitem[Cavagli{\`a} \textit{et~al.\@}(2021)Cavagli{\`a}, Gromov,
  Stefa{\'n}ski, and Torrielli]{Cavaglia:2021eqr}
A.~Cavagli{\`a}, N.~Gromov, B.~Stefa{\'n}ski,~jr. and A.~Torrielli,
\textit{Quantum spectral curve for {AdS${}_3$/CFT${}_2$}: a proposal},
\textsf{JHEP~2112,~048~(2021)},
\texttt{\arxivref{2109.05500}{arxiv:2109.05500}}.
%
\bibitem[Ekhammar and Volin(2022)]{Ekhammar:2021pys}
S.~Ekhammar and D.~Volin,
\textit{Monodromy bootstrap for {$\grpSU(2|2)$} quantum spectral curves: from
  hubbard model to {$\AdS_3/\CFT_2$}},
\textsf{JHEP~2203,~192~(2022)},
\texttt{\arxivref{2109.06164}{arxiv:2109.06164}}.
%
\bibitem[Frolov and Sfondrini(2022)]{Frolov:2021fmj}
S.~Frolov and A.~Sfondrini,
\textit{New dressing factors for {$\AdS_3/\CFT_2$}},
\textsf{JHEP~2204,~162~(2022)},
\texttt{\arxivref{2112.08896}{arxiv:2112.08896}}.
%
%%CITATION = HEP-TH/0609044;%%
\bibitem[Beisert \textit{et~al.\@}(2006)Beisert, Hern{\'a}ndez, and
  L{\'o}pez]{Beisert:2006ib}
N.~Beisert, R.~Hern{\'a}ndez and E.~L{\'o}pez,
\textit{A crossing-symmetric phase for {$\AdS_5 \times \Sphere^5$}},
\textsf{JHEP~0611,~070~(2006)},
\texttt{\arxivref{hep-th/0609044}{hep-th/0609044}}.
%
\bibitem[Frolov \textit{et~al.\@}(2023)Frolov, Polvara, and
  Sfondrini]{Frolov:2023lwd}
S.~Frolov, D.~Polvara and A.~Sfondrini,
\textit{On mixed-flux worldsheet scattering in {$\AdS_3/\CFT_2$}},
\textsf{JHEP~2311,~055~(2023)},
\texttt{\arxivref{2306.17553}{arxiv:2306.17553}}.
%
%%CITATION = HEP-TH/0610251;%%
\bibitem[Beisert \textit{et~al.\@}(2007)Beisert, Eden, and
  Staudacher]{Beisert:2006ez}
N.~Beisert, B.~Eden and M.~Staudacher,
\textit{Transcendentality and crossing},
\textsf{J.~Stat.~Mech.~0701,~P01021~(2007)},
\texttt{\arxivref{hep-th/0610251}{hep-th/0610251}}.
%
%%CITATION = 1012.3982;%%
\bibitem[Beisert \textit{et~al.\@}(2012)]{Beisert:2010jr}
N.~Beisert \textit{et~al.\@},
\textit{Review of {AdS/CFT} integrability: An overview},
\textsf{Lett.~Math.~Phys.~99,~3~(2012)},
\texttt{\arxivref{1012.3982}{arxiv:1012.3982}}.
%
%%CITATION = ARXIV:0901.4937;%%
\bibitem[Arutyunov and Frolov(2009)]{Arutyunov:2009ga}
G.~Arutyunov and S.~Frolov,
\textit{Foundations of the {$\AdS_5 \times \Sphere^5$} superstring. {P}art
  {I}},
\textsf{J.~Phys.~A~A42,~254003~(2009)},
\texttt{\arxivref{0901.4937}{arxiv:0901.4937}}.
%
%%CITATION = ARXIV:1405.6087;%%
\bibitem[Babichenko \textit{et~al.\@}(2014)Babichenko, Dekel, and
  Ohlsson~Sax]{Babichenko:2014yaa}
A.~Babichenko, A.~Dekel and O.~Ohlsson~Sax,
\textit{Finite-gap equations for strings on {$\AdS_3 \times \Sphere^3 \times
  \Torus^4$} with mixed 3-form flux},
\textsf{JHEP~1411,~122~(2014)},
\texttt{\arxivref{1405.6087}{arxiv:1405.6087}}.
%
%%CITATION = ARXIV:1211.5119;%%
\bibitem[Borsato \textit{et~al.\@}(2013)Borsato, Ohlsson~Sax, and
  Sfondrini]{Borsato:2012ud}
R.~Borsato, O.~Ohlsson~Sax and A.~Sfondrini,
\textit{A dynamic {$\algSU(1|1)^2$} {S}-matrix for {AdS${}_{3}$/CFT${}_{2}$}},
\textsf{JHEP~1304,~113~(2013)},
\texttt{\arxivref{1211.5119}{arxiv:1211.5119}}.
%
%%CITATION = ARXIV:1306.2512;%%
\bibitem[Borsato \textit{et~al.\@}(2013)Borsato, Ohlsson~Sax, Sfondrini,
  Stefa{\'n}ski, and Torrielli]{Borsato:2013hoa}
R.~Borsato, O.~Ohlsson~Sax, A.~Sfondrini, B.~Stefa{\'n}ski,~jr. and
  A.~Torrielli,
\textit{Dressing phases of {AdS${}_{3}$/CFT${}_{2}$}},
\textsf{Phys.~Rev.~D88,~066004~(2013)},
\texttt{\arxivref{1306.2512}{arxiv:1306.2512}}.
%
%%CITATION = ARXIV:1912.04320;%%
\bibitem[Ohlsson~Sax and Stefa{\'n}ski(2020)]{OhlssonSax:2019nlj}
O.~Ohlsson~Sax and B.~Stefa{\'n}ski,~jr.,
\textit{On the singularities of the {R-R} {$\AdS_3 \times \Sphere^3 \times
  \Torus^4$} {S} matrix},
\textsf{J.~Phys.~A~53,~155402~(2020)},
\texttt{\arxivref{1912.04320}{arxiv:1912.04320}}.
%
%%CITATION = HEP-TH/0608049;%%
\bibitem[Roiban(2007)]{Roiban:2006gs}
R.~Roiban,
\textit{Magnon bound-state scattering in gauge and string theory},
\textsf{JHEP~0704,~048~(2007)},
\texttt{\arxivref{hep-th/0608049}{hep-th/0608049}}.
%
%%CITATION = HEP-TH/0604175;%%
\bibitem[Dorey(2006)]{Dorey:2006dq}
N.~Dorey,
\textit{Magnon bound states and the {AdS/CFT} correspondence},
\textsf{J.~Phys.~A39,~13119~(2006)},
\texttt{\arxivref{hep-th/0604175}{hep-th/0604175}}.
%
%%CITATION = HEP-TH/0511082;%%
\bibitem[Beisert(2008)]{Beisert:2005tm}
N.~Beisert,
\textit{The {$\algSU(2|2)$} dynamic {$S$}-matrix},
\textsf{Adv.~Theor.~Math.~Phys.~12,~945~(2008)},
\texttt{\arxivref{hep-th/0511082}{hep-th/0511082}}.
%
%%CITATION = NLIN/0610017;%%
\bibitem[Beisert(2007)]{Beisert:2006qh}
N.~Beisert,
\textit{The analytic {B}ethe ansatz for a chain with centrally extended
  {$\algSU(2|2)$} symmetry},
\textsf{J.~Stat.~Mech.~0701,~P017~(2007)},
\texttt{\arxivref{nlin/0610017}{nlin/0610017}}.
%
%%CITATION = ARXIV:0710.1568;%%
\bibitem[Arutyunov and Frolov(2007)]{Arutyunov:2007tc}
G.~Arutyunov and S.~Frolov,
\textit{On string {S}-matrix, bound states and {TBA}},
\textsf{JHEP~0712,~024~(2007)},
\texttt{\arxivref{0710.1568}{arxiv:0710.1568}}.
%
%%CITATION = ARXIV:1605.00518;%%
\bibitem[Borsato \textit{et~al.\@}(2016)Borsato, Ohlsson~Sax, Sfondrini, and
  Stefa{\'n}ski]{Borsato:2016kbm}
R.~Borsato, O.~Ohlsson~Sax, A.~Sfondrini and B.~Stefa{\'n}ski,~jr.,
\textit{On the spectrum of {$\AdS_3 \times \Sphere^3 \times \Torus^4$} strings
  with {R}amond-{R}amond flux},
\textsf{J.~Phys.~A49,~41LT03~(2016)},
\texttt{\arxivref{1605.00518}{arxiv:1605.00518}}.
%
%%CITATION = ARXIV:1607.00914;%%
\bibitem[Borsato \textit{et~al.\@}(2017)Borsato, Ohlsson~Sax, Sfondrini,
  Stefa{\'n}ski, and Torrielli]{Borsato:2016xns}
R.~Borsato, O.~Ohlsson~Sax, A.~Sfondrini, B.~Stefa{\'n}ski,~jr. and
  A.~Torrielli,
\textit{On the dressing factors, {B}ethe equations and {Y}angian symmetry of
  strings on {$\AdS_3 \times \Sphere^3 \times \Torus^4$}},
\textsf{J.~Phys.~A50,~024004~(2017)},
\texttt{\arxivref{1607.00914}{arxiv:1607.00914}}.
%
\bibitem[Cavagli{\`a} and Ekhammar()]{CavagliaEkhammar}
A.~Cavagli{\`a} and S.~Ekhammar.
Unpublished. See presentation at IGST 2022.
%
\bibitem[Cavagli{\`a} \textit{et~al.\@}()Cavagli{\`a}, Ekhammar, Majumder,
  Stefa{\'n}ski, and Torrielli]{CavagliaEkhammarMajumderStefanskiTorrielli}
A.~Cavagli{\`a}, S.~Ekhammar, S.~Majumder, B.~Stefa{\'n}ski,~jr. and
  A.~Torrielli.
Unpublished.
%
%%CITATION = ARXIV:1807.07775;%%
\bibitem[Bombardelli \textit{et~al.\@}(2018)Bombardelli, Stefa{\'n}ski, and
  Torrielli]{Bombardelli:2018jkj}
D.~Bombardelli, B.~Stefa{\'n}ski and A.~Torrielli,
\textit{The low-energy limit of {AdS${}_{3}$/CFT${}_{2}$} and its {TBA}},
\textsf{JHEP~1810,~177~(2018)},
\texttt{\arxivref{1807.07775}{arxiv:1807.07775}}.
%
%%CITATION = PHRVA,101,453;%%
\bibitem[Castillejo \textit{et~al.\@}(1956)Castillejo, Dalitz, and
  Dyson]{Castillejo:1955ed}
L.~Castillejo, R.~H.~Dalitz and F.~J.~Dyson,
\textit{Low's scattering equation for the charged and neutral scalar theories},
\textsf{Phys.~Rev.~101,~453~(1956)}.
%
%%CITATION = HEP-TH/0703104;%%
\bibitem[Dorey \textit{et~al.\@}(2007)Dorey, Hofman, and
  Maldacena]{Dorey:2007xn}
N.~Dorey, D.~M.~Hofman and J.~M.~Maldacena,
\textit{On the singularities of the magnon {S}-matrix},
\textsf{Phys.~Rev.~D76,~025011~(2007)},
\texttt{\arxivref{hep-th/0703104}{hep-th/0703104}}.
%
\bibitem[Frolov \textit{et~al.\@}(2024)Frolov, Polvara, and
  Sfondrini]{Frolov:2024pkz}
S.~Frolov, D.~Polvara and A.~Sfondrini,
\textit{{Dressing Factors for Mixed-Flux $AdS_3\times S^3\times T^4$
  Superstrings}},
\texttt{\arxivref{2402.11732}{arxiv:2402.11732}}.
%
%%CITATION = HEP-TH/0405001;%%
\bibitem[Beisert \textit{et~al.\@}(2004)Beisert, Dippel, and
  Staudacher]{Beisert:2004hm}
N.~Beisert, V.~Dippel and M.~Staudacher,
\textit{A novel long range spin chain and planar {$\superN = 4$} super {Y}ang-
  {M}ills},
\textsf{JHEP~0407,~075~(2004)},
\texttt{\arxivref{hep-th/0405001}{hep-th/0405001}}.
%
\bibitem[Fabricius and McCoy(2001)]{Fabricius:2000yx}
K.~Fabricius and B.~M.~McCoy,
\textit{{B}ethe's equation is incomplete for the {XXZ} model at roots of
  unity},
\textsf{J.~Statist.~Phys.~103,~647~(2001)},
\texttt{\arxivref{cond-mat/0009279}{cond-mat/0009279}}.
%
\bibitem[Baxter(2002)]{Baxter:2001sx}
R.~J.~Baxter,
\textit{Completeness of the {B}ethe ansatz for the six and eight vertex
  models},
\textsf{J.~Statist.~Phys.~108,~1~(2002)},
\texttt{\arxivref{cond-mat/0111188}{cond-mat/0111188}}.
%
\bibitem[Braak and Andrei(2001)]{Braak:2001}
D.~Braak and N.~Andrei,
\textit{On the spectrum of the {XXZ}-chain at roots of unity},
\textsf{Journal~of~statistical~physics~105,~677~(2001)},
\texttt{\arxivref{cond-mat/0106593}{cond-mat/0106593}}.
%
%%CITATION = ARXIV:1903.10759;%%
\bibitem[Fontanella and Torrielli(2019)]{Fontanella:2019baq}
A.~Fontanella and A.~Torrielli,
\textit{Geometry of massless scattering in integrable superstring},
\textsf{JHEP~1906,~116~(2019)},
\texttt{\arxivref{1903.10759}{arxiv:1903.10759}}.
%
%%CITATION = ARXIV:1905.00757;%%
\bibitem[Fontanella \textit{et~al.\@}(2019)Fontanella, Ohlsson~Sax,
  Stefa{\'n}ski, and Torrielli]{Fontanella:2019ury}
A.~Fontanella, O.~Ohlsson~Sax, B.~Stefa{\'n}ski,~jr. and A.~Torrielli,
\textit{The effectiveness of relativistic invariance in {$\AdS_3$}},
\textsf{JHEP~1907,~105~(2019)},
\texttt{\arxivref{1905.00757}{arxiv:1905.00757}}.
%
\bibitem[Fendley and Intriligator(1992)]{Fendley:1991ve}
P.~Fendley and K.~A.~Intriligator,
\textit{Scattering and thermodynamics of fractionally charged supersymmetric
  solitons},
\textsf{Nucl.~Phys.~B~372,~533~(1992)},
\texttt{\arxivref{hep-th/9111014}{hep-th/9111014}}.
%
\bibitem[Fendley and Intriligator(1992)]{Fendley:1992dm}
P.~Fendley and K.~A.~Intriligator,
\textit{Scattering and thermodynamics in integrable {$\superN = 2$} theories},
\textsf{Nucl.~Phys.~B~380,~265~(1992)},
\texttt{\arxivref{hep-th/9202011}{hep-th/9202011}}.
%
%%CITATION = ARXIV:1208.3478;%%
\bibitem[Arutyunov \textit{et~al.\@}(2012)Arutyunov, de~Leeuw, and van
  Tongeren]{Arutyunov:2012zt}
G.~Arutyunov, M.~de~Leeuw and S.~J.~van~Tongeren,
\textit{The quantum deformed mirror {TBA} {I}},
\textsf{JHEP~1210,~090~(2012)},
\texttt{\arxivref{1208.3478}{arxiv:1208.3478}}.
%
\bibitem[Arutyunov \textit{et~al.\@}(2013)Arutyunov, de~Leeuw, and van
  Tongeren]{Arutyunov:2012ai}
G.~Arutyunov, M.~de~Leeuw and S.~J.~van~Tongeren,
\textit{The quantum deformed mirror {TBA} {II}},
\textsf{JHEP~1302,~012~(2013)},
\texttt{\arxivref{1210.8185}{arxiv:1210.8185}}.
%
%%CITATION = ARXIV:1303.1447;%%
\bibitem[Hoare \textit{et~al.\@}(2013)Hoare, Hollowood, and
  Miramontes]{Hoare:2013ysa}
B.~Hoare, T.~J.~Hollowood and J.~L.~Miramontes,
\textit{Restoring unitarity in the {q}-deformed world-sheet {S}-matrix},
\textsf{JHEP~1310,~050~(2013)},
\texttt{\arxivref{1303.1447}{arxiv:1303.1447}}.
%
%%CITATION = 0803.4323;%%
\bibitem[Arutyunov and Frolov(2008)]{Arutyunov:2008zt}
G.~Arutyunov and S.~Frolov,
\textit{The {S}-matrix of string bound states},
\textsf{Nucl.~Phys.~B804,~90~(2008)},
\texttt{\arxivref{0803.4323}{arxiv:0803.4323}}.
%
%%CITATION = ARXIV:1907.05430;%%
\bibitem[Seibold(2019)]{Seibold:2019dvf}
F.~K.~Seibold,
\textit{Two-parameter integrable deformations of the {$\AdS_3 \times \Sphere^3
  \times \Torus^4$} superstring},
\texttt{\arxivref{1907.05430}{arxiv:1907.05430}}.
%
\bibitem[Seibold \textit{et~al.\@}(2021)Seibold, van Tongeren, and
  Zimmermann]{Seibold:2021lju}
F.~K.~Seibold, S.~J.~van~Tongeren and Y.~Zimmermann,
\textit{{On quantum deformations of AdS$_{3}$ \texttimes{} S$^{3}$ \texttimes{}
  T$^{4}$ and mirror duality}},
\textsf{JHEP~2109,~110~(2021)},
\texttt{\arxivref{2107.02564}{arxiv:2107.02564}}.
%
\bibitem[Hoare \textit{et~al.\@}(2022)Hoare, Seibold, and
  Tseytlin]{Hoare:2022asa}
B.~Hoare, F.~K.~Seibold and A.~A.~Tseytlin,
\textit{Integrable supersymmetric deformations of {$\AdS_3 \times \Sphere^3
  \times \Torus^4$}},
\textsf{JHEP~2209,~018~(2022)},
\texttt{\arxivref{2206.12347}{arxiv:2206.12347}}.
%
\bibitem[Hoare \textit{et~al.\@}(2024)Hoare, Retore, and
  Seibold]{Hoare:2023zti}
B.~Hoare, A.~L.~Retore and F.~K.~Seibold,
\textit{{Elliptic deformations of the AdS$_{3}$ \texttimes{} S$^{3}$
  \texttimes{} T$^{4}$ string}},
\textsf{JHEP~2404,~042~(2024)},
\texttt{\arxivref{2312.14031}{arxiv:2312.14031}}.
%
%%CITATION = ARXIV:1512.08761;%%
\bibitem[Abbott and Aniceto(2016)]{Abbott:2015pps}
M.~C.~Abbott and I.~Aniceto,
\textit{Massless {L}{\"u}scher terms and the limitations of the {$\AdS_3$}
  asymptotic {B}ethe ansatz},
\textsf{Phys.~Rev.~D93,~106006~(2016)},
\texttt{\arxivref{1512.08761}{arxiv:1512.08761}}.
%
\bibitem[Cavagli\`a \textit{et~al.\@}(2022)Cavagli\`a, Ekhammar, Gromov, and
  Ryan]{Cavaglia:2022xld}
A.~Cavagli\`a, S.~Ekhammar, N.~Gromov and P.~Ryan,
\textit{Exploring the quantum spectral curve for {AdS${}_3$/CFT${}_2$}},
\texttt{\arxivref{2211.07810}{arxiv:2211.07810}}.
%
%%CITATION = ARXIV:1812.01007;%%
\bibitem[Eberhardt \textit{et~al.\@}(2018)Eberhardt, Gaberdiel, and
  Gopakumar]{Eberhardt:2018ouy}
L.~Eberhardt, M.~R.~Gaberdiel and R.~Gopakumar,
\textit{The worldsheet dual of the symmetric product {CFT}},
\texttt{\arxivref{1812.01007}{arxiv:1812.01007}}.
%
%%CITATION = ARXIV:1411.3676;%%
\bibitem[Ohlsson~Sax \textit{et~al.\@}(2015)Ohlsson~Sax, Sfondrini, and
  Stefa{\'n}ski]{Sax:2014mea}
O.~Ohlsson~Sax, A.~Sfondrini and B.~Stefa{\'n}ski,~jr.,
\textit{Integrability and the conformal field theory of the {H}iggs branch},
\textsf{JHEP~1506,~103~(2015)},
\texttt{\arxivref{1411.3676}{arxiv:1411.3676}}.
%
\end{thebibliography}

\end{document}